%% file: Independence-RKHS-Arxiv.tex
\documentclass{article}
\usepackage{float}
\usepackage{numprint}
\usepackage{booktabs}

\usepackage{lscape}
\usepackage{pgfplotstable}
\usepackage{booktabs}
\usepackage{multirow}
\usepackage{amsthm,amssymb, amsmath} 
\usepackage{color}
\usepackage{ifthen}
\usepackage{xspace}
\usepackage[utf8]{inputenc}
\usepackage[ruled,vlined,linesnumbered]{algorithm2e}
\usepackage{dsfont}
\usepackage{url}
\usepackage{braket}
\usepackage{tikz}
\usetikzlibrary{arrows}
\usepackage{geometry}
\usepackage{float}
\usepackage{caption}
\usepackage{subcaption}
\usepackage{enumitem}
\usepackage{graphicx}
\usepackage{centernot}
\usepackage{mathtools}

\newcommand\tr{\operatorname{trace}}
\usepackage{geometry}
\input{defsProb}

\theoremstyle{assumption}
\newtheorem{assumption}[theorem]{Assumption}

\begin{document}

\title{A kernel log-rank test of independence for right-censored data}

\author{Tamara Fern\'andez\footnote{corresponding author}\\ 
  University College London\\
    Universidad Adolfo Ib\'a\~nez\\
  \texttt{t.a.fernandez.aguilar@gmail.com} \\
   \and
   Arthur Gretton \\
  University College London\\
   \texttt{a.gretton@ucl.ac.uk} \\
   \and
   David Rindt \\
  University of Oxford\\
   \texttt{david.rindt@magd.ox.ac.uk} \\
   \and
   Dino Sejdinovic \\
  University of Oxford\\
  \texttt{dino.sejdinovic@stats.ox.ac.uk} \\
}

\maketitle

\begin{abstract}
We introduce a general non-parametric independence test between right-censored survival times and covariates, which may be multivariate. Our test statistic has a dual interpretation, first in terms of the supremum of a potentially infinite collection of weight-indexed log-rank tests, with weight functions belonging to a reproducing kernel Hilbert space (RKHS) of functions; and second, as the norm of the difference of
embeddings of certain finite measures into the RKHS, similar to the Hilbert-Schmidt Independence Criterion (HSIC) test-statistic. We study the asymptotic properties of the test, finding sufficient conditions to ensure our test correctly rejects the null hypothesis under any alternative. The test statistic can be computed straightforwardly, and the rejection threshold is obtained via an asymptotically consistent Wild Bootstrap procedure. Extensive investigations on both simulated and real data suggest that our testing procedure generally performs better than competing approaches in detecting complex non-linear dependence.
\end{abstract}

\section{Introduction}

Right-censored data appear in survival analysis and reliability theory, where the time-to-event variable one is interested in modelling may not be observed fully, but only in terms of a lower bound. This is a common occurrence in clinical trials as, usually, the follow-up is restricted to the duration of the study, or patients may decide to withdraw from the study. 

An important task when dealing with such data is to  test independence between the survival times and the covariates. For instance, in a clinical trial setting, we may wish to test if the survival times differ across treatments, e.g., chemotherapy vs radiation; or if there is dependence on other covariates, such as ages of the patients, gender, or any other measured variables. The main challenge of testing independence in this setting is that we need to deal with censored observations, where the censoring mechanism may be dependent on the covariates while the time of interest may not. E.g., patients' withdrawal times from a study can be associated to their gender even if gender is independent of the survival time. 

The problem of  testing independence has been widely studied by the statistical community. In the context of right-censored data, this problem has often been addressed through the mechanism of  two-sample tests, in which the covariate takes one out of two possible values. For the two-sample problem, the main tool is the log-rank test \cite{mantel1966evaluation,peto1972asymptotically} and its generalizations, namely weighted log-rank tests \cite{harrington1982class,Bagdonavivius2010,fernandez2019reproducing}. The more general case, in which the covariates belong to $\R^d$, is much more challenging, and most of the current approaches are ad-hoc for specific semi-parametric models, e.g. \cite{cox1972regression, gray1992,zucker1990}. In particular, the most popular of these approaches is the Cox proportional hazards model \cite{cox1972regression}, which assumes a linear effect of the covariates on the log-hazard function.  Non-parametric approaches are more scarce, however \cite{le1994association,mckeague1995omnibus,rindt2019nonparametric}. In \cite{le1994association}, a nonparametric test for independence is obtained by measuring monotonic relationships between a censored survival time and an ordinal covariate; and in \cite{mckeague1995omnibus}, the authors propose an omnibus test that can detect any type of association between a censored survival time and a 1-dimensional covariate. The recent non-parametric test in \cite{rindt2019nonparametric} was introduced to deal with general covariates on $\R^d$. This approach deals with censored data by transforming it into uncensored samples, and then  applies a well-known kernel independence test based on the Hilbert-Schmidt Independence Criterion (HSIC) \cite{GreFukTeoSonetal08,ChwGre14}.

In this paper we propose a non-parametric test which can potentially detect any type of dependence between right-censored survival times and general covariates. Our testing procedure is based on a dependence measure between survival times and covariates which is constructed using weighted log-rank tests and the theory of reproducing kernel Hilbert spaces (RKHSs). We provide asymptotic results for our test-statistic and propose an approximation of the rejection region of our test by using a Wild Bootstrap procedure. Under mild regularity conditions, we prove that both the oracle and the testing procedure based on the Wild Bootstrap approximations are asymptotically consistent, meaning that our test can detect any type of dependence.

The closest prior works to our approach are \cite{fernandez2019reproducing} and \cite{rindt2019nonparametric}, which are both based on kernel methods.  In the former, the authors specifically address the two-sample problem for right-censored data. While the present test may be seen as related,  the two-sample analysis is quite different, and heavily relies on the binary nature of the covariates: the main results apply ad-hoc theory developed for log-rank tests, which is not available in our setting. As noted above, \cite{rindt2019nonparametric} bypass the problem of right-censored data by transforming it into uncensored samples, however this comes at the cost of losing considerable information in the data. By contrast, our approach deals directly with the censored observations without loss of  information, resulting in a major  performance advantage in practice, as we demonstrate in our experiments.

The paper is structured as follows. In Section \ref{sec:Notation} we introduce relevant notation. In Section \ref{sec:testConstruction} we define the kernel log-rank test and show that it can be interpreted as i) the supremum of a collection of score tests associated to a particular family of cumulative hazard functions, and, ii) as an RKHS distance, revealing a similarity with the HSIC \cite{gretton2012kernel}. In Section \ref{sec:asymptotics} we study the asymptotic behavior of our statistic under both the null and alternative hypothesis, and we establish connections with known approaches such as the two-sample test proposed in \cite{fernandez2019reproducing}, and the Cox score test. Section \ref{sec:wildbootstrap} shows how to effectively approximate the null distribution by using Wild Bootstrap \cite{dehling1994random}. Sections \ref{sec:experiments} and \ref{sec:real_data} contain extensive experiments investigating the performance of the kernel log-rank test on a range of synthetic and real datasets.

\section{Notation}\label{sec:Notation}

\paragraph{Survival analysis notation} Let $((Z_i,C_i,X_i))_{i\in[n]}$ be a collection of random variables taking values on $\R_+\times\R_+\times\R^d$, where $Z_i$ denotes a survival time of interest, $C_i$ is a censoring time, and $X_i$ is a vector of covariates taking values on $\R^d$, and $d\geq 1$. In practice, we do not observe $(Z_i,C_i,X_i)$ directly, but instead observe triples $(T_i,\Delta_i,X_i)$, where $T_i=\min\{Z_i,C_i\}$ and $\Delta_{i}=\ind_{\{T_i=Z_i\}}$. This type of data is known as right-censored data. Additionally, we assume $Z \perp C|X$, which is known as independent right-censoring.

We denote by $F_T$, $F_Z$, $F_C$ and $F_X$, the marginal distribution functions associated to $T$, $Z$, $C$ and $X$, respectively. We use standard notation to denote joint and conditional distributions, e.g. $F_{ZC|X=x}$ denotes the joint distribution of $Z$ and $C$ conditional on $X=x$. We denote by $S_T(t)=1-F_T(t)$, $S_Z(t)=1-F_{Z}(t)$ and $S_C(t)=1-F_{C}(t)$ the marginal survival functions associated to $T$, $Z$ and $C$, respectively, and by $S_{T|X=x}(t)=1-F_{T|X=x}(t)$, $S_{Z|X=x}(t)=1-F_{Z|X=x}(t)$ and $S_{C|X=x}(t)=1-F_{C|X=x}(t)$, the respective survival functions conditioned on $X=x$. In this work, we assume that $Z|X=x$ and $C|X=x$ are continuous random variables for almost all $x\in\R^d$, with densities denoted by $dF_{Z|X=x}(t)$ and $dF_{C|X=x}(t)$ respectively. We further assume that $Z$ and $C$ are proper random variables, meaning that  $\Prob(Z<\infty|X=x)=1$ and $\Prob(C<\infty|X=x)=1$ for almost all $x\in\R^d$. The marginal cumulative hazard function of $Z$ is defined as $\Lambda_{Z}(t)=\int_0^t S_Z(s)^{-1}dF_{Z}(s)$. Similarly, the  conditional cumulative hazard of $Z$ given $X=x$ is $\Lambda_{Z|X=x}(t)=\int_0^t S_{Z|X=x}(s)^{-1}dF_{Z|X=x}(s)$. We define $\tau_n=\max\{T_1,\ldots,T_n\}$, $\tau_x=\sup\{t:S_{T|X=x}(t)>0\}$ and $\tau=\sup \left\{t:S_T(t)>0\right\}$; note that $\tau_n\overset{a.s.}{\to}\tau$.  

\paragraph{Counting processes notation} We use standard survival analysis/counting processes notation. For $i\in[n]$, we define the individual and pooled counting processes by $N_i(t)=\Delta_i\ind_{\{T_i\leq t\}}$ and  $N(t)=\sum_{i=1}^nN_i(t)$, respectively. Similarly, we define the individual and pooled risk functions by $Y_i(t)=\ind_{\{T_i\geq t\}}$ and $Y(t)=\sum_{i=1}^n Y_i(t)$.

We assume that all our random variables take values on a common filtrated probability space $(\Omega,\mathcal{F},(\mathcal{F}_t)_{t\geq 0},\Prob)$, where the sigma-algebra $\mathcal{F}_t$ is generated by $\left\{\ind_{\{T_i\leq s,\Delta_i=0\}},\ind_{\{T_i\leq s,\Delta_i=1\}},X_i:s\leq t,i\in [n]\right\},$ and the $\Prob$-null sets of $\mathcal{F}$. Under the null hypothesis, for $i\in [n]$, we define the individual and pooled $(\mathcal{F}_t)$-martingales, $M_i(t)=N_i(t)-\int_{(0,t]}Y_i(s)d\Lambda_Z(s)$ and $M(t)=N(t)-\int_{(0,t]}Y(s)d\Lambda_Z(s)$, respectively. Finally, we denote by $d\widehat{\Lambda}(t)=dN(t)/Y(t)$ the Nelson Aalen estimator of $d\Lambda_Z(t)$ under the null hypothesis. For more information about counting processes martingales, we refer the reader to Fleming and Harrington  \cite[Chapters 1 and 2]{Flemming91}.

In this work $\int_a^b$ means integration over $(a,b]$ unless $b=\tau$, in which case we integrate over $(a,\tau)$. Due to the simple nature of the martingales that appear in this work (which arise from counting processes), properties such as (squared-)integrability of these processes are trivial, and thus we state them without  proof. Also, note that for any $t>\tau_n$, it holds that $N(t)=N(\tau_n)$ and $M(t)=M(\tau_n)$. Hence $\int_{\R_+}g(t)dN(t)=\int_{0}^{\tau}g(t)dN(t)=\int_{0}^{\tau_n}g(t)dN(t)$; the same holds for the martingale $M$.

For simplicity of exposition and notation we assume $X\in \R^d$, however our results also apply straightforwardly to general covariate spaces, as our statistic is based on kernel functions that may be defined on more general domains: see next section. 

\section{Construction of  the test}\label{sec:testConstruction}
We are interested in testing if the failure times $Z$ are independent of the covariates $X$. Specifically, we would like to test the null hypothesis,
\begin{align*}
H_0: F_{ZX}=F_ZF_X,\quad\text{against}\quad H_1: F_{ZX}\neq F_ZF_X.
\end{align*}

One of the most popular approaches to solve this problem is the log-rank test for proportional hazard functions. This test can be obtained as a score test from a partial likelihood function for the Cox's proportional hazards model given by $\Lambda_{Z|X=x}(t)=e^{\beta^\intercal x}\Lambda_Z(t)$. This approach fails in many scenarios, however, since it only considers a linear effect of the covariates on the log hazard, which is given by the term  $\beta^\intercal x$. 

Our method generalizes the previous method by defining a general collection of log-rank tests in which the association between time and covariates is modeled through general functions $\omega(t,x)$, instead of the simple expression $\beta^\intercal x$. 

\paragraph{General score test} We obtain a general log-rank test, for a fixed function $\omega:\R_+\times\R^d\to\R$, by computing the score test associated to the model defined in terms of the conditional cumulative hazard function, 
\begin{align}
\Lambda_{Z|X=x}(t;\theta,\omega)&=\int_0^{t}e^{\theta\omega(s,x)}d\Lambda_Z(s)\quad \theta\in \Theta,\label{eqn:model}
\end{align}
where $\omega:\R_+\times\R^d\to \R$ is some non-zero fixed function, $\Theta$ is an open subset of $\R$ containing $\theta=0$, and $\Lambda_Z(s)$ is the marginal (baseline) cumulative hazard function associated to the failure time $Z$. Under the assumption that our data is generated by this model for some fixed function $\omega(t,x)$, testing the null hypothesis $H_0: Z\perp X$ is equivalent to testing $H_0:\theta=0$,  which can be done using a score test.

A score test is a hypothesis test used to check whether a restriction imposed on a model estimated by maximum likelihood is violated by the data. The score test assesses the gradient of the log-likelihood function, known as score function, evaluated at some parameter $\theta$ under the null hypothesis. Intuitively,  if the maximizer of the log-likelihood function is close to 0, the score, evaluated at $\theta=0$, should not differ from zero by more than sampling error. 

The likelihood function associated to the right-censored data $(T_i,\Delta_i,X_i)_{i=1}^n$ can be computed as follows. Given $X_i$, the contribution to the likelihood of an uncensored observation $(T_i,\Delta_i)$ (that is, for $\Delta_i=1)$ is $dF_{Z|X_i}(T_i)=d\Lambda_{Z|X_i}(T_i)S_{Z|X_i}(T_i)$. When $(T_i,\Delta_i)$ is censored, $\Delta_i=0$, the contribution corresponds to  $S_{Z|X_i}(T_i)$. The latter follows from the fact that when $T_i$ is censored, we only know that $Z_i> T_i$. 

Thus, given the covariates $(X_i)_{i=1}^n$, the likelihood function for the data $(T_i,\Delta_i)_{i=1}^n$ under the model in Equation \eqref{eqn:model} corresponds to 
\begin{align*}
L_n(\theta;\omega)&=\prod_{i=1}^nd\Lambda_{Z|X_i}(T_i)^{\Delta_i}S_{Z|X_i}(T_i)\\
&=\prod_{i=1}^n e^{\theta\Delta_i\omega(T_i,X_i)}d\Lambda_{Z}(T_i)^{\Delta_i}\exp\left\{-\int_{0}^{T_i}e^{\theta\omega(s,X_i)}d\Lambda_{Z}(s)\right\},
\end{align*}
where the second  equality follows since $S_{Z|X}(t)=\exp\left\{-\int_{0}^td\Lambda_{Z|X}(s)\right\}$. 

The score function is then defined as
\begin{align*}
U_n(\theta;\omega)&=\frac{d}{d\theta}\log L_n(\theta;\omega)=\sum_{i=1}^n\left(\Delta_i\omega(T_i,X_i)-\int_0^{T_i}\omega(t,X_i)e^{\theta\omega(t,X_i)}d\Lambda_Z(t)\right),
\end{align*}
and $U_n(0,\omega)$ is the score statistic associated to the null hypothesis, $H_0:\theta=0$. A normalized version of $U_n(0,\omega)$ can be obtained using the variance/covariance matrix of $U_n(\theta;\omega)$, written as $\Sigma(\theta;\omega)=\E(-\frac{\partial^2}{\partial\theta^2}\log L_n(\theta;\omega))$, and then writing $S_n(0;\omega)=U_n(0;\omega)^{\intercal}\Sigma(0;\omega)^{-1}U_n(0;\omega)$. By the Neyman-Pearson Lemma \cite{neyman1933ix}, it follows that the test based on $S_n(0;\omega)$ is the most powerful test for small deviations from the null under the model defined in Equation \eqref{eqn:model}. 

In general the marginal hazard function $d\Lambda_Z(s)$ is unknown, and thus $U_n(0;\omega)$ cannot be evaluated in practice. However, under the null, $d\Lambda_{Z}(s)$ can be estimated from the data using the Nelson-Aalen estimator \cite{aalen1978nonparametric} $d\widehat{\Lambda}_Z(t)=dN(t)/Y(t)$, yielding
\begin{align}
\widehat{U}_n(0;\omega)&=\sum_{i=1}^n\int_{\R_+}(\omega(t,X_i)-\bar\omega_n(t))dN_i(t),\label{eqn:score}
\end{align}
where  $\bar\omega_n(t)=\sum_{j=1}^n\omega(t,X_j)Y_j(t)/Y(t)$.

\paragraph{Log-rank formulation}
The expression for the un-normalized score statistic given in Equation \eqref{eqn:score} can be written as a discrepancy between two empirical measures with respect to the weight function $\omega(t,x)$. In survival analysis terminology, this is known as weighted log-rank test, and, in our scenario, it takes the form
\begin{align}
\text{LR}_n(\omega)=\frac{1}{n}\widehat{U}_n(0;\omega)&=\int_{\R_+}\int_{x\in\R^d}\omega(t,x)\left(d\nu_1^n(t,x)-d\nu_0^n(t,x)\right)\label{eqn:logrank},
\end{align}
where  $\nu_1^n$ and $\nu_0^n$ are empirical measures defined as
\begin{align}
d\nu_1^n(t,x)=\frac{1}{n}\sum_{i=1}^ndN_{i}(t)\delta_{X_i}(x)=\frac{1}{n}\sum_{i=1}^n\Delta_i\delta_{T_i,X_i}(t,x)
\label{eqn:nu_n measures}
\end{align}
and 
\begin{align}
d\nu_0^n(t,x)&=\frac{dN(t)}{n}\sum_{i=1}^n\frac{Y_i(t)}{Y(t)}\delta_{X_i}(x)=\frac{1}{n}\sum_{j=1}^n\Delta_j\delta_{T_j}(t)\sum_{i=1}^n\frac{Y_i(t)}{Y(t)}\delta_{X_i}(x).\label{eqn:nu_n measures0}
\end{align}

The next theorem gives a consistency limit result for $\text{LR}(\omega)$.
\begin{theorem}\label{thm:Logrankconistency}
Let $\omega:\R_+\times\R^d\to\R$ be a bounded measurable function. Then
\begin{align*}
\text{LR}_n(\omega)&\overset{\Prob}{\to} \int_{\R^d} \int_{0}^{\tau} \omega(s,x)(d\nu_1(s,x)-d\nu_0(s,x)),
\end{align*}
where $\nu_0$ and $\nu_1$ are defined by $d\nu_1(t,x)=S_{C|X=x}(t)dF_{ZX}(t,x)$, $d\nu_0(t,x)=S_{T|X=x}(t)d\alpha(t)dF_X(x)$ and $\alpha(I)=\int_{I}\int_{\R^d}S_{C|X=x}(t)/S_T(t)dF_{ZX}(t,x)$ for any measurable $I\subseteq (0,\tau)$.
\end{theorem}
Simple algebra shows that, under the null hypothesis (i.e., $H_0:Z\perp X$), $\nu_1=\nu_0$, and consequently $\text{LR}_n(\omega)\overset{\Prob}{\to} 0$ for any weight function $\omega(t,x)$. Under some regularity conditions which we state in Assumption \ref{assumption:no_complete_censoring}, we prove that $\nu_1=\nu_0$ implies $Z\perp X$. 

\begin{assumption}\label{assumption:no_complete_censoring}
For almost all $x\in\R^d$, $S_{C|X=x}(t)=0$ implies $S_{Z|X=x}(t)=0$. 
\end{assumption}
\begin{proposition}\label{Prop: nu iff indep} Under Assumption \ref{assumption:no_complete_censoring}, it holds that $\nu_1= \nu_0$  if and only if $Z\perp X$.
\end{proposition}  
Note that $\text{LR}_n(\omega)\overset{\Prob}{\to} 0$ does not necessarily imply $\nu_0=\nu_1$, since, if we choose $\omega$ equal to the zero function, then $\text{LR}_n(\omega)=0$ trivially. Thus, when using log-rank tests, it is very important to use a relevant weight function for the problem at hand.  Instead of choosing a single weight function, we propose to optimize over a large collection of candidate functions.

\paragraph{RKHS approach}
While normalized log-rank tests exhibit good statistical properties for small deviations from alternatives belonging to the model in Equation $\eqref{eqn:model}$, this good behavior is only guaranteed for a single weight function $\omega$ at a time.  In practice, it is very unlikely that the dependence structure of $Z$ and $X$ is known beforehand, and thus choosing the correct weight $\omega(t,x)$ (if it exists) seems hard. 

In order to avoid choosing a particular weight $\omega(t,x)$ in advance, we consider a family of weighted log-rank statistics, and compute  
\begin{align}
\Psi_n^2&=\left(\sup_{\omega\in\mathcal{H}:\|\omega\|^2_{\mathcal{H}}\leq 1}\text{LR}_n(\omega)\right)^2,\label{eqn:RKHS test}
\end{align} 
where the function $\omega(t,x)$ is allowed to take values in a potentially infinite-dimensional space of functions $\mathcal{H}$. We refer to $\Psi_n^2$ as the \emph{kernel log-rank} statistic. 

In particular, we choose $\mathcal{H}$ as a reproducing kernel Hilbert space (RKHS) of functions. One of the main advantages of choosing this particular space, is that it gives a simple closed-form solution for the optimization problem of Equation \eqref{eqn:RKHS test}. For general spaces of functions, finding the function $\widehat\omega$ that maximizes the likelihood function, or solving the optimization problem of Equation \eqref{eqn:RKHS test}, might be much harder problem, as it is likely that $\widehat\omega$ does not have a closed-form solution. We will prove that, under some mild regularity assumptions, a sufficiently rich choice of RKHS will be able to detect any type of dependencies. Comparing with works that consider a maximum among normalised log-rank statistics (i.e., divided by the standard deviation) \cite{kossler2010max,tarone1981distribution,gares2015omnibus}, our test will use the un-normalised statistic $\text{LR}_n(\omega)$. This is fundamental to our result, as the linearity in $\omega$ of $\text{LR}_n(\omega)$, combined with the properties of the RKHS, leads to a simple closed formula to evaluate $\Psi^2_n$. This being said, note that we are indirectly normalizing by choosing $\omega$ in the unit ball of $\mathcal H$.

\paragraph{Reproducing Kernel Hilbert Spaces} An  RKHS $(\mathcal{H},\langle\cdot,\cdot\rangle_{\mathcal{H}})$ is a Hilbert space of functions in which the evaluation operator is continuous. By the Riesz representation theorem, for all $(t,x)\in\R_+\times\R^d$ there exists a unique element $\mathfrak{K}_{(t,x)}\in\mathcal{H}$ such that, for all $\omega\in\mathcal{H}$, it holds $\omega(t,x)=\langle\omega,\mathfrak{K}_{(t,x)}\rangle_{\mathcal{H}}$; this property is known as the \emph{reproducing property}. We define the so-called reproducing kernel $\mathfrak{K}:(\R_+\times\R^d)^2\to\R$ as $\mathfrak{K}((t,x),(t',x'))=\langle \mathfrak{K}_{(t',x')},\mathfrak{K}_{(t,x)}\rangle_{\mathcal{H}}$ for any $(t,x),(t',x')\in \R_+\times\R^d$. By the Moore-Aronszajn theorem, for any symmetric positive-definite kernel $\mathfrak{K}$, there exists a unique RKHS for which $\mathfrak{K}$ is its reproducing kernel. Finally, for any finite signed Radon measure (not necessarily a probability measure), we define its embedding into $\mathcal{H}$ as 
\begin{align*}
\phi_{\nu}(\cdot)&=\int_{\R_+}\int_{\R^d} \mathfrak{K}((t,x),\cdot)d\nu(t,x)\in\mathcal{H},
\end{align*}
which existence is guaranteed by $\int_{\R_+}\int_{\R^d}\sqrt{\mathfrak{K}((t,x),(t,x))}d\nu(t,x)<\infty$, see \cite{berlinet2011reproducing}.

\paragraph{RKHS distance}  We define the embeddings of the empirical measures $\nu_1^n$ and $\nu_0^n$ (introduced in Equations \eqref{eqn:nu_n measures} and \eqref{eqn:nu_n measures0}) into $\mathcal{H}$ with reproducing kernel $\mathfrak{K}$ as
\begin{align}
\phi_1^n(\cdot)=\int_{0}^{\tau}\int_{\R^d} \mathfrak{K}((t,x),\cdot)d\nu^n_1(t,x)\quad\text{and}&\quad\phi_0^n(\cdot)=\int_{0}^{\tau}\int_{\R^d} \mathfrak{K}((t,x),\cdot)d\nu^n_0(t,x),\label{eqn:emmbeddings}
\end{align}
respectively. Notice both $\phi_1^n$ and $\phi_0^n$ are well-defined elements of $\mathcal{H}$, as they are finite sums of elements of $\mathcal{H}$. 

The next Theorem gives a closed-form expression for the kernel log-rank statistic in terms of the distance (induced by the norm) of the embeddings $\phi_0^n$ and $\phi_1^n$. 
\begin{theorem}\label{Thm:closed form}
\begin{align} \label{eqn:kernel_logrank_rkhs_distance}
\Psi_n^2
&=\|\phi_0^n-\phi_1^n\|_{\mathcal{H}}^2=\frac{1}{n^2}\sum_{i=1}^n\sum_{j=1}^n\Delta_i\Delta_j\bar{\mathfrak{K}}_n((T_i,X_i),(T_j,X_j)),
\end{align}
where
\begin{align*}
\bar{\mathfrak{K}}_n((t,x),(t',x'))
&=\mathfrak{K}((t,x),(t',x'))-\sum_{k=1}^n \mathfrak{K}((t,x),(t',X_k))\frac{Y_k(t')}{Y(t')}\\
&\quad-\sum_{j=1}^n \mathfrak{K}((t,X_j),(t',x'))\frac{Y_j(t)}{Y(t)}+\sum_{j,k=1}^n\mathfrak{K}((t,X_j),(t',X_k))\frac{Y_j(t)}{Y(t)}\frac{Y_k(t')}{Y(t')}.
\end{align*}
Moreover,  if $\mathfrak{K}((t,x),(t',x'))=L(t,t')K(x,x')$, then
\begin{align}
\Psi_n^2=\|\phi_0^n-\phi_1^n\|_{\mathcal{H}}^2&=\frac{1}{n^2}\tr(\boldsymbol{L^{\Delta}}(\boldsymbol{I}-\boldsymbol{A})\boldsymbol{K}(\boldsymbol{I}-\boldsymbol{A})^{\intercal})\label{eqn:Psi^2}
\end{align}
where $\boldsymbol{K}, \boldsymbol{L^{\Delta}}$ and $\boldsymbol A$ are $(n \times n)$-dimensional matrices whose entries $(i,j)$ are defined as $(\boldsymbol{K})_{i,j}=K(X_i,X_j)$, $(\boldsymbol{L^\Delta})_{i,j}=\Delta_i\Delta_jL(T_i,T_j)$ and $(\boldsymbol{A})_{i,j}=A_{ij}=\frac{Y_j(T_i)}{Y(T_i)}$, and $\boldsymbol{I}$ denotes the identity matrix.
\end{theorem}

\section{Asymptotic Analysis}\label{sec:asymptotics}
\paragraph{Asymptotic null distribution}
We study the asymptotic null distribution of $n\Psi_n^2$ which is fundamental to construct a testing procedure.  The key step is to show that we can rewrite $\Psi_n^2$ as a V-statistic plus an asymptotically negligible term. The asymptotic null  distribution of $n\Psi_n^2$ then follows from the standard theory of V-statistics. We refer to \cite[Section 5.5.2]{serfling2009approximation} for a discussion of V-statistics. 

\begin{proposition}\label{Prop:martingalerepre} Under the null hypothesis $H_0: Z\perp X$, the kernel log-rank statistic can be written as
\begin{align}
\Psi^2_n&=\frac{1}{n^2}\sum_{i=1}^n\sum_{j=1}^nJ_n((T_i,\Delta_i,X_i),(T_j,\Delta_j,X_j)),\label{eqn:MartingaleReprePsi}
\end{align}
where $J_n:(\R_+\times\{0,1\}\times\R^d)^2\to\R$ is a symmetric random function defined as
\begin{align*}
J_n((s,c,x),(s',c',x'))&=\int_{\R_+}\int_{\R_+}\bar{\mathfrak{K}}_n((t,x),(t',x'))dm_{s,c}(t)dm_{s',c'}(t),
\end{align*}
and $dm_{s,c}(t)=c\delta_s(t)-\ind_{\{s\geq t\}}d\Lambda_{Z}(t)$. 
\end{proposition}

The expression in Equation \eqref{eqn:MartingaleReprePsi} suggests a V-statistic representation for the kernel log-rank statistic. In the next result, we prove that $n\Psi_n^2$ can, indeed, be approximated by a $V$-statistic, by showing that $J_n$ can be replaced by its population version $J$, which follows from replacing the random kernel $\bar{\mathfrak{K}}_n$ by its corresponding population version $\bar{\mathfrak{K}}$, given by 
\begin{align}
&\bar{\mathfrak{K}}((t,x),(t',x'))\label{eqn:popkernel}\\
&=\int_{\R^d}\int_{\R^d}\left(\mathfrak{K}((t,x),(t',x'))-\mathfrak{K}((t,x),(t',y'))\frac{S_{C|X=y'}(t')}{S_C(t')}\right.\nonumber\\
&-\mathfrak{K}((t,y),(t',x'))\frac{S_{C|X=y}(t)}{S_C(t)}+\left.\mathfrak{K}((t,y),(t',y'))\frac{S_{C|X=y}(t)S_{C|X=y'}(t')}{S_C(t)S_C(t')}\right)dF_X(y)dF_X(y'),\nonumber
\end{align}
which is valid under the null  as $S_T(t)=S_Z(t)S_C(t)$.

\begin{assumption}
$\mathfrak{K}((t,x),(t',x'))=L(t,t')K(x,x')$, and both $K$ and $L$ are bounded.\label{Assu:BoundedKernelSep}
\end{assumption}

\begin{lemma}\label{lemma:rkhs approx}
Under Assumption \ref{Assu:BoundedKernelSep} and the null hypothesis, it holds
\begin{align*}
\Psi^2_n&=\frac{1}{n^2}\sum_{i=1}^n\sum_{j=1}^nJ((T_i,\Delta_i,X_i),(T_j,\Delta_j,X_j))+o_p(n^{-1}),
\end{align*}
where
\begin{align}
J((s,c,x),(s',c',x'))&=\int_{\R_+}\int_{\R_+}\bar{\mathfrak{K}}((t,x),(t',x'))dm_{s,c}(t)dm_{s',c'}(t).\label{eqn:Jkernel}
\end{align}
\end{lemma}
It can be easily checked that $\E(J((t,c,x),(T_1,\Delta_1,X_1)))=0$ for any $(t,c,x)\in \R_+\times\{0,1\}\times\R^d$ under the null (since $dm_{T_i,\Delta_i}(t)=dM_i(t)$). The statistic $\Psi_n^2$ is then approximately a degenerate V-statistic, and thus we deduce its limit distribution from the classical theory of degenerate V-statistics \cite[Section 5.5.2]{serfling2009approximation}.

\begin{theorem}\label{thm:rkhsLimit}
Under Assumption \ref{Assu:BoundedKernelSep} and the null hypothesis, it holds that
\begin{align*}
n\Psi^2_n\overset{\mathcal{D}}{\to}\int_{x\in\R^d}\int_{0}^\tau \bar{\mathfrak{K}}((t,x),(t,x))S_{C|X=x}(t)dF_Z(t)dF_X(x)+\mathcal{Y}
\end{align*}
where $\mathcal{Y}=\sum_{i=1}^\infty\lambda_i(\xi_i^2-1)$, $\xi_1,\xi_2,\ldots$ are i.i.d. standard normal random variables, and $\lambda_1,\lambda_2,\ldots$ are non-negative constants which depend on the distribution of the random variables $(Z,C,X)$ and the kernel $\mathfrak{K}$. 
\end{theorem}

The next result states that if we directly replace $\bar{\mathfrak{K}}_n$ by its limit $\bar{\mathfrak{K}}$ in Equation \eqref{eqn:kernel_logrank_rkhs_distance}, the resulting test-statistic has the same asymptotic null distribution as $n\Psi^2_n$. 

\begin{theorem}\label{Thm:SimplerKernel} $n\Psi_n^2$ and $\frac{1}{n}\sum_{i=1}^n\sum_{j=1}^n\Delta_i\Delta_j\bar{\mathfrak{K}}((T_i,X_i),(T_j,X_j))$ have the same asymptotic distribution under the null hypothesis. 
\end{theorem}

\paragraph{Power under alternatives}
We next analyze the asymptotic behavior of $\Psi_n^2$ under the alternative hypothesis, i.e., $H_1:F_{ZX}\neq F_{Z}F_{X}$. To this end, we first establish a consistency result for $\Psi^2_n$.

\begin{lemma}\label{lemma:consistency}
Under Assumption \ref{Assu:BoundedKernelSep}, it holds $\Psi_n^2\overset{\Prob}{\to}\|\phi_0-\phi_1\|^2_{\mathcal{H}}$, where
\begin{align*}
\phi_1(\cdot)=\int_0^\tau\int_{\R^d}K(\cdot,x)L(\cdot,t)d\nu_1(t,x)&\quad\text{and}\quad \phi_0(\cdot)=\int_0^\tau\int_{\R^d}K(\cdot,x)L(\cdot,t)d\nu_0(t,x),
\end{align*}
and $\nu_1$ and $\nu_0$ are the population measures defined in Theorem \ref{thm:Logrankconistency}. 
\end{lemma}
The next step is to ensure that $\|\phi_0-\phi_1\|^2_{\mathcal{H}}$ is zero if and only if the null hypothesis holds. This result will follow from assuming conditions on the kernel $\mathfrak{K}$ that ensure the embeddings of the measures $\nu_0$ and $\nu_1$ onto $\mathcal{H}$ are injective, and from Proposition \ref{Prop: nu iff indep}, which proves that $\nu_0=\nu_1$ if and only if the null hypothesis holds.  

\begin{theorem}\label{Thm:Power} Let $\gamma>0$ be any constant. Suppose that both $L$ and $K$, are bounded, continuous, characteristic \cite{sriperumbudur2011universality}, translation invariant, and $c_0$-kernels. Then, under the alternative hypothesis, and under Assumptions \ref{assumption:no_complete_censoring} and \ref{Assu:BoundedKernelSep}, $n\Psi^2_n\to \infty$ as $n$ grows to infinity, and thus 
\begin{align*}
\limsup_{n\to\infty}\Prob\left(n\Psi^2_n>\gamma\right)=1.
\end{align*}
\end{theorem}

In the previous result we say $K$ is a $c_0$-kernel if $K(x,\cdot)\in \mathcal{C}_0(\R^d)$, where $\mathcal{C}_0(\R^d)$ denotes the class of continuous functions in $\R^d$ that vanish at infinity. An example of a kernel that satisfies the conditions stated in the previous Theorem is the exponentiated quadratic kernel, given by $K(x,y)=\exp\{-(x-y)^\intercal \Sigma^{-1} (x-y)\}$.

Under the assumptions of Theorem \ref{Thm:Power}, our testing procedure has asymptotic power tending to one for any alternative, and thus \emph{it is able to  detect any type of dependency} between survival times and covariates, given enough observations. Even if the kernel does not satisfy the properties stated in Theorem \ref{Thm:Power}, however, we can guarantee the power of the test for alternatives following the model of Equation \eqref{eqn:model}, that is, for alternatives of the form $\Lambda_{Z|X=x}(t;\theta)=\int_0^t e^{\theta\omega^\star(s,x)}d\Lambda_Z(s)$ for some $\omega^\star\in\mathcal{H}$ on the unit ball and $\theta\neq 0$, as the log-rank statistic $\text{LR}_n(\omega^{\star})^2\to c>0$, with $c$ a positive constant. Then,  
\begin{align*}
\Psi_n^2=\left(\sup_{\omega\in\mathcal{H}:\|\omega\|^2_{\mathcal{H}}\leq 1}\text{LR}_n(\omega)\right)^2\geq \text{LR}_n(\omega^{\star})^2\to c,
\end{align*}
and thus when re-scaling by $n$, it holds that $n\Psi^2_n\to\infty$. 
\paragraph{Recovering existing tests} 
We show our approach can also recover certain known tests for specific choices of the kernel function. 

\begin{example}[Two-sample weighted log-rank test]\label{sec:twoSampleWeightedLogRank}
Consider $X\in\{0,1\}$, i.e., the two-sample problem. We can recover the standard weighted log-rank test with arbitrary weight function $\tilde{\omega}:\R_+\to\R$, by choosing $\omega(t,1)=-\omega(t,0)$ and $\omega(t,0)=\tilde\omega(t)/2$. Then, by replacing $\omega$ into Equation \eqref{eqn:logrank}, we obtain
\begin{align}
\text{LR}_n(\omega)&=\frac{1}{n}\int_0^\tau\tilde{\omega}(t)L(t)(d\widehat\Lambda^0(t)-d\widehat\Lambda^1(t)),
\end{align}
 where $d\widehat\Lambda^j$ denotes the Nelson-Aalen estimator for each group $j\in\{0,1\}$. Furthermore,  $\tilde{\Psi}_n=\sup_{\omega:\|\omega\|_{\mathcal{H}}= 1}\text{LR}_n(\omega)$ recovers the general test proposed in \cite{fernandez2019reproducing}.
\end{example}
\begin{example}[Cox proportional hazards model]\label{sec:fisherkernel}
Consider the Hilbert space of functions $\omega(t,x)=V^{1/2}\beta^\intercal x$, where $\beta\in \R^d$ and $V$ is a positive-definite matrix of length-scales.  By using this space of functions, our kernel log-rank statistic becomes
\begin{align}
\Psi_n=\sup_{\beta\in\R^d:\|V^{1/2}\beta\|^2\leq 1}\text{LR}_n(\beta),\label{eqn:Psi2Cox}
\end{align} 
and it can be computed using Equation \eqref{eqn:Psi^2} with a linear kernel on the covariates, $K(x,x')=(V^{1/2}x)^\intercal (V^{1/2}x')$, and a constant kernel on times, $L(t,t')=1$. Then
\begin{align*}
n\Psi^2_n&=\frac{1}{n}\sum_{i=1}^n\sum_{l=1}^n \int_{\R_+}\int_{\R_+} \bar{ \mathfrak K}_n((t,X_i),(t',X_l))dM_i(t)dM_l(t')=U_{\text{Cox}}(0)^\intercal V U_{\text{Cox}}(0),
\end{align*} 
where $U_{\text{Cox}}(0)=\left.\frac{d}{d\beta}l_{\text{Cox}}(\beta)\right|_{\beta=0}$ is the score function associated to the so-called \emph{Cox partial likelihood}  $l_{\text{Cox}}(\beta)$.  By choosing $V$ equal to the inverse of the Fisher information matrix, the Cox score test and our $\Psi_n$ are asymptotically equivalent.
\end{example}

\section{Wild Bootstrap}\label{sec:wildbootstrap}
In practice, the asymptotic null distribution is unknown, and thus we propose to use a Wild Bootstrap approximation to it. The Wild Bootstrap test-statistic $(\Psi_n^W)^2$ is given by
\begin{align*}
(\Psi_n^W)^2&
=\frac{1}{n^2}\sum_{i=1}^n\sum_{j=1}^nW_iW_j\Delta_i\Delta_j\bar{\mathfrak{K}}_n((T_i,X_i),(T_j,X_j)),
\end{align*}
where $W=(W_1\ldots,W_n)$ are a collection of i.i.d. Rademacher random variables, which are independent of the data $\mathcal{D}=\{(T_i,\Delta_i,X_i)\}_{i=1}^n$. 

In this section we prove two main results. The first result establishes that, under the null hypothesis, the asymptotic distribution of $n(\Psi_n^W)^2$ coincides with the asymptotic distribution of the kernel log-rank test-statistic $n\Psi_n^2$. The second result is analogous to Theorem \ref{Thm:Power}, but replacing $\gamma$ by the $1-\alpha$ quantile obtained by the Wild Bootstrap procedure.

\begin{lemma}\label{lemma:Wild1}
Under Assumption \ref{Assu:BoundedKernelSep}, it holds that
\begin{align}
(\Psi_n^W)^2=\frac{1}{n^2}\sum_{i=1}^n\sum_{j=1}^nW_iW_j\Delta_i\Delta_j\bar{\mathfrak{K}}((T_i,X_i),(T_j,X_j))+o_p(n^{-1}).
\label{eqn:WBPsi2N}
\end{align}
\end{lemma}

Our first main result is given in the following Theorem.
\begin{theorem}\label{thm:Wild1} Suppose that the null hypothesis and Assumption \ref{Assu:BoundedKernelSep} hold true, and let $\mathcal{L}$ denote the asymptotic distribution of $n\Psi^2_n$.  Then, for almost all sequences $(t_i,\delta_i,x_i)_{i\geq 1}$ sampled from $(T_i,\Delta_i,X_i)_{i\geq 1}$, 
$n(\Psi_n^W)^2\overset{\mathcal{D}}{\to}\mathcal{L},$
as $n\to\infty$.
\end{theorem}
The previous result guarantees $\lim_{n\to\infty}\Prob(n(\Psi^W)^2_n>Q_{1-\alpha})=\alpha$, where $Q_{1-\alpha}$ is the $1-\alpha$ quantile of $\mathcal{L}$. Note that $\mathcal L$ depends on the distribution $\theta$ of the triple $(X,Z,C)$ that defines the null. Changing this distribution to a distribution $\theta'$ (that still satisfies the null) will lead to a potentially different asymptotic distribution $\mathcal L'$ of the test-statistic. Therefore, the speed of convergence of the above result may depend on $\theta$. It is thus important to emphasize that our result ensures a pointwise asymptotic level, but not uniformly asymptotic level.

Our second main result is given in the following theorem.
\begin{theorem}\label{Thm:PowerW} Consider Assumptions \ref{assumption:no_complete_censoring} and \ref{Assu:BoundedKernelSep}, and assume that  both $L$ and $K$, are bounded, continuous, characteristic, translation invariant, and $c_0$-kernels. Let $\alpha\in(0,1)$, and let $Q^{W}_{n,M}$ denote the $1-\alpha$ quantile obtained from a  sample of fixed size $M$ of the Wild Bootstrap test-statistic, $n(\Psi^W_n)_1^2,\ldots,n(\Psi^W_n)_M^2$.  Then, under the alternative hypothesis
\begin{align*}
\Prob\left(n\Psi^2_n>Q_{n,M}^{W}\right)\to 1\quad\text{ as }n\to\infty.
\end{align*}
\end{theorem}

From the previous result, we deduce that, under the Assumptions of Theorem \ref{Thm:PowerW}, the test based on the Wild Bootstrap approximation of the null distribution is able to \emph{detect any type of dependency between survival times and covariates} asymptotically, as long as censoring does not hide the regions in which the dependence occurs.

\paragraph{Implementation} 

Under Assumption \ref{Assu:BoundedKernelSep}, the Wild Bootstrap test-statistic can be easily evaluated  as follows,
\begin{align}
n(\Psi_n^2)^{W}&=\frac{1}{n^2}\tr(\boldsymbol{L^{\Delta,W}}(\boldsymbol{I}-\boldsymbol{A})\boldsymbol{K}(\boldsymbol{I}-\boldsymbol{A})^{\intercal}),\label{eqn:Psi^2WB}
\end{align}
where $\boldsymbol{L^{\Delta,W}}$ is a $(n \times n)$-matrix defined as  $(\boldsymbol{L^{\Delta,W}})_{i,j}=(\Delta_iW_i)(\Delta_jW_j)L(T_i,T_j)$, and $\boldsymbol{K}, \boldsymbol{A}$ and $\boldsymbol{I}$ are defined in Theorem \ref{Thm:closed form}. Algorithm \ref{Algorithm} below describes the implementation of our testing procedure. 

\paragraph{Computational time}

By the following Proposition, our algorithm has the same computational complexity as the HSIC based permutation test of \cite{gretton2012kernel}. 

\begin{proposition}\label{proposition:computational_time}
$n\Psi_n^2$ and $n(\Psi_n^W)^2$ can be computed in $\mathcal{O}(n^2)$ time.
\end{proposition}

Using a simple Python implementation that does not use a GPU, running on a CPU with 4 cores at 1.6GHz, computation of the kernel log-rank statistic takes about 10 seconds for a sample of size 10000, and about 0.1 second for a sample of size 1000. If faster computation is required, we may adopt the large-scale approximations proposed in \cite{largescaleindependence}. Moreover, Wild Bootstrap statistics can be computed in parallel, and matrix computations can be done on a GPU.

\begin{flushleft}
\scalebox{0.9}{
\begin{algorithm}[H]
  \KwIn{data $\{T_i,\Delta_i,X_i\}_{i=1}^n$,  $\alpha$ and $M$}
  \For{k in 1 $\to$ M}{
  Sample $W=(W_1,\ldots,W_n)\overset{i.i.d.}{\sim}$ Rademacher \\
  Compute $(\Psi_n^W)^{2}_k$ as in equation \eqref{eqn:Psi^2WB}
}
Denote by $Q_{n,M}^{W}$ the $1-\alpha$ quantile of the sample $n(\Psi_n^W)^{2}_1,\ldots,n(\Psi_n^W)^{2}_M$\\
Compute  $n\Psi_n^2$ as in Equation \eqref{eqn:Psi^2}  \\
Reject if $n\Psi_n^2>Q_{n,M}^{W}$
\caption{Wild Bootstrap.}\label{Algorithm}
\end{algorithm}
}
\end{flushleft}

\section{Experiments}\label{sec:experiments}

We study the performance of the proposed kernel log-rank test for various choices of kernels.  We choose the kernels to be products of a kernel on the covariates, $K$, and a kernel on the times, $L$. We denote the product kernel by $(K,L)$. We study the following four cases: 1. $(K=\text{Lin},L=1)$, 2. $(\text{Gau},1)$, 3. $(\text{Fis},1)$ and 4. $(\text{Gau},\text{Gau})$, where $\text{``Lin"}$ denotes the linear kernel, $\text{``Gau"}$ the exponentiated quadratic (Gaussian) kernel, $\text{``Fis"}$ the linear kernel scaled by the Fisher information (see Example \ref{sec:fisherkernel}) and ``1" denotes the constant kernel, i.e. $L=1$ implies $L(t,s)=1$ for all $t,s\in\R_+$. In all experiments we use the \emph{median heuristic} to select the bandwidth of the Gaussian kernel: we  choose $\sigma^2=\text{median}\{||x_i-x_j ||^2 \ : i\neq j\}/2$. We discuss the sensitivity of the test to different choices of bandwidth later this section.  We set the level of the test to $\alpha=0.05$ and use Algorithm \ref{Algorithm} of Section \ref{sec:wildbootstrap} to perform the test with $M=2000$ (or $M=5000$ to estimate type 1 error) Wild Bootstrap samples to estimate the rejection region. We compare the kernel log-rank test with the traditional Cox likelihood ratio (Cox LR) test \cite{cox1972regression}, denoted by Cph in the legends, and the optHSIC test \cite{rindt2019nonparametric}, denoted by Opt in the legends. As in \cite{rindt2019nonparametric}, we use the Brownian covariance kernel in optHSIC. The assessment of the Type I error can be found in Appendix A.1. Code to implement the kernel log-rank test and reproduce the experiments below can be found at \url{https://github.com/davidrindt/kernel_logrank_python_code}.

\paragraph{Power for 1-dimensional covariates}\label{sec:power1dimensional}

\begin{figure}[t]
\centering
\begin{subfigure}{.5\textwidth}
  \centering
  \includegraphics[width=1\linewidth]{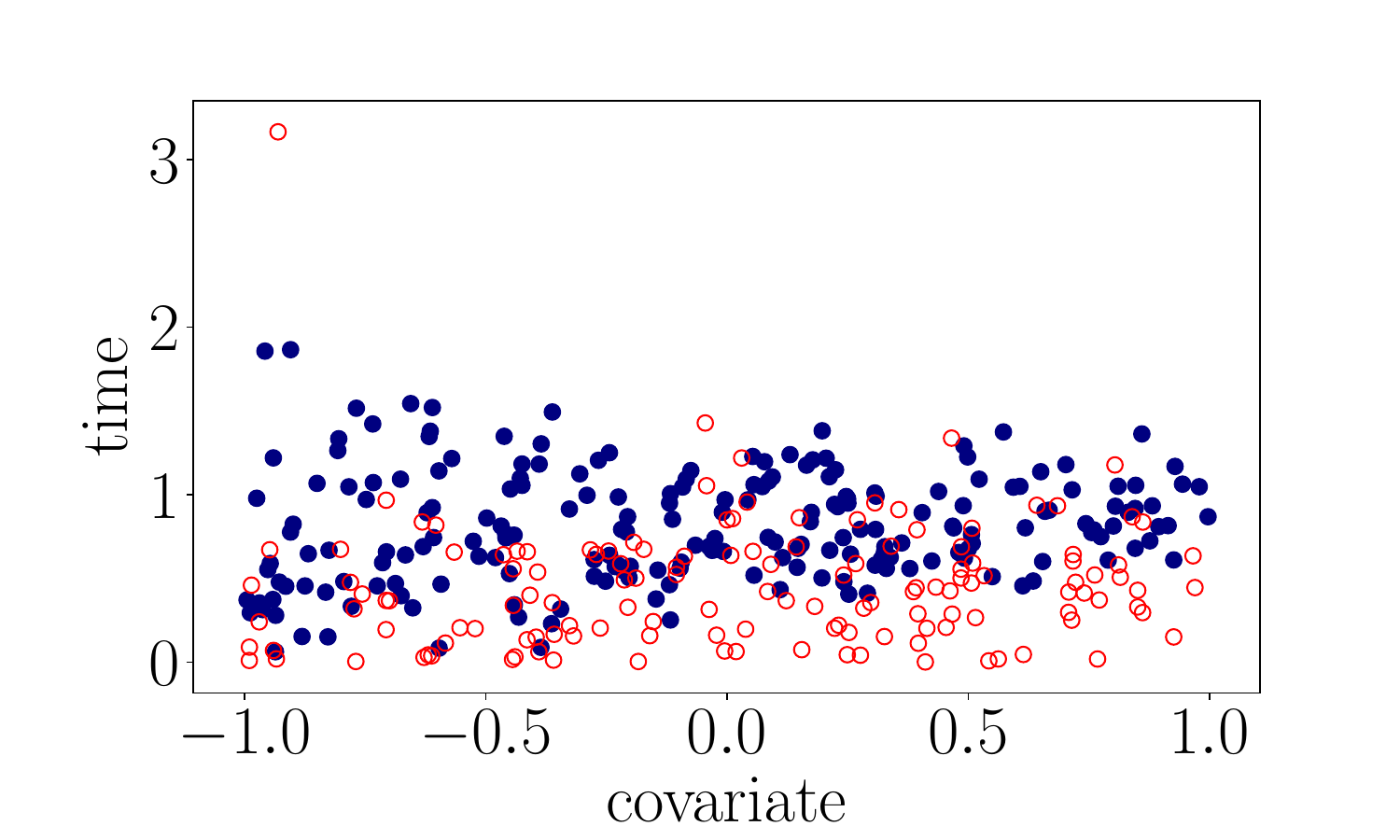}
\end{subfigure}%
\begin{subfigure}{.5\textwidth}
  \centering
  \includegraphics[width=1\linewidth]{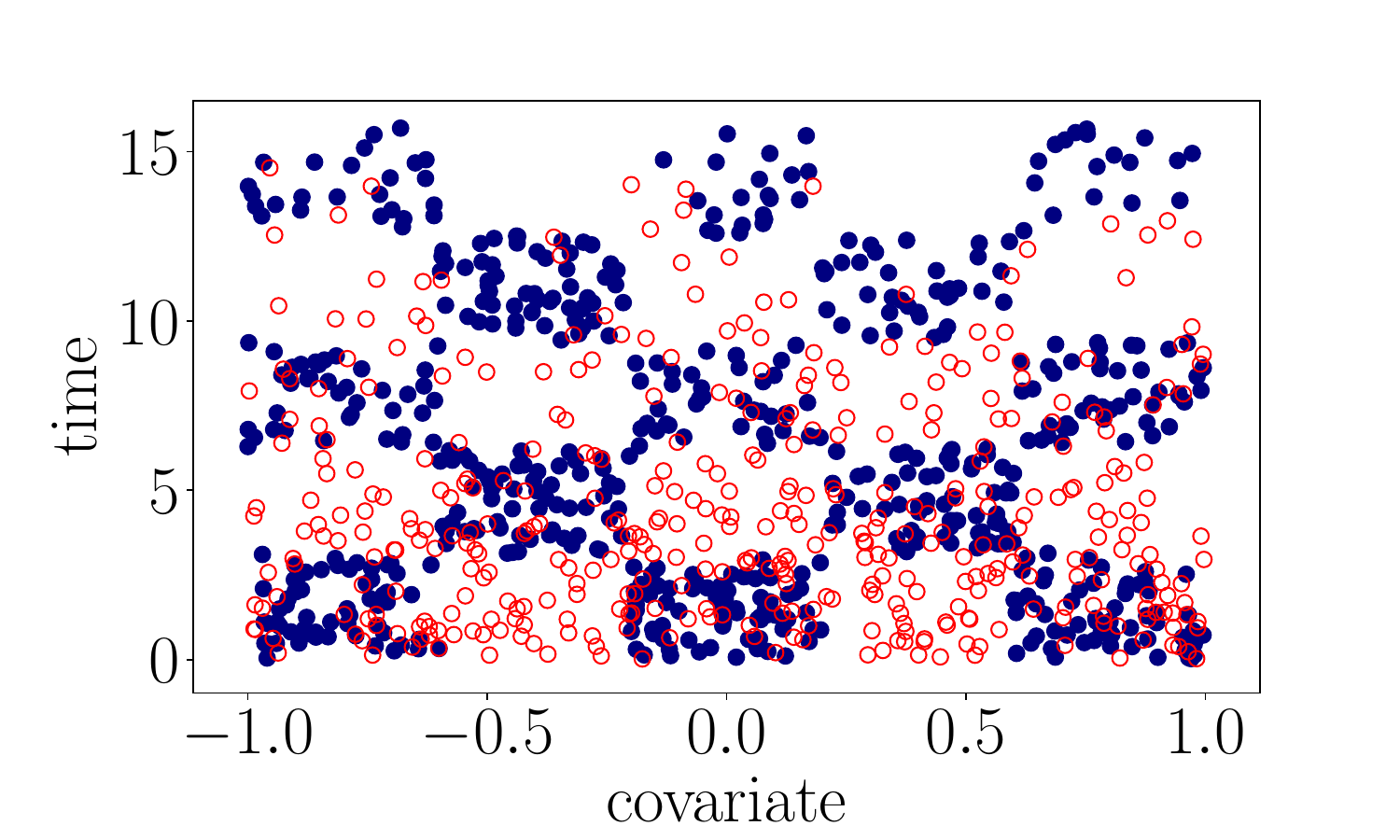}
\end{subfigure}%

\begin{subfigure}{.4\textwidth}
  \centering
  \includegraphics[width=1\linewidth]{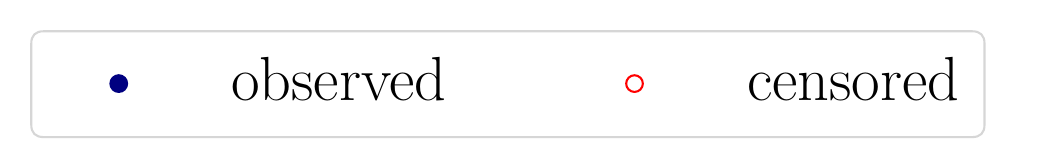}
\end{subfigure}
\caption{\label{plot:scatterplots} Scatterplots of samples from D.3 (left) and D.4 (right).}
\end{figure}

In this section we investigate the power of the different tests using data simulated from distributions in which $X \not \perp Z$. In each case, we use an exponential censoring distribution in which $C\perp X$ and the mean of $C$ is chosen such that $60\%$ of the events are observed. Throughout this subsection we let $    X \sim \text{Unif}[-1,1]$. Consider the following four distributions.

\begin{figure}[t]
\centering
\begin{subfigure}{.495\textwidth}
  \centering
  \includegraphics[width=1\linewidth]{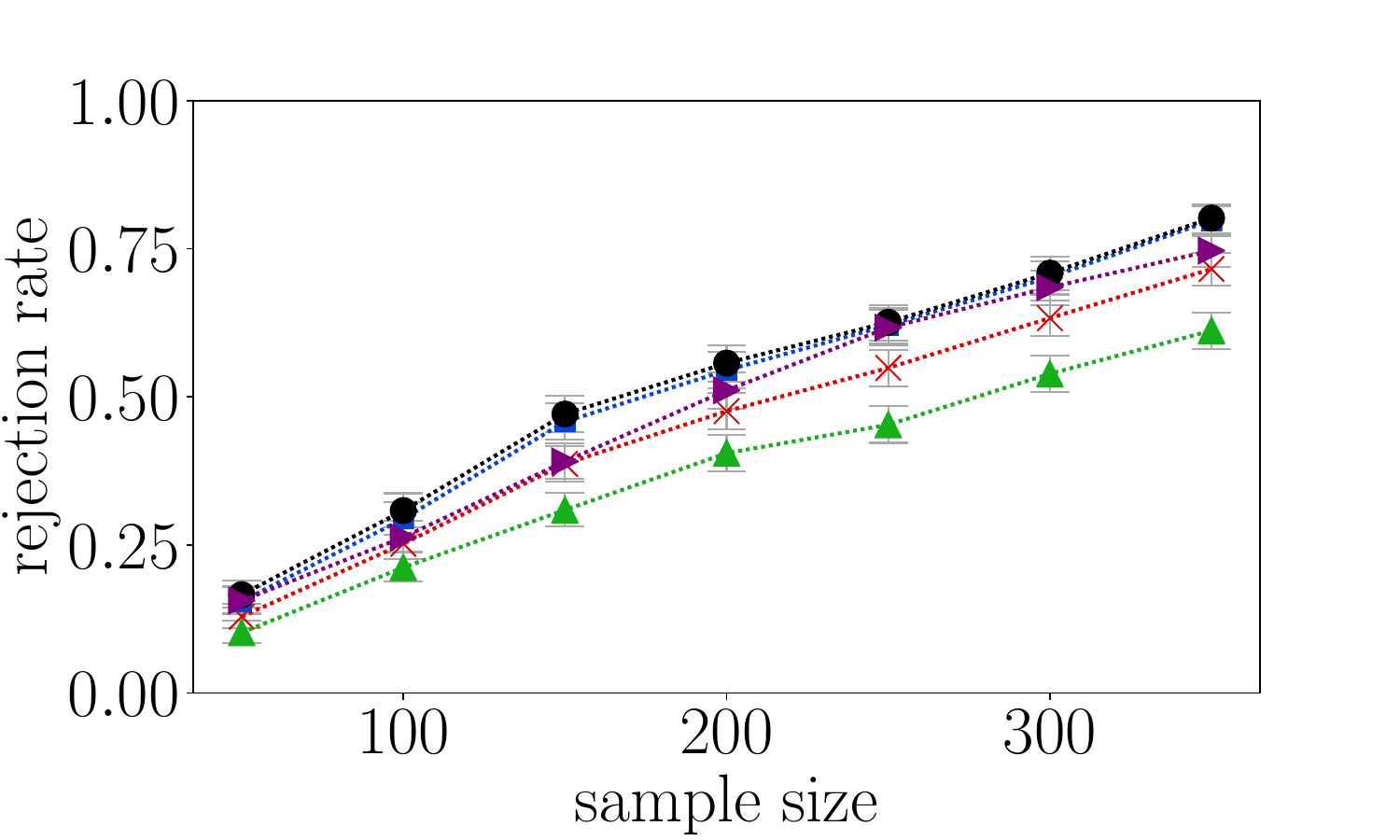}
\end{subfigure}
\begin{subfigure}{.495\textwidth}
  \centering
  \includegraphics[width=1\linewidth]{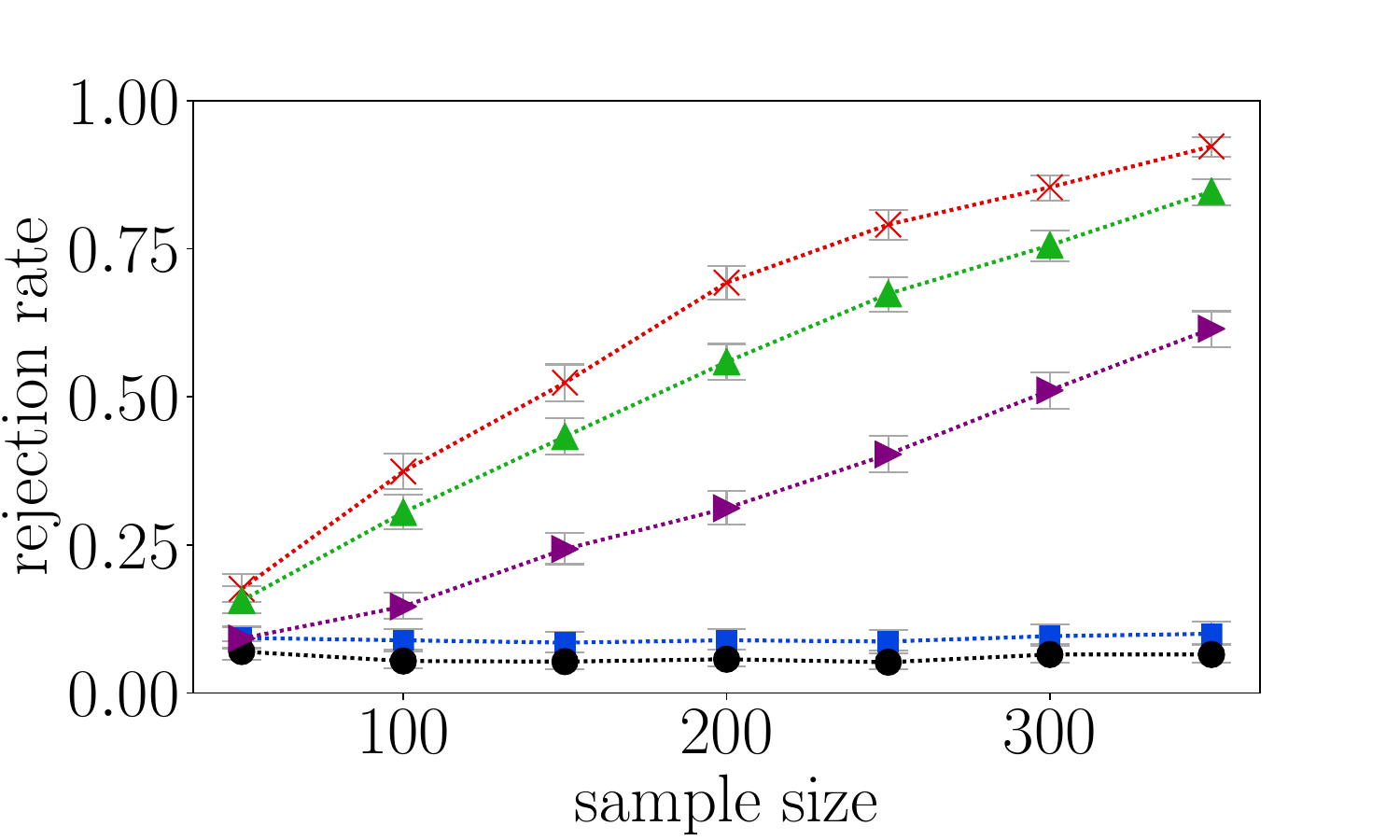}
\end{subfigure}
\begin{subfigure}{.495\textwidth}
  \centering
  \includegraphics[width=1\linewidth]{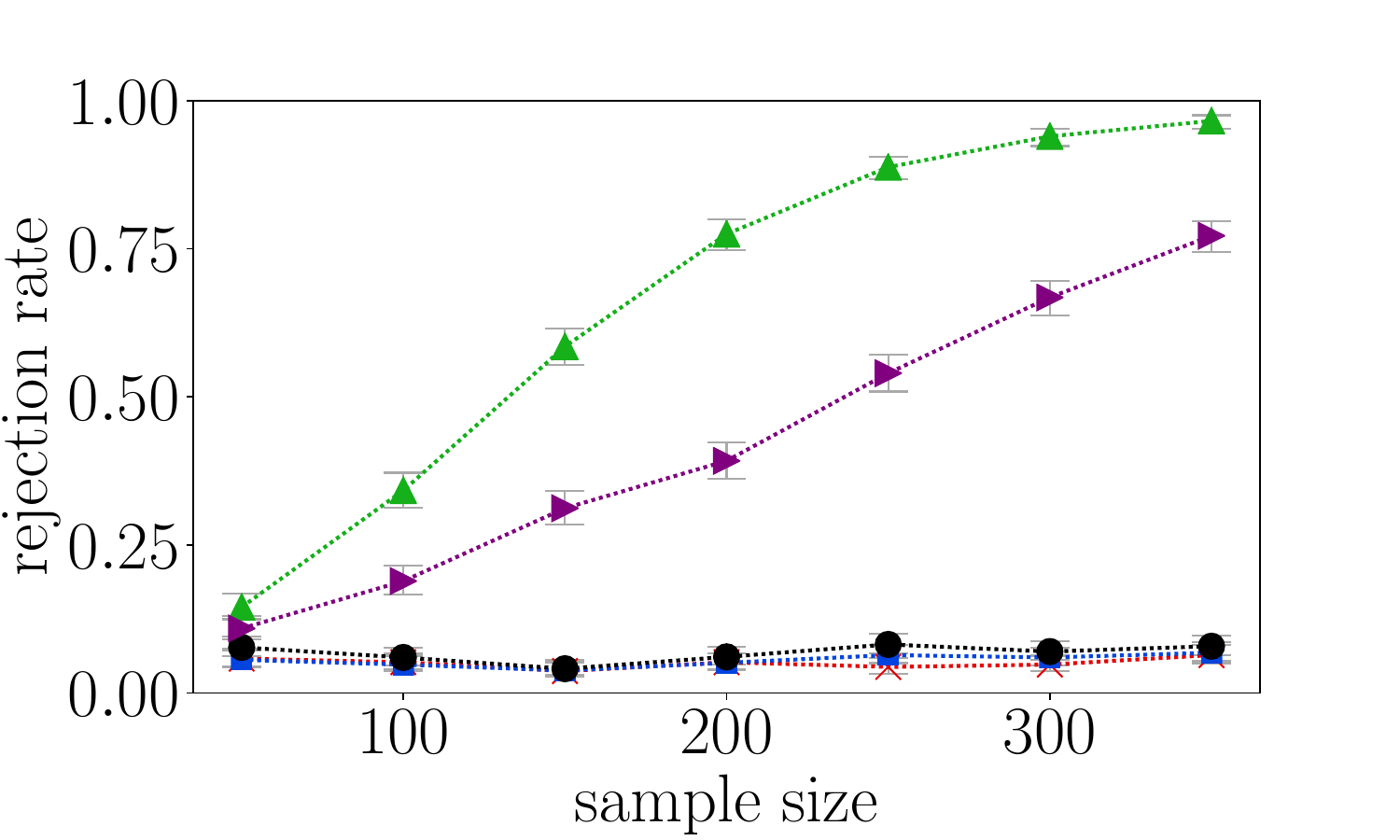}
\end{subfigure}
\begin{subfigure}{.495\textwidth}
  \centering
  \includegraphics[width=1\linewidth]{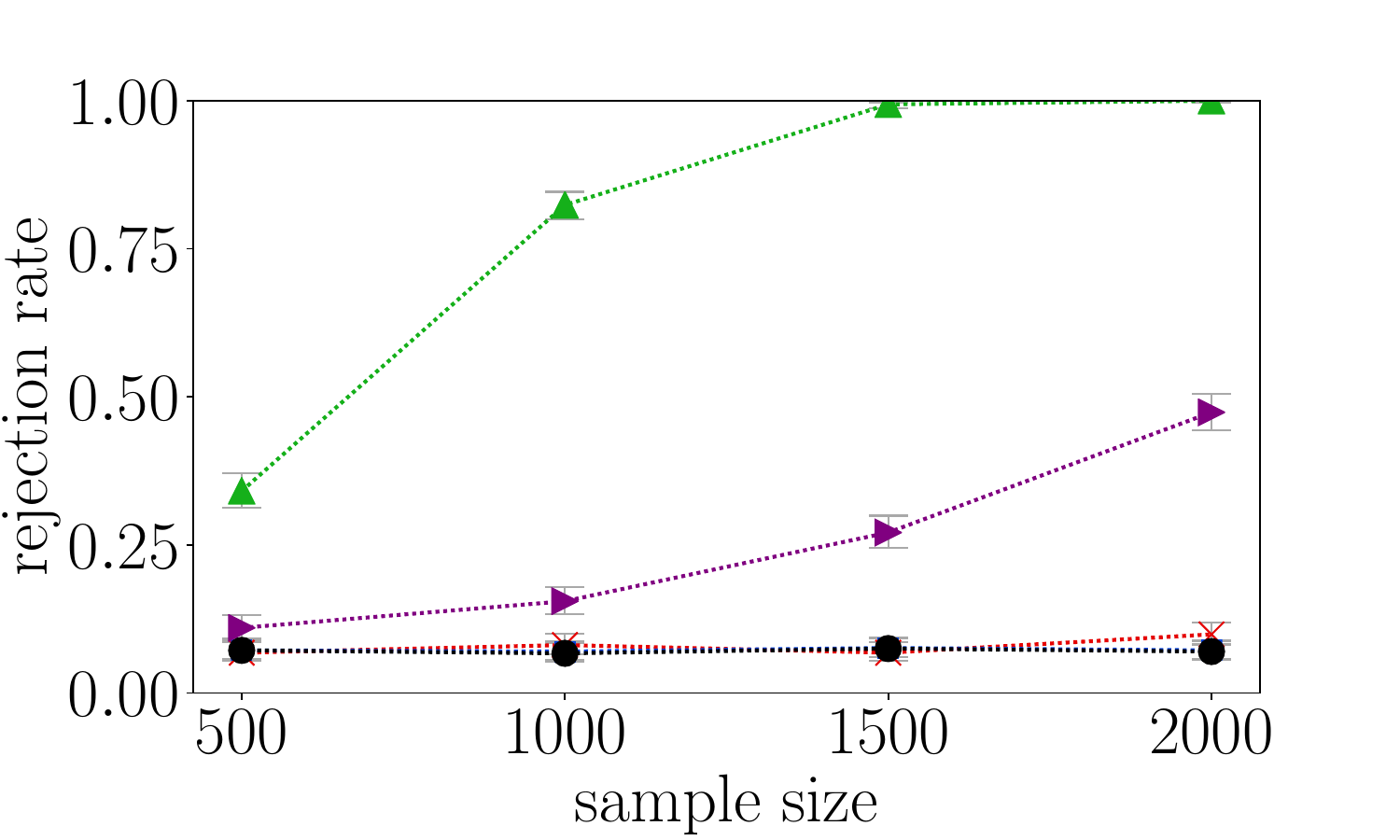}
\end{subfigure}
\begin{subfigure}{.5\textwidth}
  \centering
  \includegraphics[width=1\linewidth]{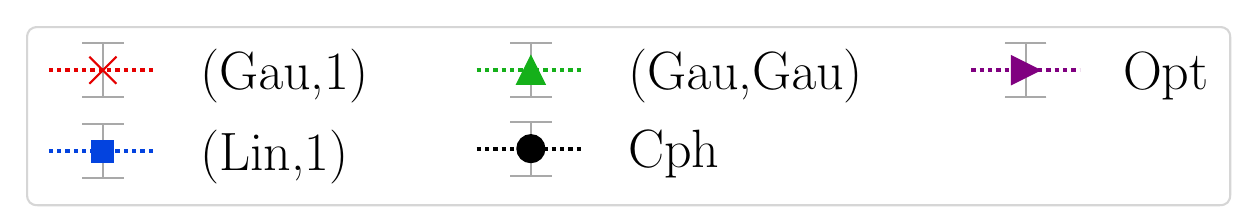}
\end{subfigure}
\caption{\label{plot:1d_rejection_rates} Rejection rates of the various test for D.1 (top left), D.2 (top right), D.3 (bottom left) and D.4 (bottom right).}
\end{figure}

\begin{enumerate}[nolistsep]
    \item[\textit{D.1}] \emph{A CPH distribution:} $Z \vert X=x \sim \text{Exp}( \text{mean} = \exp\{x/3\} )$ and $C \vert X=x \sim \text{Exp}( \text{mean} = 1.5 )$.
    \item[\textit{D.2}] \emph{A non-linear log-hazard:} $Z \vert X=x \sim \text{Exp}( \text{mean} = \exp\{x^2\} )$ and $C \vert X=x \sim \text{Exp}( \text{mean} = 2.25 )$.
    \item[\textit{D.3}] \emph{A family of Weibull distributions:}  $Z \vert X=x \sim \text{Weib}  (\text{shape}=3.35+1.75 \cdot x,\text{scale}=1)$ and $C \vert X=x \sim \text{Exp}( \text{mean} = 1.75 )$.
    \item[\textit{D.4}] \emph{A checkerboard pattern:} See Figure \ref{plot:scatterplots}. Because the pattern is more complicated, we let the sample size range from 500 to 2000 in steps of 500.
\end{enumerate}
The top row of Figure \ref{plot:1d_rejection_rates} displays the rejection rates of the tests for samples from D.1 and D.2. Note that in D.1 the kernel log-rank test with linear kernel performs  roughly equivalently to the LR test, which is ideally suited for this distribution as the CPH assumption holds. While the kernel log-rank tests with the rich kernels $(\text{Gau}, 1)$ and $(\text{Gau}, \text{Gau})$ do not lose much power in detecting this CPH dependency, we do observe a small tradeoff between richness of the kernels and power: the $(\text{Gau}, \text{Gau})$ kernel has slightly less power than the $(\text{Gau}, 1)$ kernel, which in turn has slightly less power than the $(\text{Lin}, 1)$ kernel and CPH LR test in D.1. In D.2 the quadratic term in the log-hazard violates the CPH assumption. The top right panel of Figure \ref{plot:1d_rejection_rates} shows that the linear kernel and the CPH LR test are unable to detect the dependency. Again, we find that the least rich kernel that is still able to model the dependency has the highest power, which in this case is the $(\text{Gau},1)$ kernel. In D.3 and D.4 the hazard function does not factorize into a function of covariates and a function of time, and a kernel is needed on time to model the dependency. Rejection rates are displayed in the bottom of Figure \ref{plot:1d_rejection_rates}, confirming that indeed the kernel log-rank test with kernel $(\text{Gau},\text{Gau})$ is the only test able to correctly reject the null hypothesis.

\begin{figure}[htp]
\centering
\begin{subfigure}{.45\textwidth}
  \centering
  \includegraphics[width=1\linewidth]{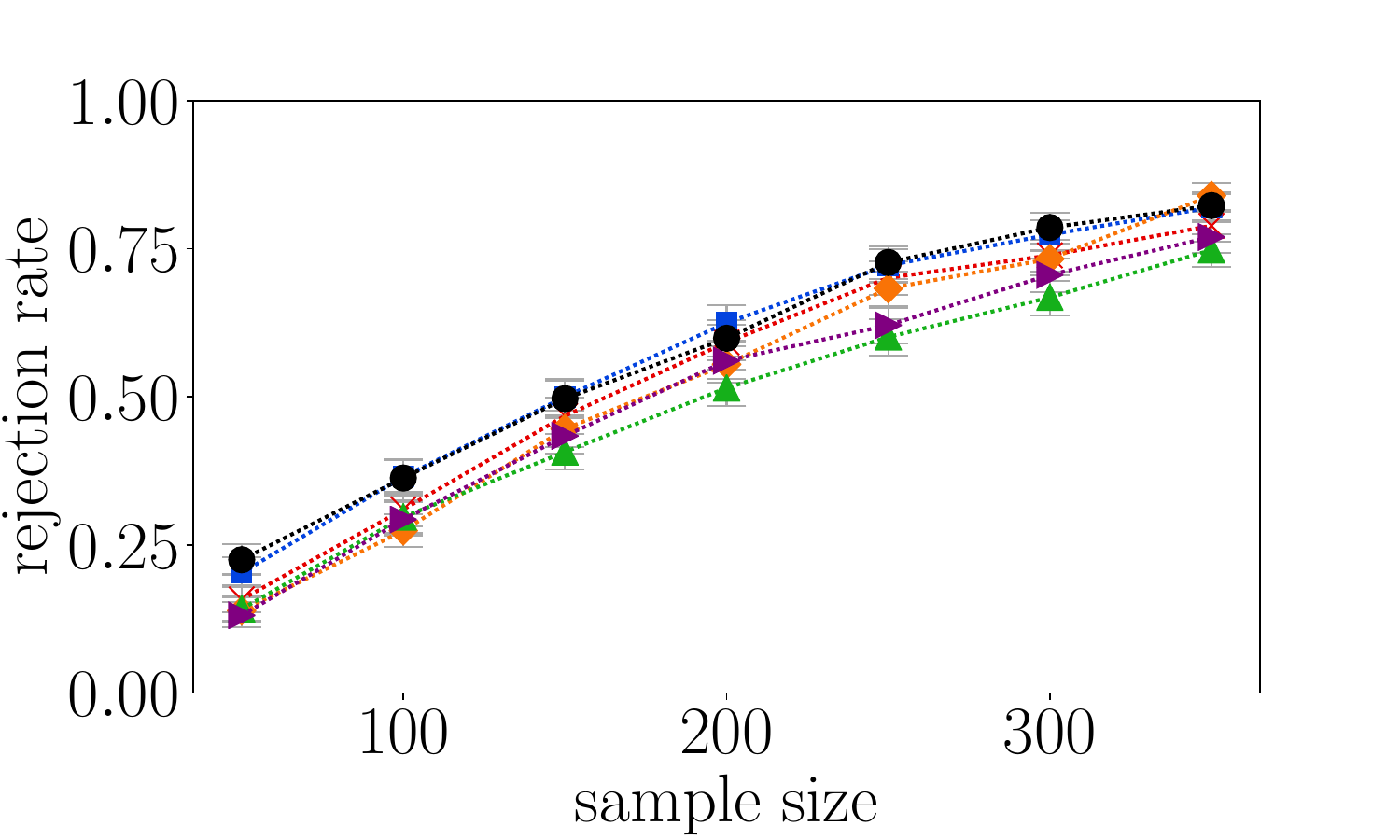}
\end{subfigure}%
\begin{subfigure}{.45\textwidth}
  \centering
  \includegraphics[width=1\linewidth]{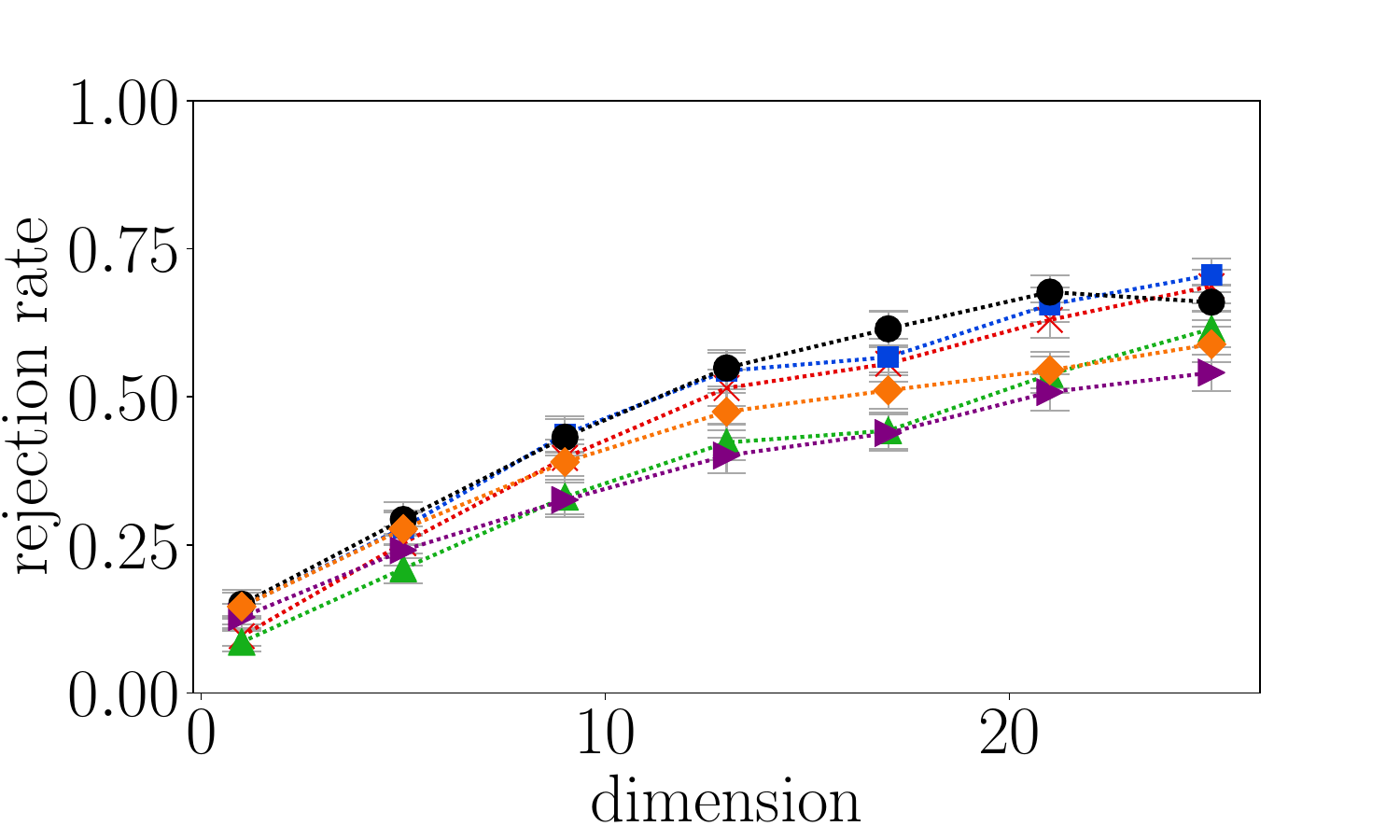}
\end{subfigure}

\begin{subfigure}{.45\textwidth}
  \centering
  \includegraphics[width=1\linewidth]{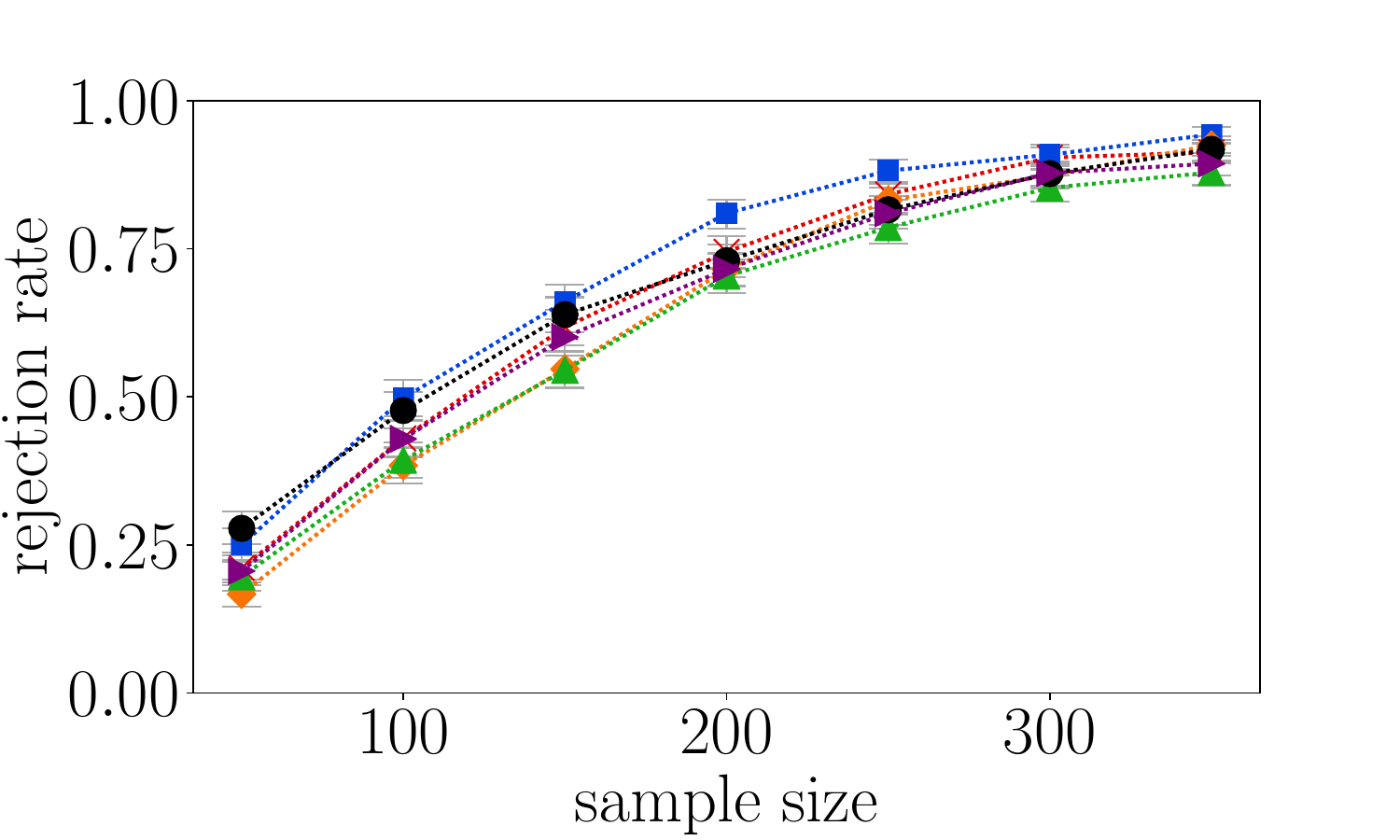}
\end{subfigure}%
\begin{subfigure}{.45\textwidth}
  \centering
  \includegraphics[width=1\linewidth]{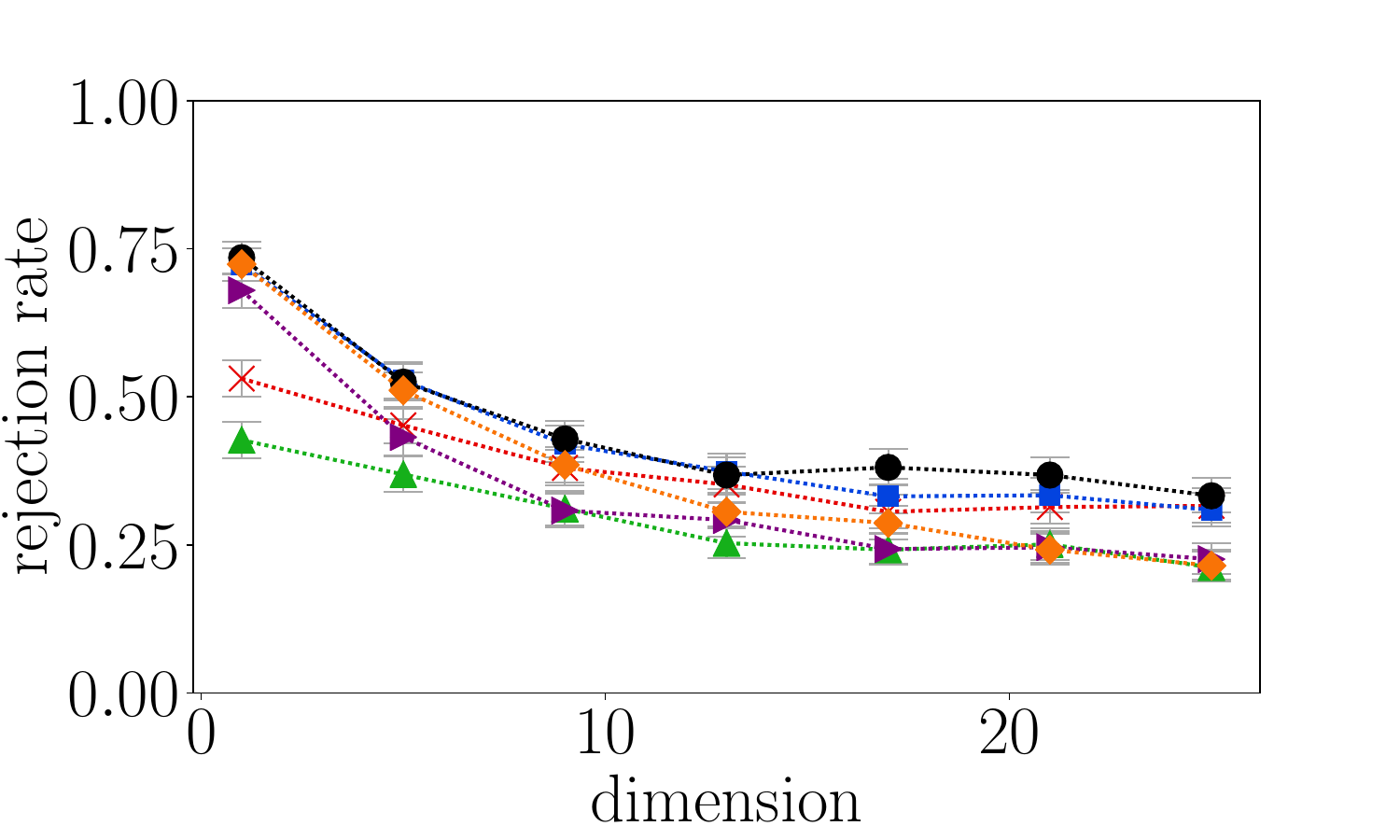}
\end{subfigure}

\begin{subfigure}{.45\textwidth}
  \centering
  \includegraphics[width=1\linewidth]{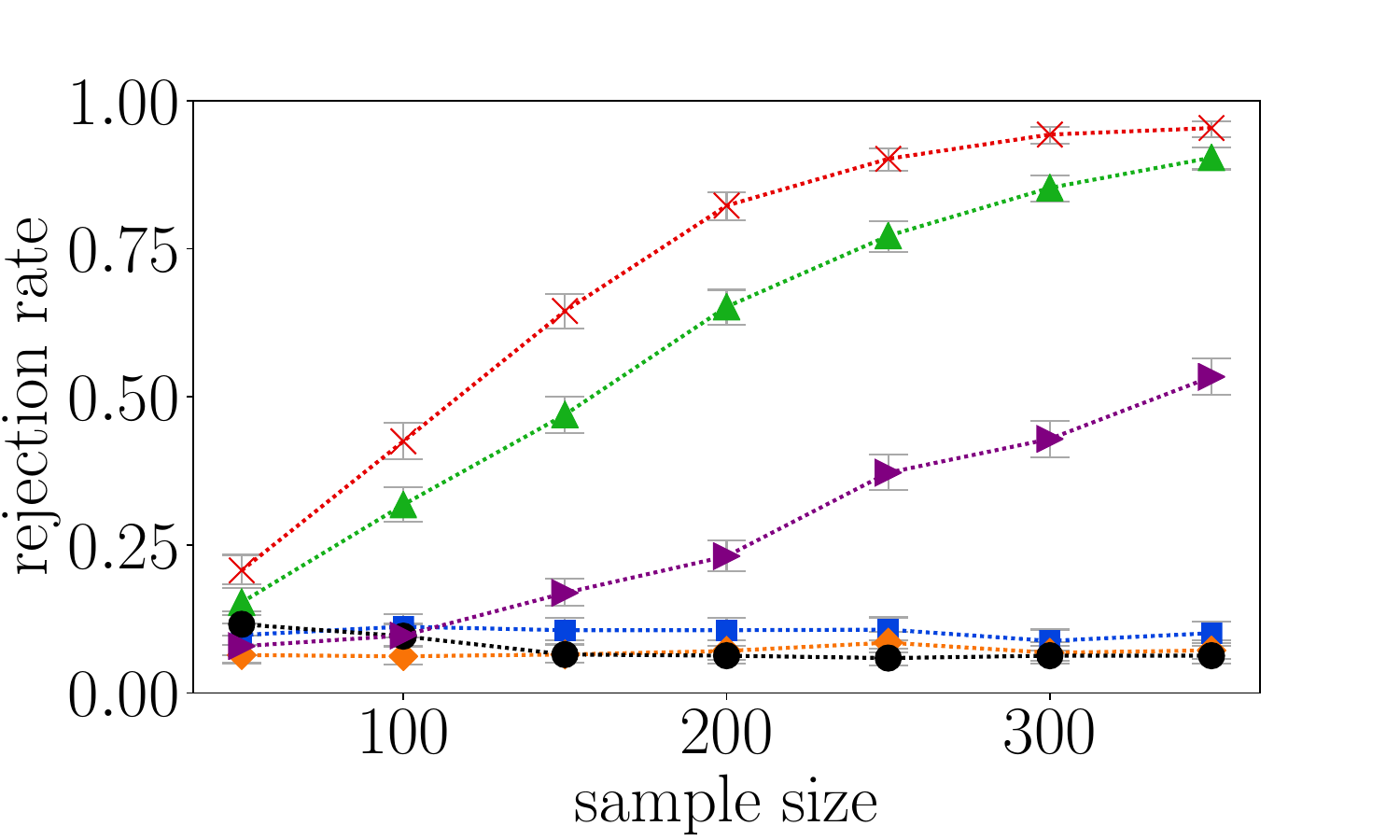}
\end{subfigure}%
\begin{subfigure}{.45\textwidth}
  \centering
  \includegraphics[width=1\linewidth]{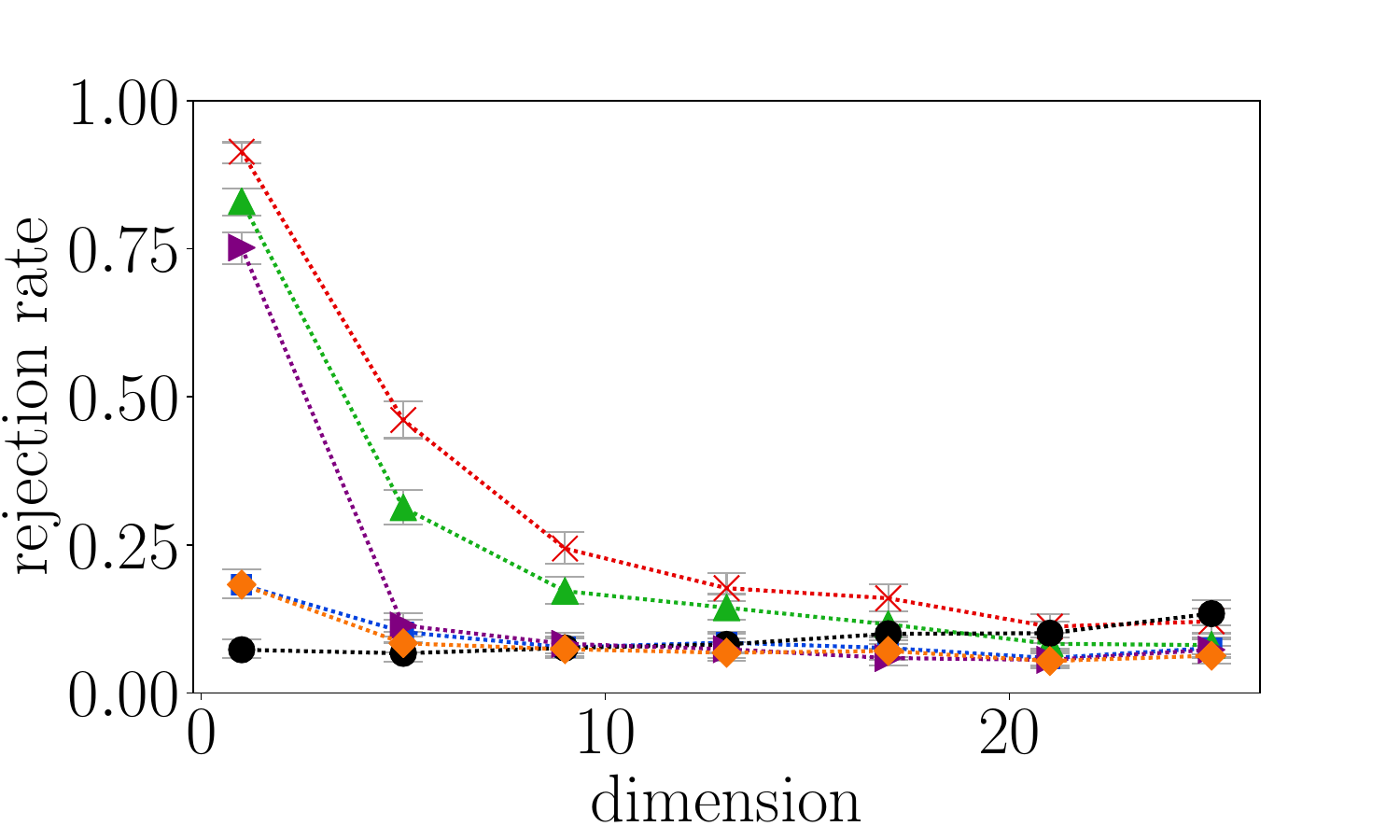}
\end{subfigure}

\begin{subfigure}{.45\textwidth}
  \centering
  \includegraphics[width=1\linewidth]{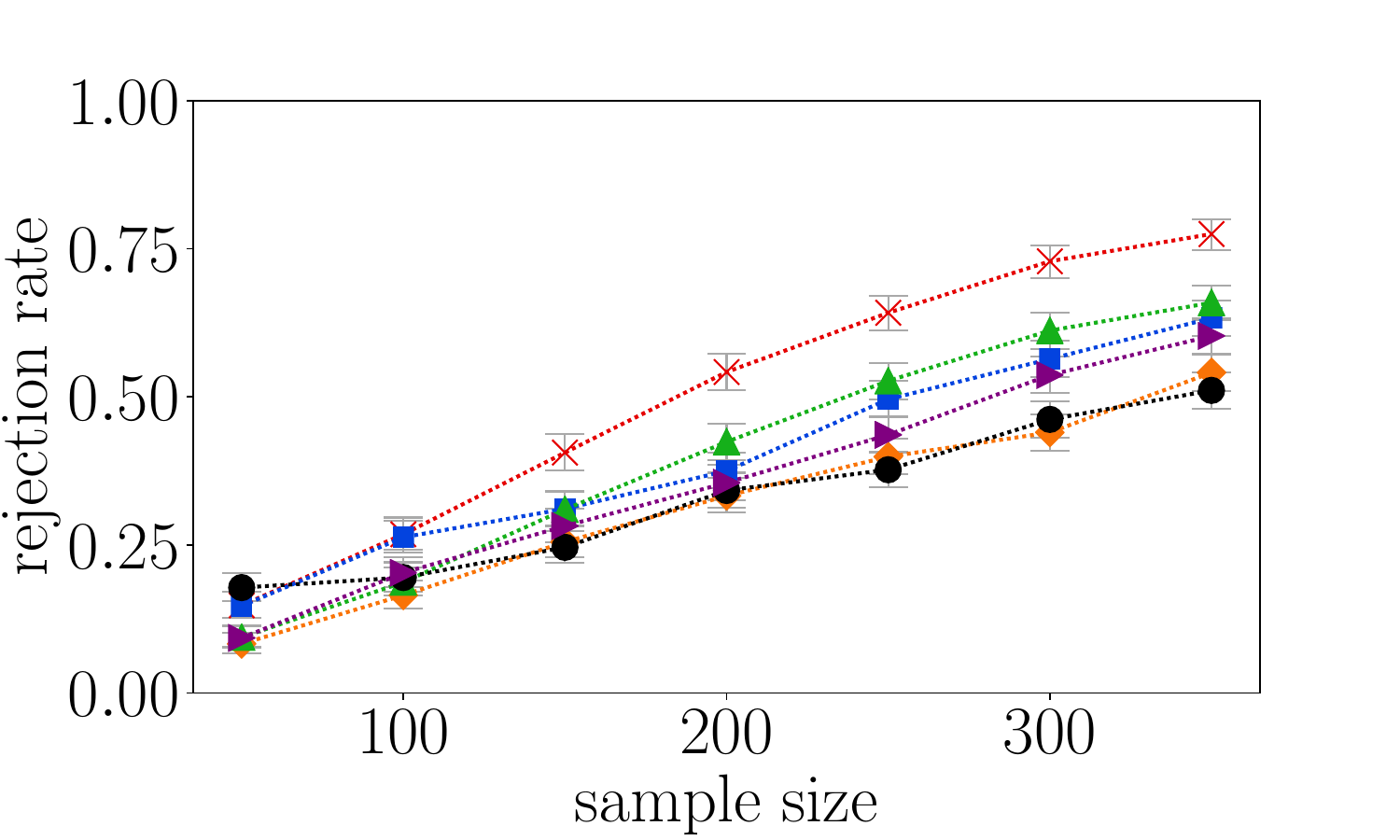}
\end{subfigure}%
\begin{subfigure}{.45\textwidth}
  \centering
  \includegraphics[width=1\linewidth]{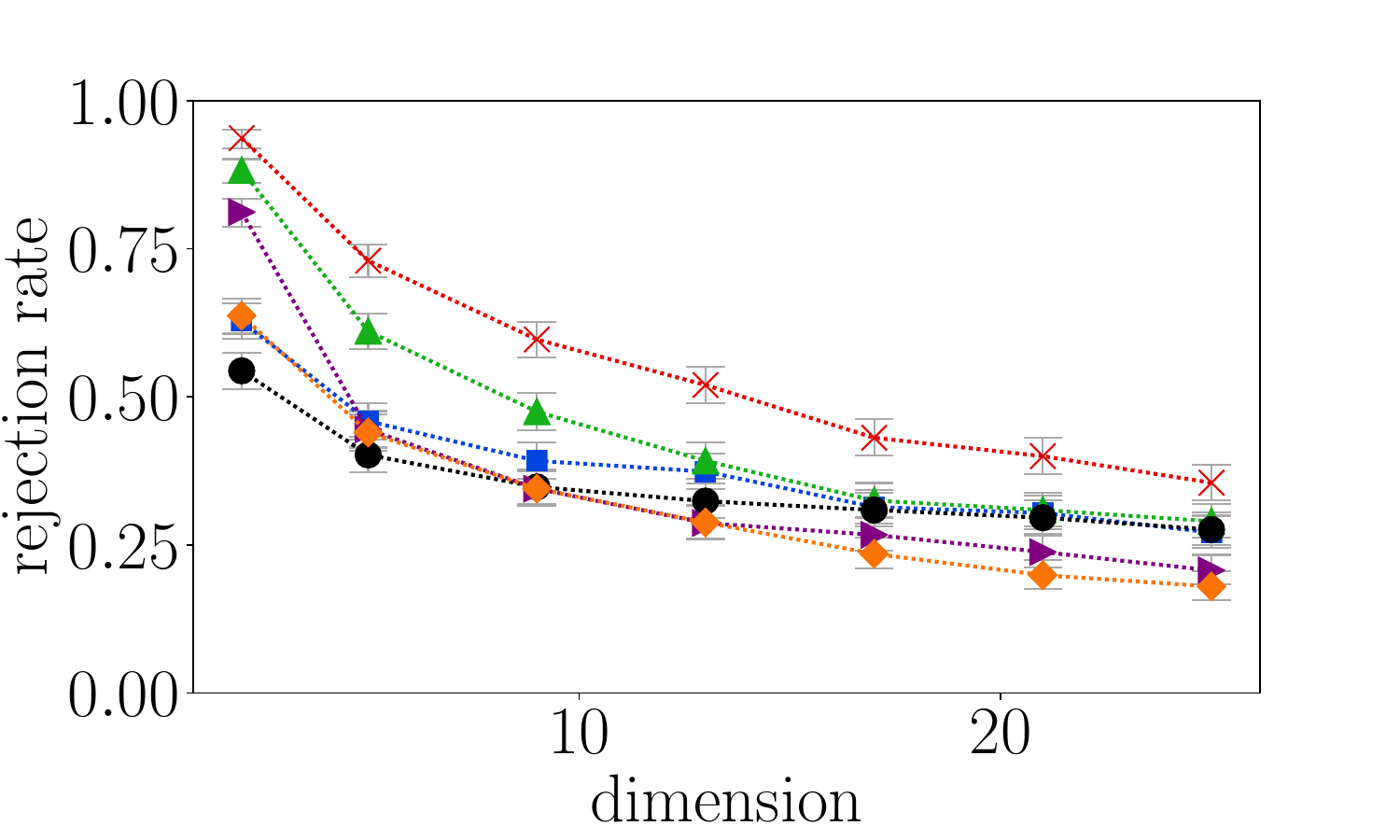}
\end{subfigure}%

\begin{subfigure}{.5\textwidth}
  \centering
  \includegraphics[width=1\linewidth]{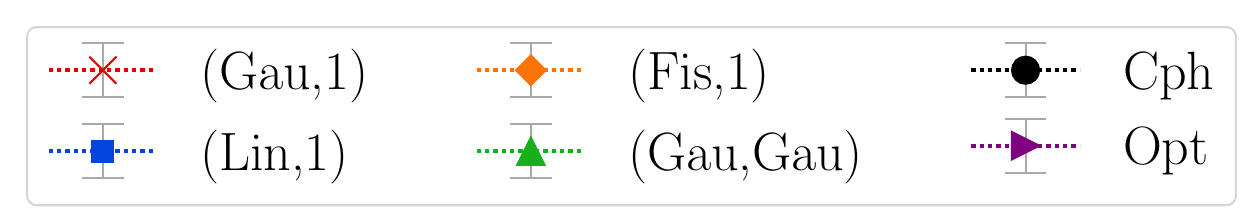}
\end{subfigure}%
\caption{Rejection rates for D.5 (top) - D.8 (bottom). Left: rejection rates as the sample size increases. Right: rejection rates as the dimension increases.  \label{figure:highdimensionrejectionrate} }
\end{figure}

\paragraph{\label{section:powerhighdimension}Power for multidimensional covariates}
Let $ X \sim \text{Normal}(\text{mean}=0_d$ and $\text{cov}=\Sigma_d)$ where $0_d=(0,\dots,0)\in \R_d$, $\Sigma_d=MM^T$ and $M$ is a $d \times d$ matrix of independent $\text{Normal}(0,1)$ entries. Consider the following four distributions:

\begin{enumerate}[nolistsep]
\item[\textit{D.5:}] \emph{A CPH dependence on all covariates:} $Z \vert X=x \sim \text{Exp}(\text{mean}=  \exp\{1_d^T x/20\})$ and $C \vert X=x \sim \text{Exp}(\text{mean}= 1.5)$  where $1_d=(1,\dots,1)\in\R^d$.

\item[\textit{D.6:}] \emph{A CPH dependence on single covariate:} $Z \vert X=x \sim \text{Exp}(\text{mean}= \exp\{ x_1/60\})$ and $C \vert X=x \sim \text{Exp}(\text{mean}=1.5)$.

\item[\textit{D.7:}] \emph{A non-CPH dependence on single covariate:}  $Z \vert X=x \sim \text{Exp}(\text{mean} = \exp\{ x_1^2/60\})$ and $C \vert X=x \sim \text{Exp}(\text{mean}= 1.5)$.

\item[\textit{D.8:}] \emph{A mixed dependence on 2-covariates:} $Z \vert X=x \sim \text{Exp}(\text{mean} = \exp\{ (x_1^2+3x_2)/60\})$ and $C \vert X=x \sim \text{Exp}(\text{mean}= 2)$.
\end{enumerate}
Figure \ref{figure:highdimensionrejectionrate} displays rejection rates for both varying dimension and sample sizes. As our first finding, we observe that, while the CPH assumption holds for D.5 and D.6, the power of the kernel log-rank test is similar to that of the CPH LR test, which is an encouraging result. In D.7 and D.8 we again observe that in the presence of a non-CPH dependence, the kernel log-rank test has good power.

\paragraph{Varying censoring rates and distributions} We study the rejection rates in cases where censoring depends on the covariate, and where the percentage of observed $(\Delta=1)$ events is $15,30,45,60,75,90$ or $100\%$. The experiments are described in Appendix A.3. Under these varying censoring percentages, the main findings from the previous section remain true: the kernel log-rank test shows competitive power when the CPH assumption holds, is able to detect non-CPH dependencies, and achieves correct type 1 error. A final important observation is that optHSIC loses power compared to the kernel log-rank test for higher censoring rates.

\paragraph{Sensitivity to choice of bandwidth} 
In the experiments presented thus far we set the bandwidth of the Gaussian kernel to be $\sigma^2=\text{median}\{||x_i-x_j ||^2 \ : i\neq j\}/2$. We now study the effect of varying the bandwidth of the Gaussian kernel on the rejection rate. Details of the experiments are in Appendix A.4. We find that, while for most scenarios a bandwidth can be selected that slightly outperforms the median heuristic, the median heuristic has a consistently good performance across all scenarios.  

\paragraph{Choice of kernel}
When using the kernel log-rank test it is important to choose an appropriate kernel. While in principle we can choose $\mathfrak{K}$ to be any kernel, choosing a kernel that factorizes into a kernel on time and a kernel on the covariates has the important advantage that it gives the simple closed-form expression for our test-statistic in Theorem \ref{Thm:closed form}. In Theorems \ref{Thm:Power} and \ref{Thm:PowerW}, we prove that our testing procedure is consistent for sufficiently expressive kernels $K$ and $L$.

Our experiments show that in some scenarios there is a trade off between richness of the RKHS and the power of the test. For example, when the data is sampled from a distribution satisfying the CPH assumption, the kernel $(\text{Lin},1)$ is generally more powerful than the richer $(\text{Gau},1)$ or $(\text{Gau},\text{Gau})$ kernels. Hence, if we have prior knowledge about the relationship, we may use this knowledge to choose the appropriate kernel. On the whole, however, we found the richer kernels to have competitive power, even in cases where less rich kernels were optimal.

We may also consider improvements on the median heuristic when selecting kernel bandwidth. In \cite{gretton2012optimal,SutTunStretal17,LiuXuLuZhaGreSut20}, which deal with the case of two-sample testing in the uncensored setting, part of the data is used to select parameters that result in the lowest approximate asymptotic  $p$-value, and the test is then performed with the selected parameters on the remaining data. In \cite{meynaoui19Adaptive}, which addresses independence testing in  the uncensored setting, an aggregation procedure is proposed over kernel bandwidths, which does not require data splitting, and is adaptive in a minimax sense  over Sobolev classes of alternatives. It will be an interesting topic for future research to extend these kernel selection strategies to the censored setting.

\section{Application to real-world datasets}\label{sec:real_data}

We next apply our proposed method to two real-world datasets. We compare the $p$-values obtained by the kernel log-rank test with kernels  $(\text{Gau},1)$ and $(\text{Gau},\text{Gau})$ to the $p$-values obtained by the CPH likelihood ratio test. We use $10000$ Wild Bootstrap samples in these examples. 

 \begin{table}[H]
\centering
\resizebox{0.7\textwidth}{!}{\begin{tabular}{l l l l l }
\toprule
& & \multicolumn{3}{l}{ $p$-value } \\
\cmidrule(r){3-5} 

Data & Covariates & Cph \ \ \ \ \   & $(\text{Gau},1)$ & $(\text{Gau},\text{Gau})$ \\ \hline
Biofeedback  & (Trt, Recov) & 0.496 & 0.014 & 0.023 \\
  & Trt & 0.462 & 0.458 & 0.050 \\
  & Recov & 0.301 & 0.007 & 0.029 \\
  & (Trt, $\log ($Recov$)$) & 0.027 & 0.013 & 0.025 \\ \hline
Colon & Age & 0.627 & 0.080 & 0.097 \\
 & (Age, Perfor, Adhere) & 0.102 & 0.017 & 0.018\\ 
 \bottomrule
\end{tabular}}
\caption{$p$-values obtained by the various tests for the Biofeedback and Colon data. \label{table:p_values_real_data}}
\end{table}

\paragraph{Biofeedback data}


The biofeedback treatment data studies the time until patients suffering from aspiration after head and neck surgery achieve full swallowing rehabilitation. The study is presented in \cite{denk1997videoendoscopic} and the data was made available in the R package Coxphw \cite{dunkler2018weighted}. In our presentation we name the event-time variable Rehab. Covariates in the dataset are a binary variable indicating biofeedback treatment, denoted Trt, and the time after the surgery until treatment could be started, denoted Recov. This dataset contains 33 individuals, of whom 70\% were observed to fully rehabilitate (coded as $\delta=1$). 

\begin{figure}
\centering

\begin{subfigure}{.6\textwidth}
  \centering
  \includegraphics[width=1\linewidth]{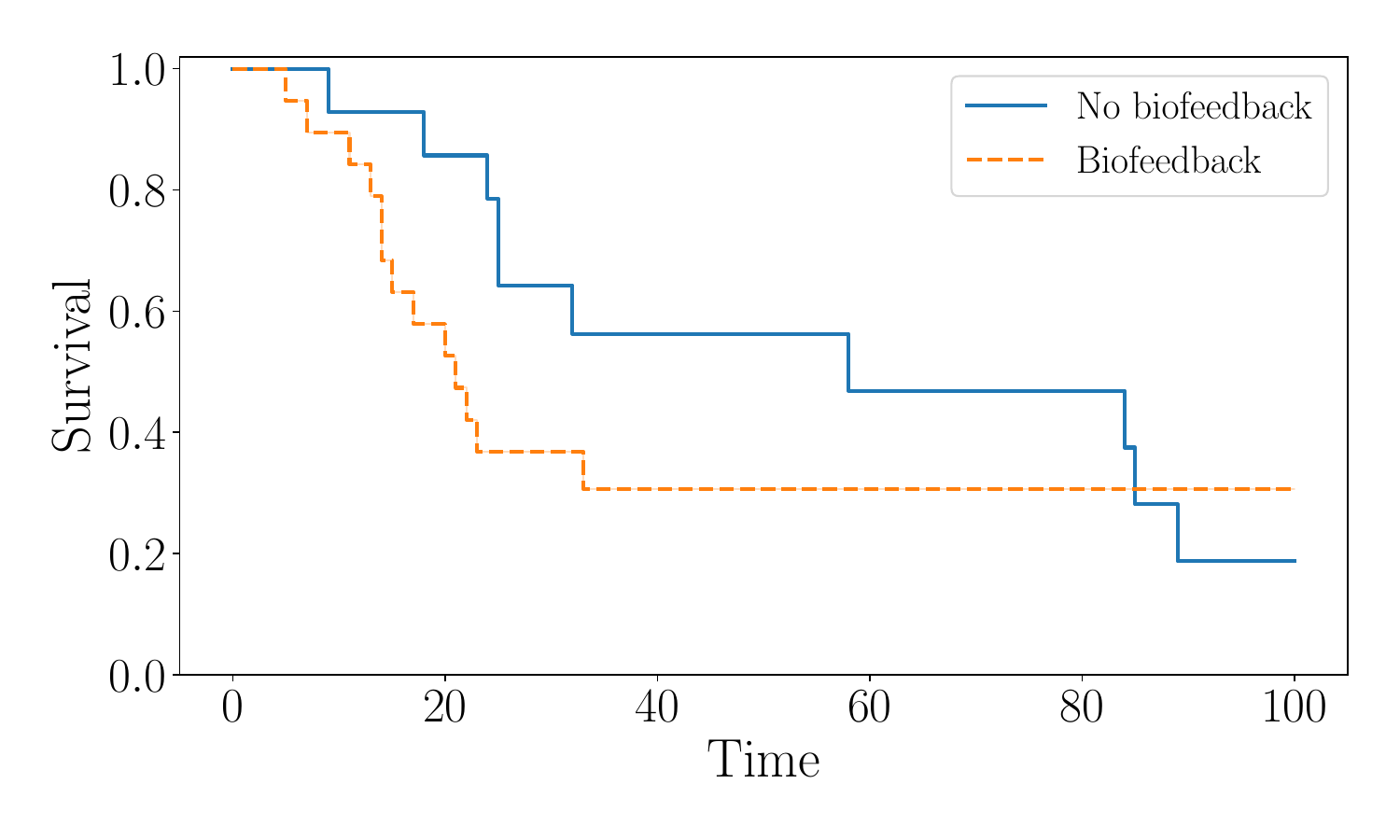}
\end{subfigure}
\caption{Kaplan-Meier estimates of the survival curves associated to the group that receives the Biofeedback treatment and one that does not receive the treatment. We observe that both curves cross which violets the CPH assumption. \label{fig:Biofeed}}
\end{figure}

In the first row of Table \ref{table:p_values_real_data}, we observe that the kernel log-rank test results in the $p$-values $0.014$ for the kernel $(\text{Gau},1)$ and $0.023$ for the kernel $(\text{Gau},\text{Gau})$ when including both covariates. In contrast, the CPH likelihood ratio test results in the higher $p$-value of $0.496$. This  suggests the possibility of a non-linear relationship between the event-time of interest, Rehab, and the covariates Trt and Recov. This interpretation is strengthened by the following two observations. First, when we apply a logarithm transformation to the covariate Recov, as suggested in \cite{denk1997videoendoscopic}, the CPH likelihood ratio test (shown in the fourth row of Table \ref{table:p_values_real_data}) results in a $p$-value of $0.027$. The main advantage of the kernel log-rank test is that it does not require to manually transform the data to detect potential non-linear dependencies. Second, we observe that the kernel log-rank test using the $(\text{Gau},\text{Gau})$ kernel is the only test that rejects the null hypothesis of independence between Rehab and Trt (see the second row of Table \ref{table:p_values_real_data}) at a significance level of $0.05$. This result may be attributed to the fact that the survival curves associated to the two treatment groups cross (see Figure \ref{fig:Biofeed}). 

\paragraph{Colon data}
\begin{figure}
\centering

\begin{subfigure}{.6\textwidth}
  \centering
  \includegraphics[width=1\linewidth]{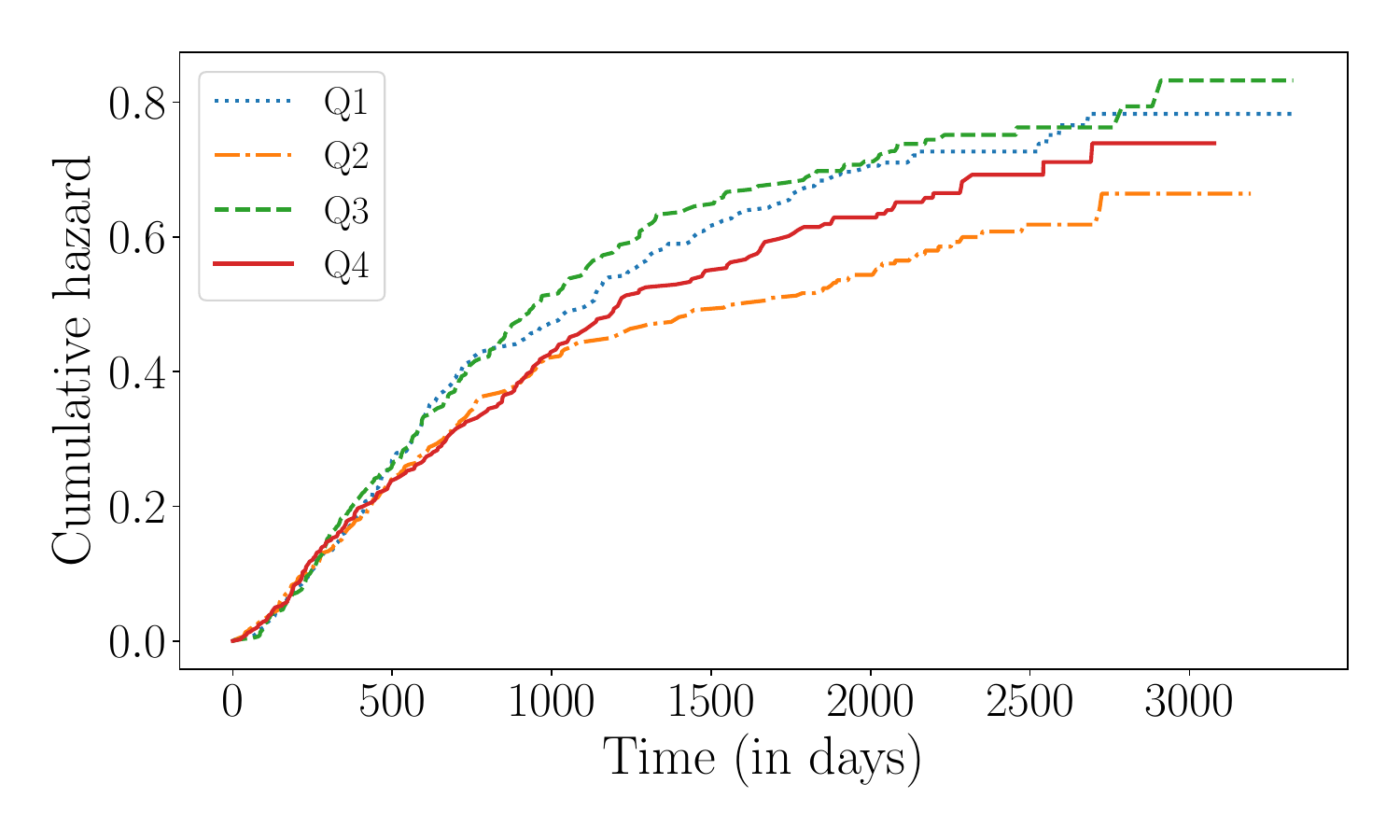}
\end{subfigure}
\caption{Nelson-Aalen estimates of the cumulative hazard for the four quartiles of age. Q1 denotes the youngest quartile and Q4 the oldest. Note that the curves do not satisfy the CPH assumption. \label{fig:colon_nelson_aalen}}
\end{figure}
The Colon dataset studies the recurrence of tumors and survival in patients undergoing treatment for stage B/C colon cancer. The study is presented in \cite{laurie1989surgical} and \cite{moertel1995fluorouracil}. Data from the 929 patients in the study is available in the R package Survival \cite{therneau2015package}. Each individual has 11 covariates. We consider the time of death in our analysis as the event-time, which was observed, i.e. $\delta=1$, for 49\% of the individuals. We focus our analysis on the variables Age, Perfor (a binary variable indicating the perforation of the colon) and Adhere (a binary variable indicating adherence of the tumor to nearby organs).

The $p$-values of different tests of independence are given in Table \ref{table:p_values_real_data}. Testing for independence between the age covariate and the event-time, the kernel log-rank test results in $p$-values of $0.080$ and $0.097$ for the $(\text{Gau},1)$ and $(\text{Gau},\text{Gau})$ kernels respectively. In contrast, the CPH likelihood ratio test produces a much higher $p$-value of $0.627$. This suggests the possibility of a non-linear dependence between the age covariate and the event-time of interest.  This interpretation is strengthened by Figure \ref{fig:colon_nelson_aalen}, which displays cumulative hazard functions for different age quartiles. Indeed, these curves could suggest a non-linear effect of age, as the curves are not ordered by age. In particular we observe that first age quartile has a higher cumulative hazard function than the second quartile, but lower than the third. Finally, in the last row of Table \ref{table:p_values_real_data}, we observe differences in the conclusions of the kernel log-rank test (for both kernels) and the CPH likelihood ratio test at a significance of $\alpha=0.05$ when we include the binary variables Perfor and Adhere to the model. This difference in $p$-values suggests there may be a non-linear relationship between the event-time of interest and these covariates.

\section{Conclusions}
We have introduced a novel non-parametric independence test between right-censored survival times $Z$ and covariates $X$. Our approach uses an infinite-dimensional exponential family of cumulative hazard functions,  which are parameterized by functions in a reproducing kernel Hilbert space. By choosing an expressive Hilbert space of functions, we show that our testing procedure is able to detect any type of dependence, while for simpler Hilbert spaces, we recover ubiquitous approaches such as the Cox score test. The test statistic furthermore has an easily computed closed form. We provide a simple testing procedure based on the Wild Bootstrap, and demonstrate strong performance on a range of synthetic and real datasets. 

\paragraph{Acknowledgements}
We thank David Steinsaltz for the helpful discussions early in the project. Also, we thank Nicol\'as Rivera for the helpful discussions and suggestions for this paper. Finally, we thank the reviewers for their helpful advice that led to the final version of this manuscript. Tamara Fernandez was supported by Biometrika Trust.

\begin{center}
{\large\bf SUPPLEMENTARY MATERIAL}
\end{center}

\begin{description}
\item[A. Additional experiments] In Section A.1 we study the Type 1 error, in Section A.2 we show additional experiments regarding the power of tests, in Section A.3 we show experiments for varying censoring percentages, and in Section A.4 we show experiments for varying bandwidths of the Gaussian kernel. 
\item[B. Preliminary results:] In this section, and in order for this paper to be self-contained, we review some preliminary results that will be used in our proofs.
\item[C. Auxiliary results:] In this section we state some auxiliary results used for the proofs of the main results of the paper.
\item[D. Main results:] In this section we prove the main results of the paper.
\item[E. Proofs of auxiliary results] In this section we prove the auxiliary results introduced in Section C.
\end{description}

\bibliography{ref}
\bibliographystyle{plain}

\newpage
\appendix

\section{Additional experiments \label{sec:appendix:experiments}}
Here we present some experiments that were not described in detail in the main text.

\subsection{Type 1 error rate}\label{sec:appendix:type1error}
\begin{figure}
\centering
\begin{subfigure}{.5\textwidth}
  \centering
  \includegraphics[width=1\linewidth]{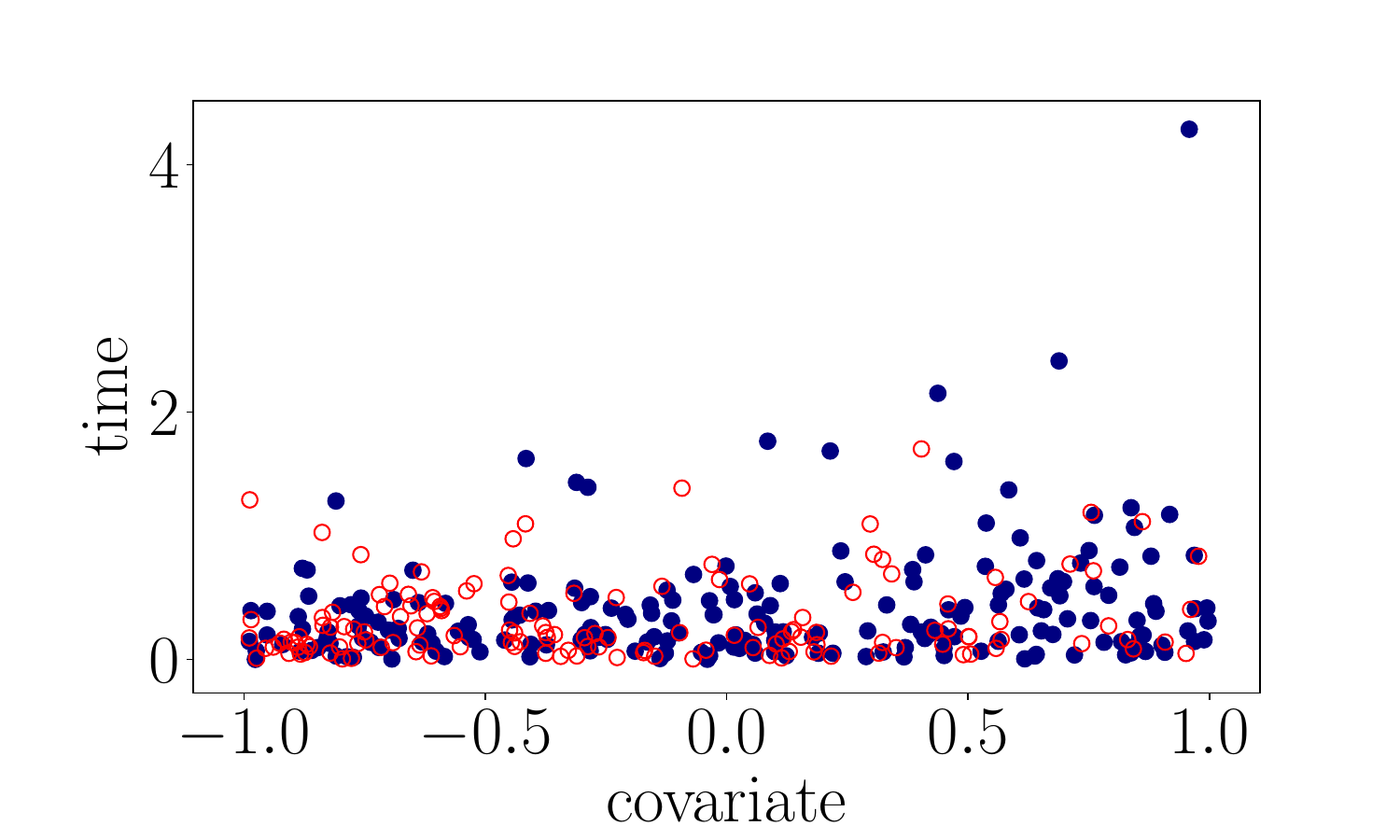}
\end{subfigure}%
\begin{subfigure}{.5\textwidth}
  \centering
  \includegraphics[width=1\linewidth]{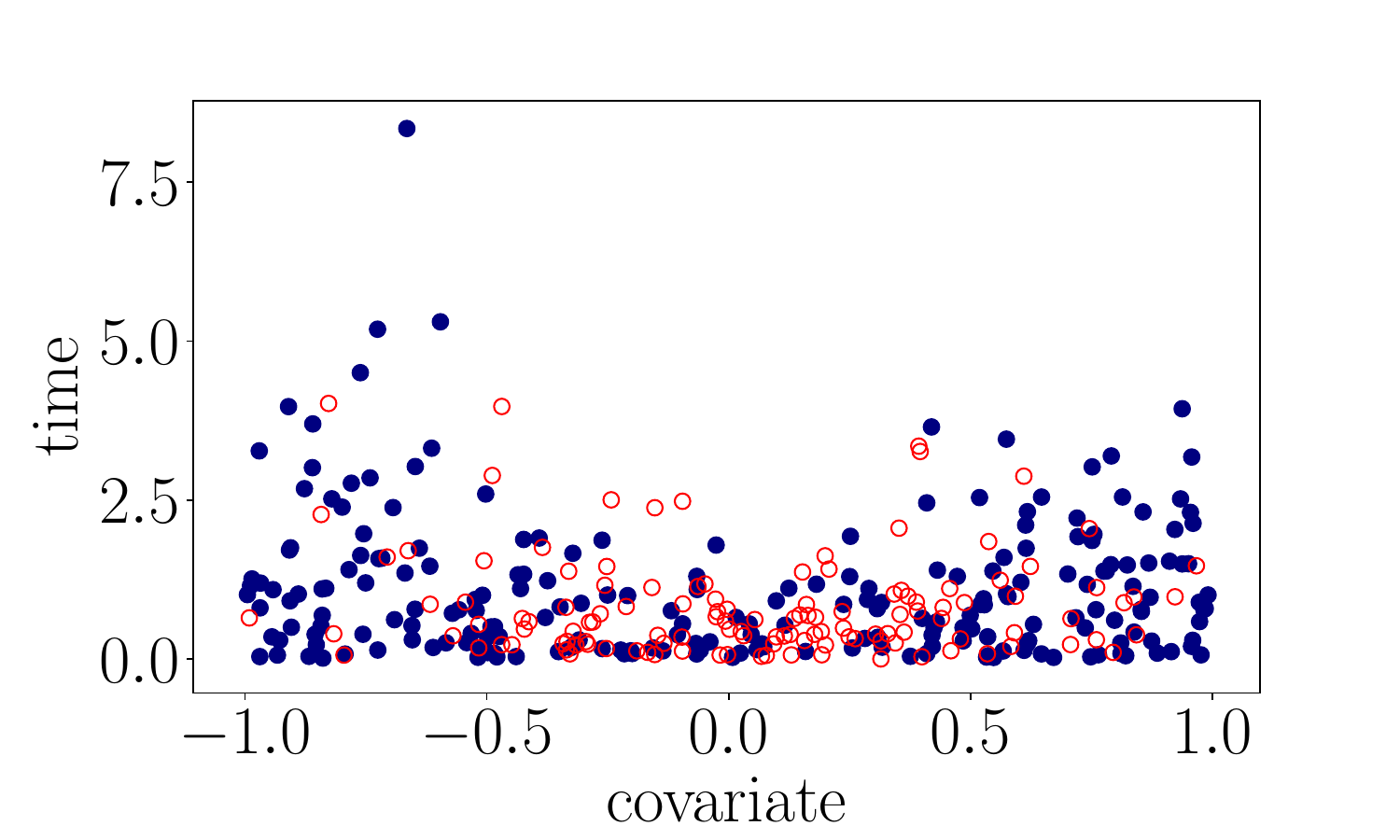}
\end{subfigure}

\begin{subfigure}{.5\textwidth}
  \centering
  \includegraphics[width=1\linewidth]{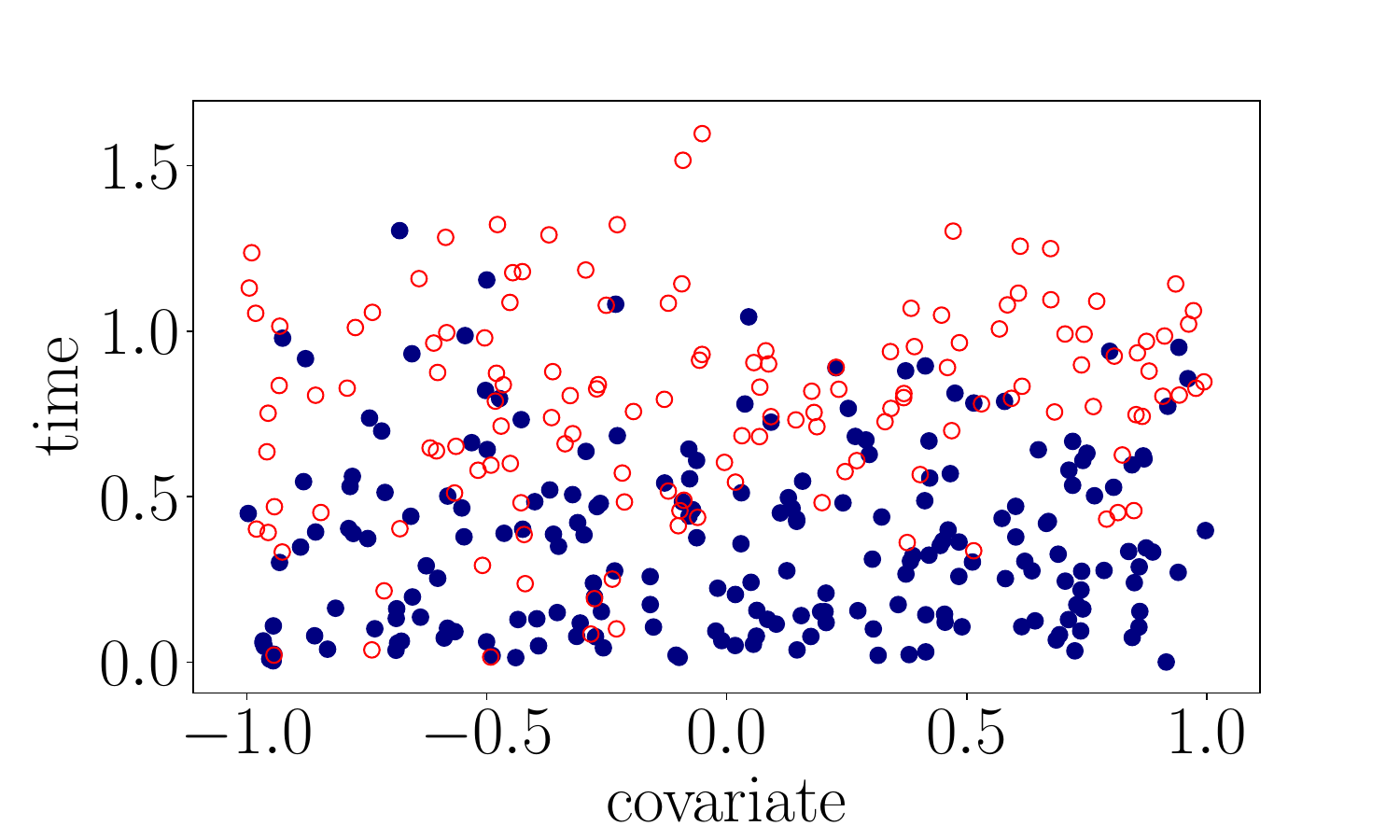}
\end{subfigure}%
\begin{subfigure}{.5\textwidth}
  \centering
  \includegraphics[width=1\linewidth]{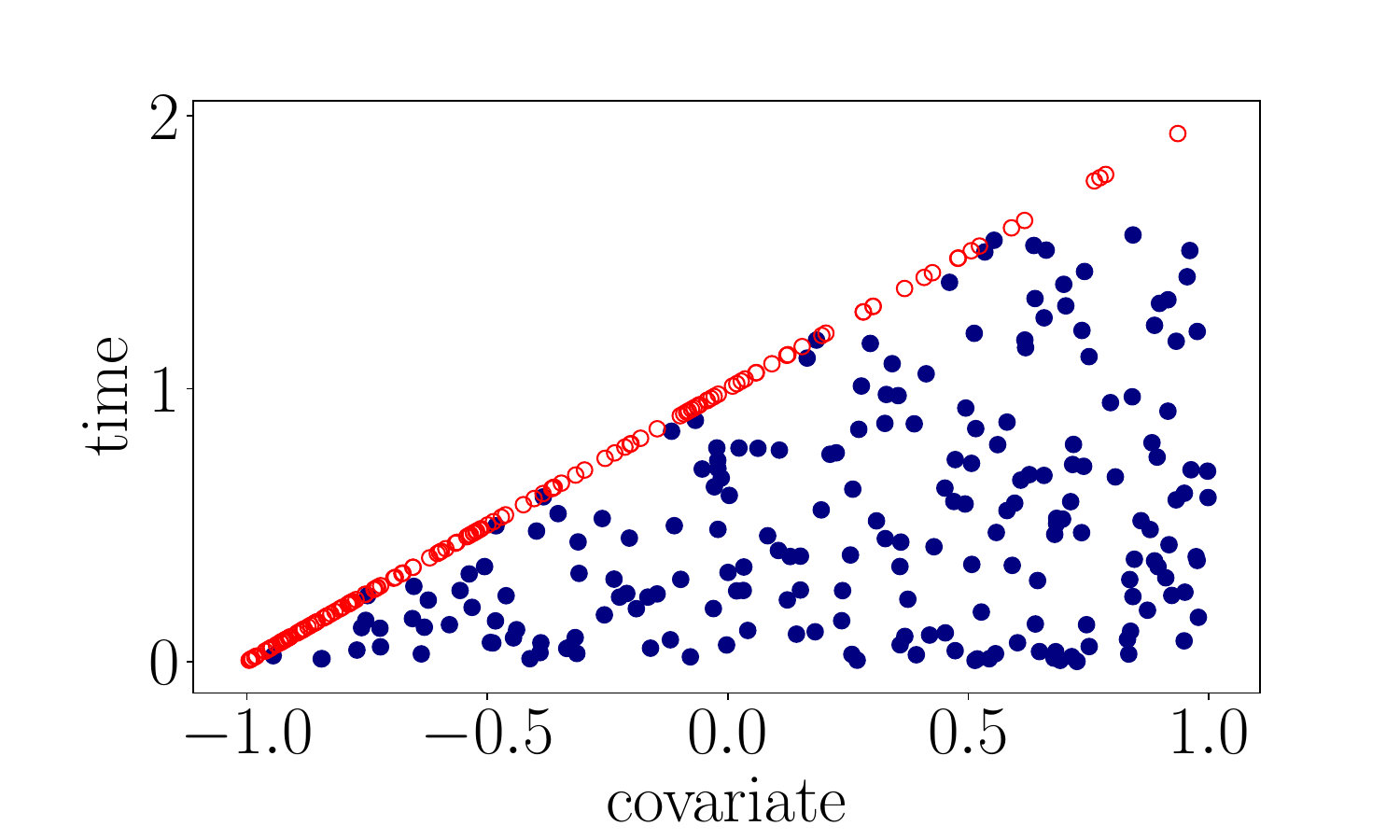}
\end{subfigure}
\begin{subfigure}{.5\textwidth}
  \centering
  \includegraphics[width=.9\linewidth]{Figures/legend_scatterplot_flat.pdf}
\end{subfigure}
\caption{\label{figure:appendix:type_1_error_scatterplots}Top: Scatterplots of samples from D.1 (top left), D.2 (top right) D.3 (bottom left) and D.4 (bottom right) defined in Table \ref{table:appendix:type1error_distributions}.}
\end{figure}

To estimate the false rejection rate of our proposed test, we repeatedly take samples from distributions in which $Z\perp X$ and we count the proportion of experiments in which the kernel log-rank test rejects the null hypothesis. We set the significance level at $\alpha=0.05$  and let the sample sizes range from $50$ to $350$ in steps of $50$. For each distribution and sample size, we take 5000 samples. Table \ref{table:appendix:type1error_distributions} shows the distributions we sample from.  Figure \ref{figure:appendix:type_1_error_scatterplots} shows scatterplots of samples from distributions 1-4. 

\begin{table}[H]
\centering
\begin{tabular}{  l l l l  }
\toprule
    D. &  $ Z \vert X$  & $C \vert X$  & X \\ \hline
    1   & $ \text{Exp}(\text{mean}=0.66) $ & $ \text{Exp}(\text{mean}=\exp(X/3)) $  & \text{Unif}[-1,1] \\ 
     2  & $ \text{Exp}(\text{mean}=0.9)$ & $ \text{Exp}(\text{mean}=\exp(X^2)) $ & \text{Unif}[-1,1]  \\ 
     3 &$\text{Exp}(\text{mean}=0.9)$ & $\text{Weib}(\text{mean}=3.25+1.75X)$& \text{Unif}[-1,1] \\ 
    4  & $ \text{Exp}(\text{mean}=0.9) $ & $ 1+X$ & \text{Unif}[-1,1] \\ 
    5  &$\text{Exp}(\text{mean}=0.6)$ & $\text{Exp}(\text{mean}=\exp(1^T X))$ & $\mathcal{N}_{10}(0,\text{cov}=\Sigma_{10})$  \\
        6  &$\text{Exp}(\text{mean}=0.6)$ & $\text{Exp}(\text{mean}=\exp(X_1))$ & $\mathcal{N}_{10}(0,\text{cov}=\Sigma_{10})$   \\ \bottomrule
\end{tabular}\caption{\label{table:appendix:type1error_distributions}The distributions used to estimate the type 1 error rate. $\text{Exp}(\text{mean})=\mu$ denotes the exponential distribution with mean $\mu$. We define $\Sigma_{10}=MM^T$ where $M$ is a $10\times 10$ matrix of i.i.d. standard normal entries. Parameters are chosen such that approximately $60\%$ of events are observed $(\Delta=1)$.}    
\end{table}

We estimate the bandwidth of the Gaussian kernel from the same data to which we apply the test. Similarly we estimate the Fisher information matrix on the same data to which we apply the kernel log-rank test. Due to this dependence of the kernel on the data, the arguments of Section \ref{sec:wildbootstrap} do not apply directly. The obtained rejection rates show that this does not lead to increased type 1 error rate. The resulting rejection rates are displayed in Tables \ref{table:appendix:type1_1d}
 and \ref{table:appendix:type1_md}. We can see that for each distribution the kernel log-rank test has valid level for all combinations of kernels and sample size. Remarkably the type 1 error rate is correct even in cases in which the censoring distribution depends strongly on the covariate. 
 
 \begin{table}[H]
\centering
\begin{tabular}{lllllllll}
\toprule
& & \multicolumn{7}{l}{ Sample size $n=$} \\ 
\cmidrule(r){3-9} 

D. &   Method & 50  & 100 & 150 & 200 & 250 & 300 & 350 \\ \hline
1 & Cph & 0.053 & 0.051 & 0.049 & 0.051 & 0.055 & 0.048 & 0.053 \\
 & (Lin,1) & 0.037 & 0.049 & 0.050 & 0.047 & 0.055 & 0.048 & 0.044 \\
 & (Gau,1)  & 0.046 & 0.046 & 0.045 & 0.050 & 0.045 & 0.052 & 0.051 \\
 & (Gau,Gau) & 0.047 & 0.044 & 0.044 & 0.048 & 0.045 & 0.054 & 0.052 \\
 & Opt & 0.050 & 0.051 & 0.055 & 0.046 & 0.051 & 0.053 & 0.054 \\ \hline
2 & Cph & 0.049 & 0.060 & 0.051 & 0.049 & 0.056 & 0.046 & 0.050 \\
 & (Lin,1) & 0.044 & 0.044 & 0.050 & 0.047 & 0.052 & 0.045 & 0.047 \\
 & (Gau,1) & 0.038 & 0.044 & 0.048 & 0.047 & 0.052 & 0.057 & 0.049 \\
 & (Gau,Gau) & 0.043 & 0.045 & 0.051 & 0.049 & 0.047 & 0.050 & 0.051 \\
 & Opt & 0.053 & 0.050 & 0.051 & 0.052 & 0.047 & 0.050 & 0.054 \\ \hline
3 & Cph & 0.045 & 0.052 & 0.052 & 0.052 & 0.049 & 0.052 & 0.056 \\
 & (Lin,1) & 0.039 & 0.047 & 0.048 & 0.048 & 0.053 & 0.046 & 0.046 \\
 & (Gau,1) & 0.039 & 0.047 & 0.041 & 0.052 & 0.052 & 0.051 & 0.051 \\
 & (Gau,Gau) & 0.048 & 0.046 & 0.051 & 0.049 & 0.051 & 0.052 & 0.049 \\
 & Opt & 0.057 & 0.051 & 0.055 & 0.049 & 0.046 & 0.053 & 0.050 \\ \hline
4 & Cph & 0.050 & 0.051 & 0.051 & 0.049 & 0.052 & 0.052 & 0.047 \\
 & (Lin,1) & 0.047 & 0.049 & 0.052 & 0.049 & 0.049 & 0.051 & 0.052 \\
 & (Gau,1) & 0.045 & 0.046 & 0.048 & 0.045 & 0.044 & 0.053 & 0.047 \\
 & (Gau,Gau) & 0.043 & 0.051 & 0.049 & 0.045 & 0.049 & 0.052 & 0.048 \\
 & Opt & 0.050 & 0.050 & 0.046 & 0.046 & 0.046 & 0.047 & 0.052 \\
 \bottomrule
\end{tabular}
\caption{The type 1 error rates of the various methods for D.1-4 defined in Table \ref{table:appendix:type1error_distributions}. The covariates are 1-dimensional. Rejection rates above 0.057 are displayed in italics.  \label{table:appendix:type1_1d}}
\end{table}

 \begin{table}[H]
\centering
\begin{tabular}{lllllllll}
\toprule
& & \multicolumn{7}{l}{ Sample size $n=$} \\ 
\cmidrule(r){3-9} 

D. &   Method & 50  & 100 & 150 & 200 & 250 & 300 & 350 \\ \hline
5 & Cph &  \textit{0.122} &  \textit{0.080} &  \textit{0.064} &  \textit{0.068} &  \textit{0.067} &  \textit{0.060} &  \textit{0.060} \\
 & (Lin,1) & 0.050 & 0.051 & 0.053 & 0.056 & 0.050 & 0.054 & \textit{0.058} \\
 & (Fis,1) & 0.054 & 0.045 & 0.051 & 0.051 & 0.050 & 0.048 & 0.045 \\
 & (Gau,1) & 0.046 & 0.041 & 0.047 & 0.053 & 0.053 & 0.054 & 0.054 \\
 & (Gau,Gau) & 0.044 & 0.043 & 0.040 & 0.049 & 0.053 & 0.048 & 0.049 \\
 & Opt & 0.052 & 0.054 & 0.054 & 0.047 & 0.049 & 0.052 & 0.054 \\ \hline
6 & Cph  & \textit{0.123} & \textit{0.076} & \textit{0.065} & \textit{0.065} & \textit{0.06}0 & \textit{0.061} & 0.056 \\
 & (Lin,1) & 0.054 & 0.051 & 0.051 & 0.053 & 0.046 & 0.054 & \textit{0.058} \\
 & (Fis,1) & 0.050 & 0.048 & 0.054 & 0.050 & 0.050 & 0.049 & 0.047 \\
 & (Gau,1) & 0.044 & 0.046 & 0.049 & 0.051 & 0.049 & 0.054 & 0.049 \\
 & (Gau,Gau) & 0.047 & 0.045 & 0.044 & 0.053 & 0.048 & 0.045 & 0.054 \\
 & Opt & 0.049 & 0.049 & 0.053 & 0.050 & 0.054 & 0.056 & 0.047\\
 \bottomrule
\end{tabular}
\caption{The type 1 error rates of the various methods for D.5-6 defined in Table \ref{table:appendix:type1error_distributions}. The covariates are 10-dimensional. Rejection rates above $0.057$ are in italics. \label{table:appendix:type1_md}}
\end{table}

\subsection{Additional experiments on test power \label{sec:appendix:additional_experiments_power}}

\begin{figure}[t]
\centering
\begin{subfigure}{.5\textwidth}
  \centering
  \includegraphics[width=1\linewidth]{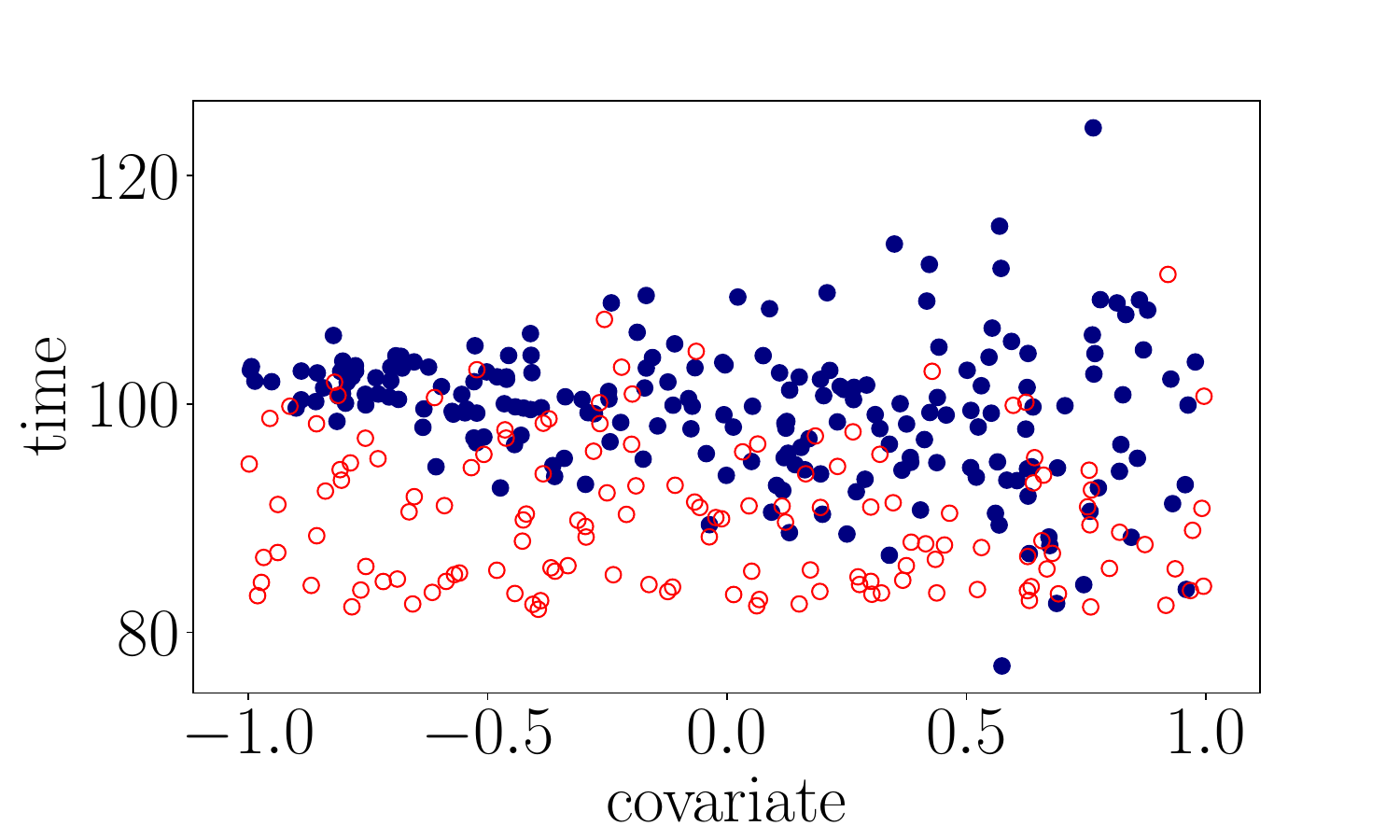}
\end{subfigure}%
\begin{subfigure}{.5\textwidth}
  \centering
  \includegraphics[width=1\linewidth]{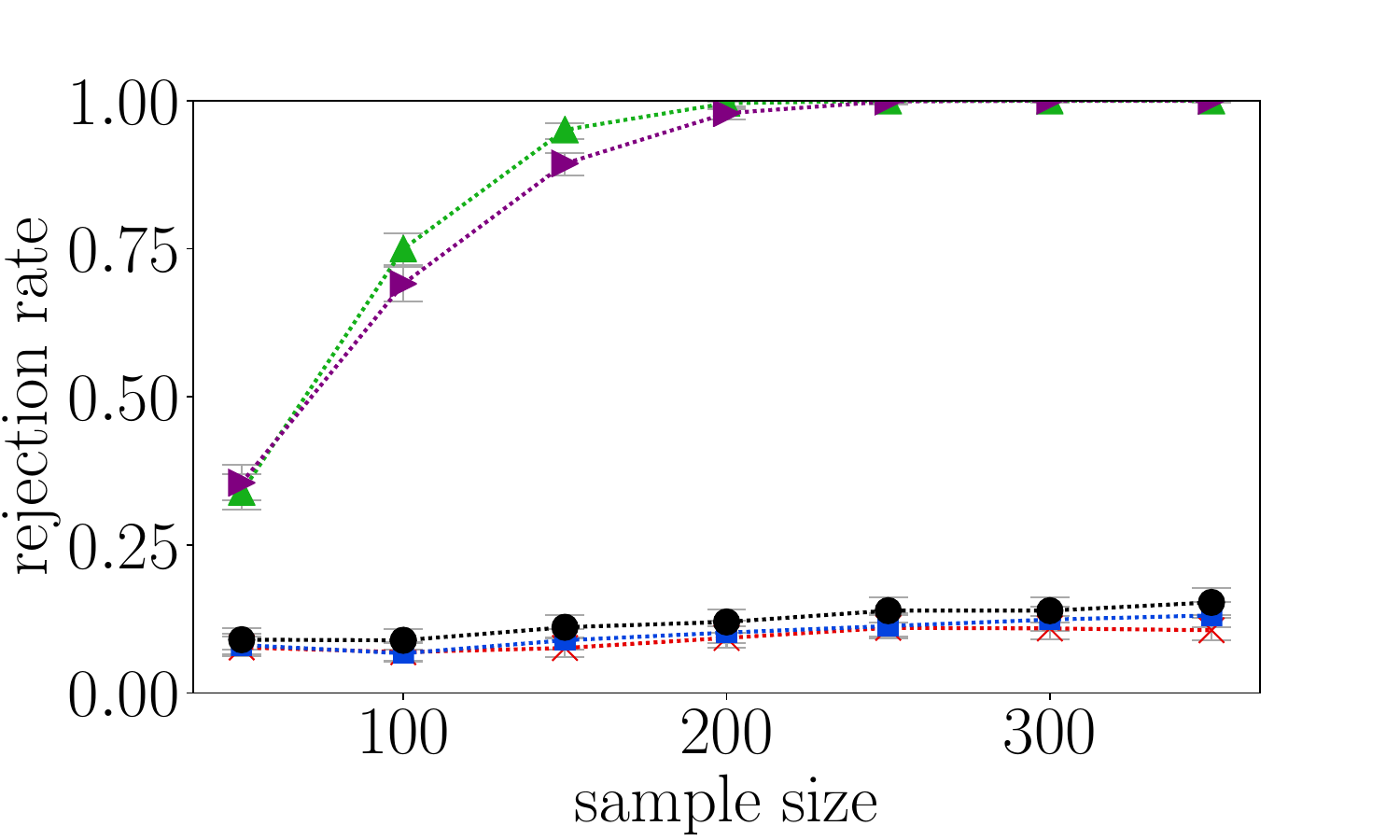}
\end{subfigure}%

\begin{subfigure}{.49\textwidth}
  \centering
  \includegraphics[width=0.4\linewidth]{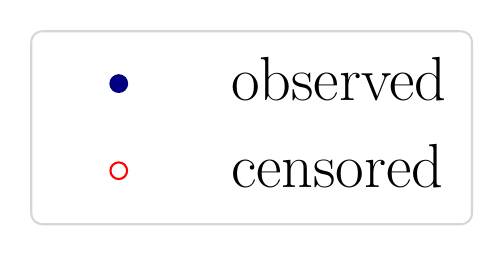}
\end{subfigure}
\begin{subfigure}{.49\textwidth}
  \centering
  \includegraphics[width=0.7\linewidth]{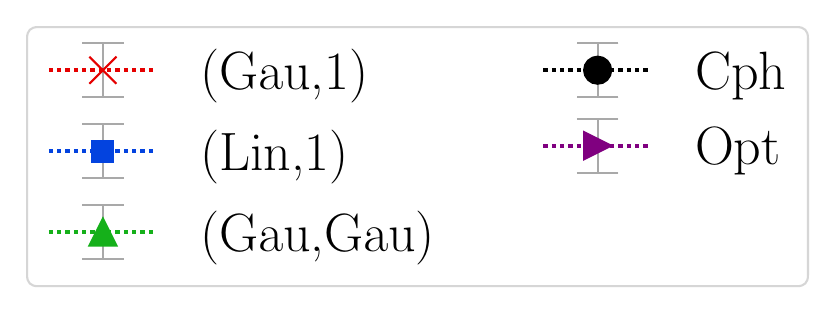}
\end{subfigure}

\caption{A scatterplot and the rejection rates for the normal distribution defined in Section \ref{sec:appendix:additional_experiments_power}.\label{fig:normals_appendix} }
\end{figure}

This section contains some experiments that were omitted from Section \ref{sec:power1dimensional}. First we investigate the power of the tests for a family of normal distributions. Second, we visualize the hazard rates of those normal distributions, and of the Weibull and Checkerboard patterns considered in the main paper. Third, we investigate the power of the different tests against deviations from the null hypothesis, and lastly we give an example of the effect of the Fisher kernel.  

\begin{figure}[H]
\centering
\begin{subfigure}{.5\textwidth}
  \centering
  \includegraphics[width=1\linewidth]{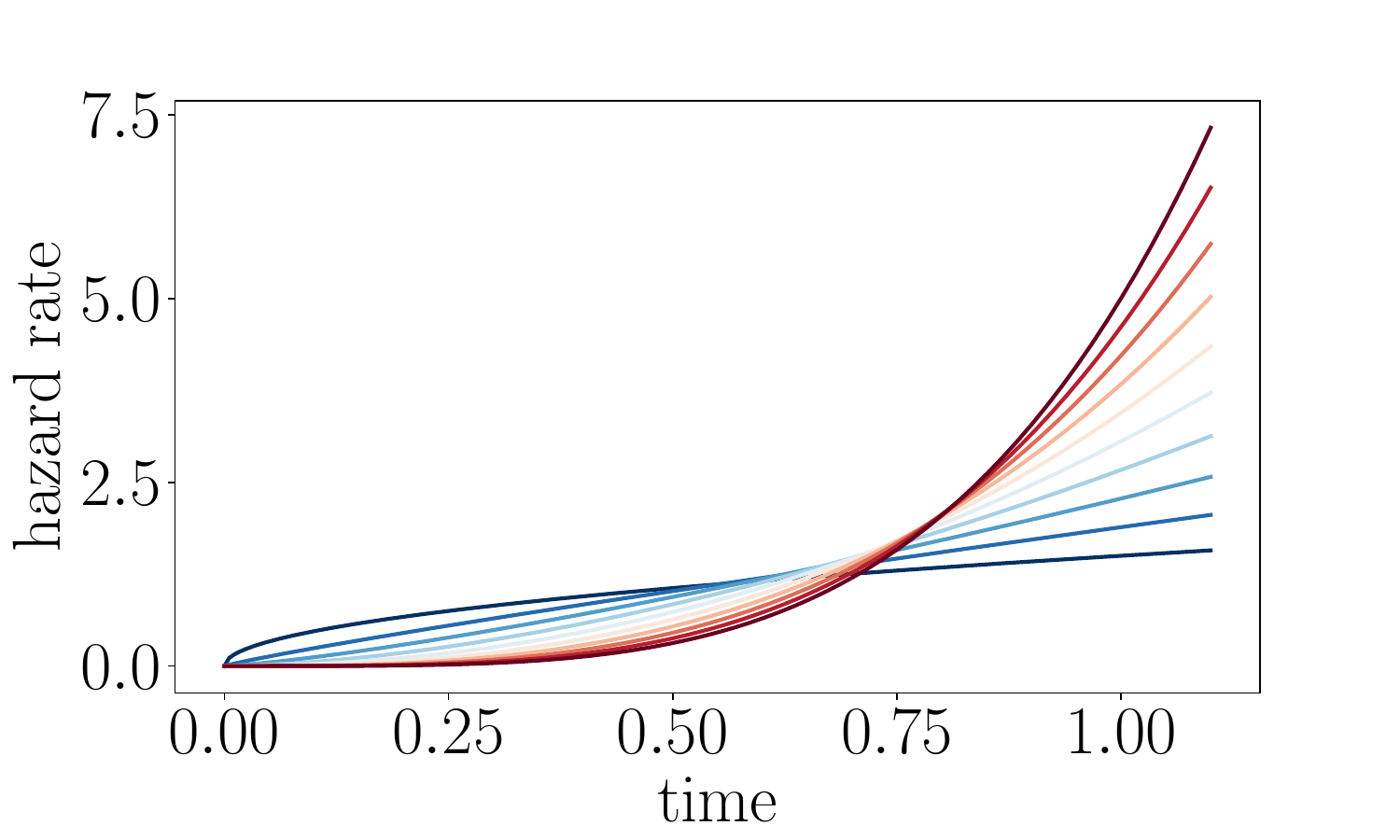}
\end{subfigure}%
\begin{subfigure}{.5\textwidth}
  \centering
  \includegraphics[width=1\linewidth]{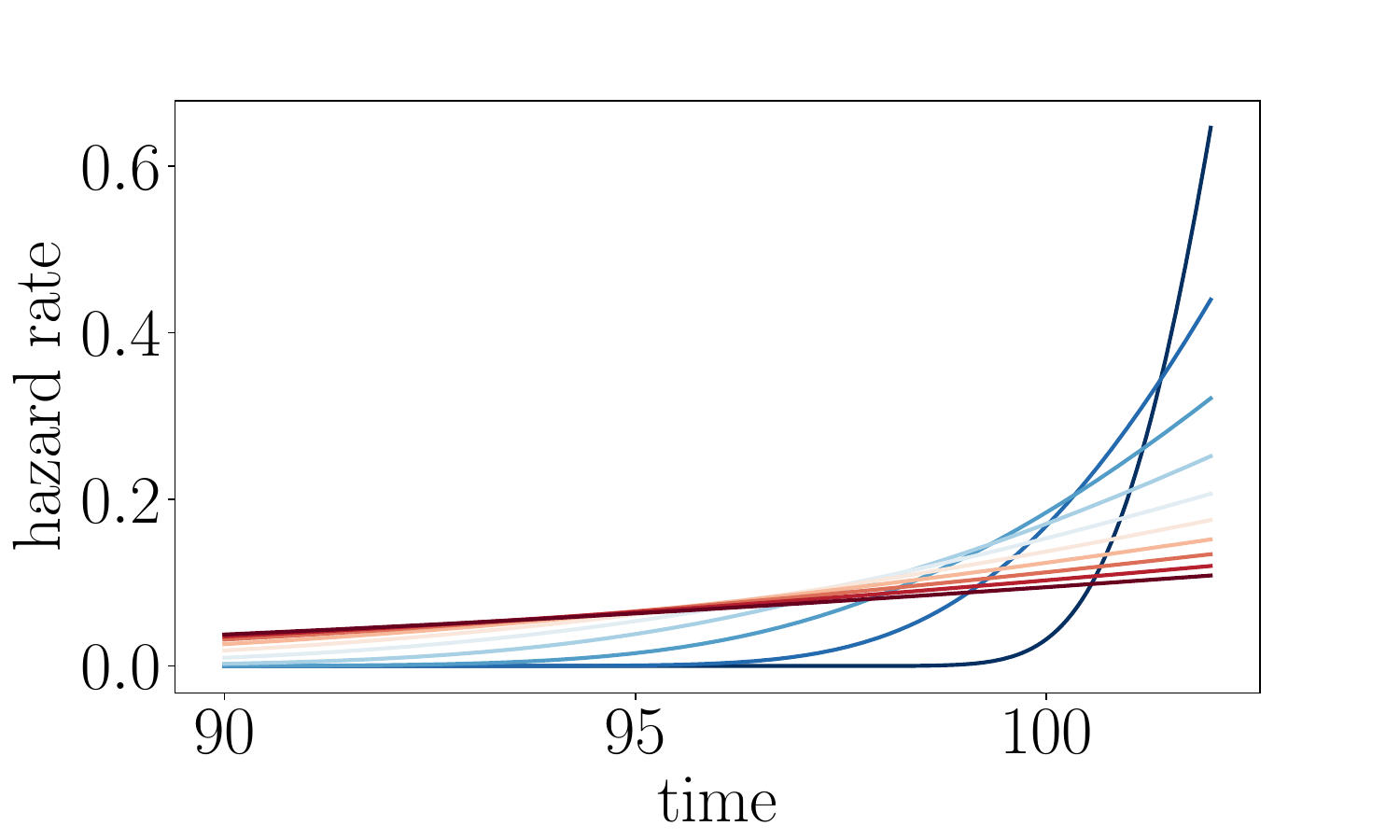}
\end{subfigure}
\centering
\begin{subfigure}{.5\textwidth}
  \centering
  \includegraphics[width=1\linewidth]{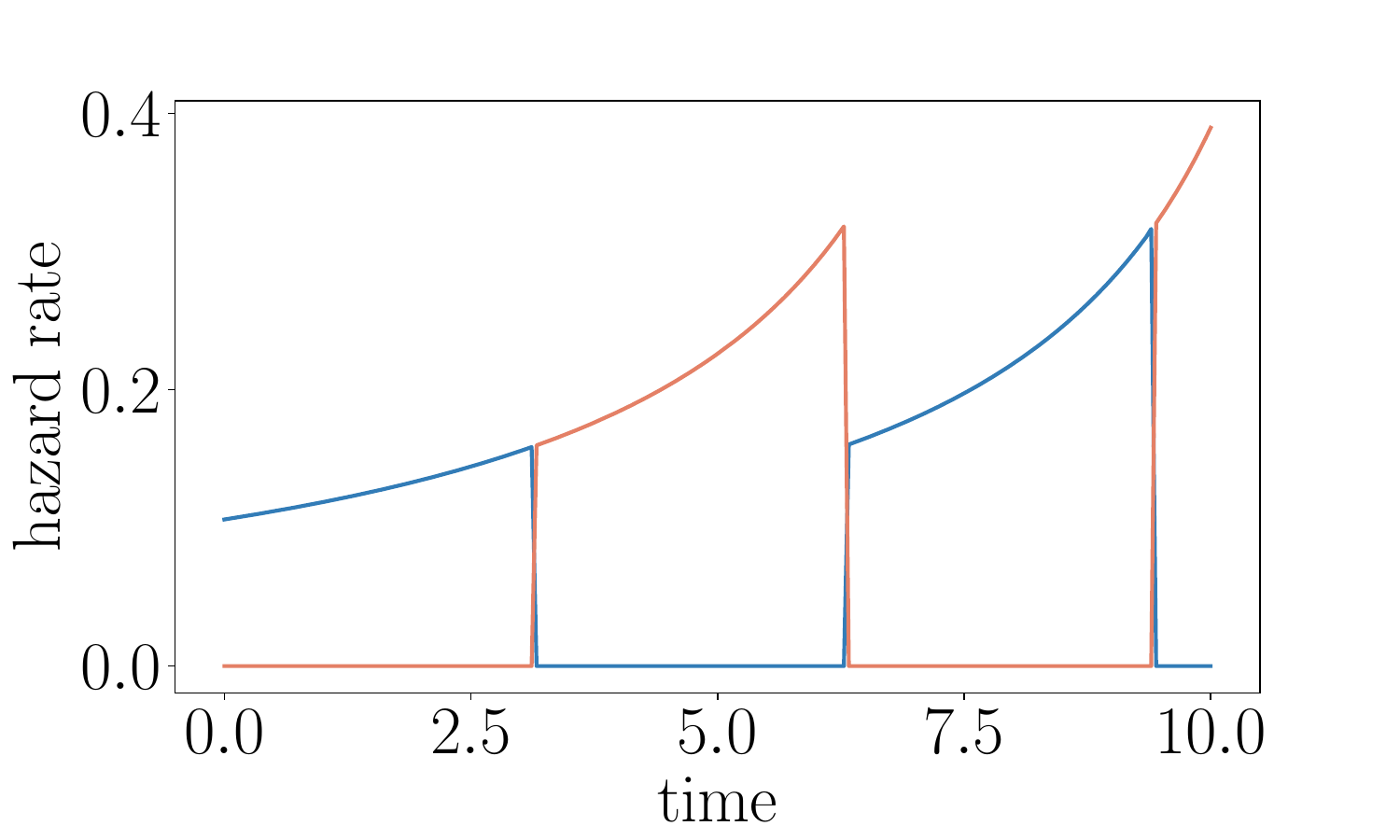}
\end{subfigure}%
\begin{subfigure}{.11\textwidth}
  \centering
  \includegraphics[width=1\linewidth]{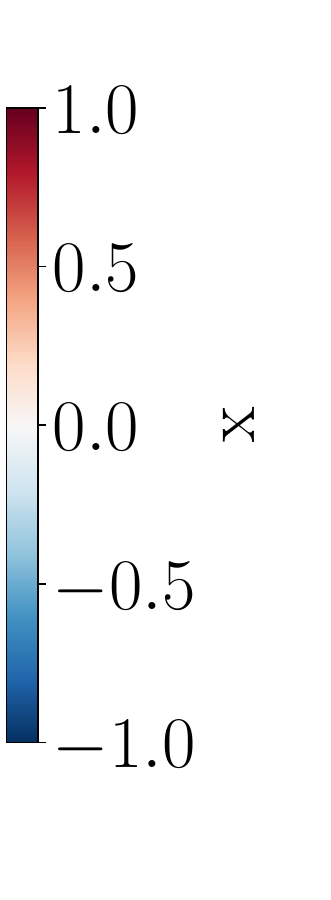}
\end{subfigure}%
\caption{\label{figure:appendix:hazardplots}Top row left: the hazard rate as a function of time for the Weibull distribution (D.3 of the main text). Top row right: the hazard rate of the normal distributions with varying bandwidths presented in Figure \ref{fig:normals_appendix}. Bottom row: the hazard rate of the checkerboard pattern (D.4 of the main text).}
\end{figure}

Firstly we consider the following distribution. 
\begin{enumerate}
    \item[] \emph{Normal distributions with different variances:} $Z \vert X=x \sim \text{Normal}(\text{mean}=100-2.25 \cdot x,\text{var}=5.5+x \cdot 4.5)$ and $C \vert X=x \sim 82+\text{Exp}(\text{mean} =35)$.
\end{enumerate}
A scatter plot and rejection rates are displayed in Figure \ref{fig:normals_appendix}. The hazard rates of this distribution, as well as of the Weibull and Checkerboard distributions in the main text, are given in Figure \ref{figure:appendix:hazardplots}.

While in the main text we investigated the power performance at different sample sizes, we now fix the sample size at 100 and study the power of the various tests for small deviations from the null hypothesis. To do so, we sample $X\sim \text{Normal}(\mu=0,\sigma^2=1)$ and consider two different settings in which $\theta$ varies from $-0.4$ to $0.4$ in steps of $0.1$. Notice that, for the two settings described below, $\theta=0$ recovers the null hypothesis. 
\begin{enumerate}
\item[] \emph{Cph distributions:} $Z \vert X=x \sim \text{Exp}(\text{mean}= \exp\{\theta \cdot x\})$ and $C \vert X=x \sim \text{Exp}(\text{mean}=1.5)$
    
\item[] \emph{Non-linear log-hazard distributions:} $Z \vert X=x \sim \text{Exp}(\text{mean} =\exp\{\theta \cdot x^2\})$ and $C \vert X=x \sim \text{Exp}(\text{mean}= 1.5)$
\end{enumerate}

Figure \ref{plot:deviationsfromnull} below shows the rejection rates against deviations from the null for each method.
\begin{figure}[H]
\centering
\begin{subfigure}{.5\textwidth}
  \centering
  \includegraphics[width=1\linewidth]{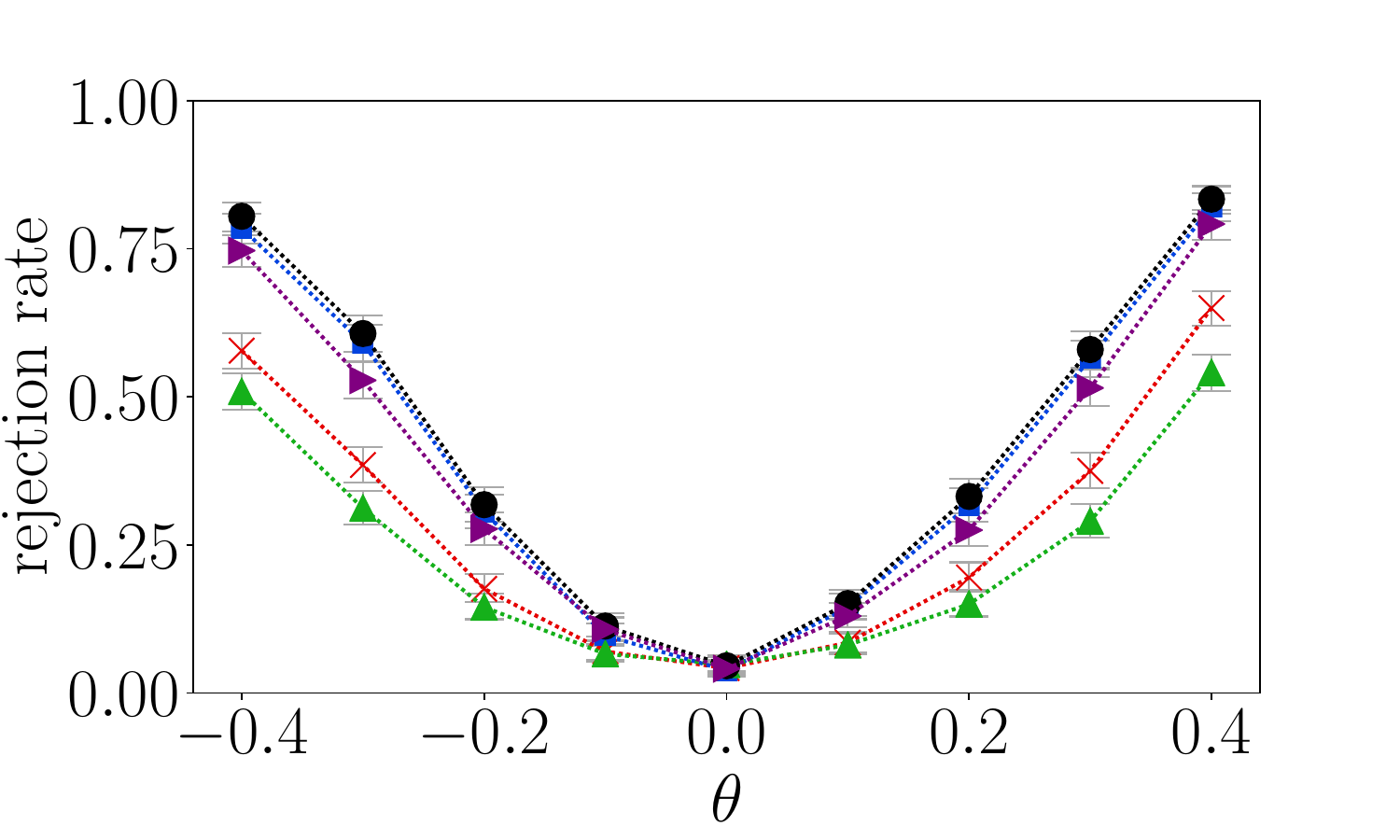}
\end{subfigure}%
\begin{subfigure}{.5\textwidth}
  \centering
  \includegraphics[width=1\linewidth]{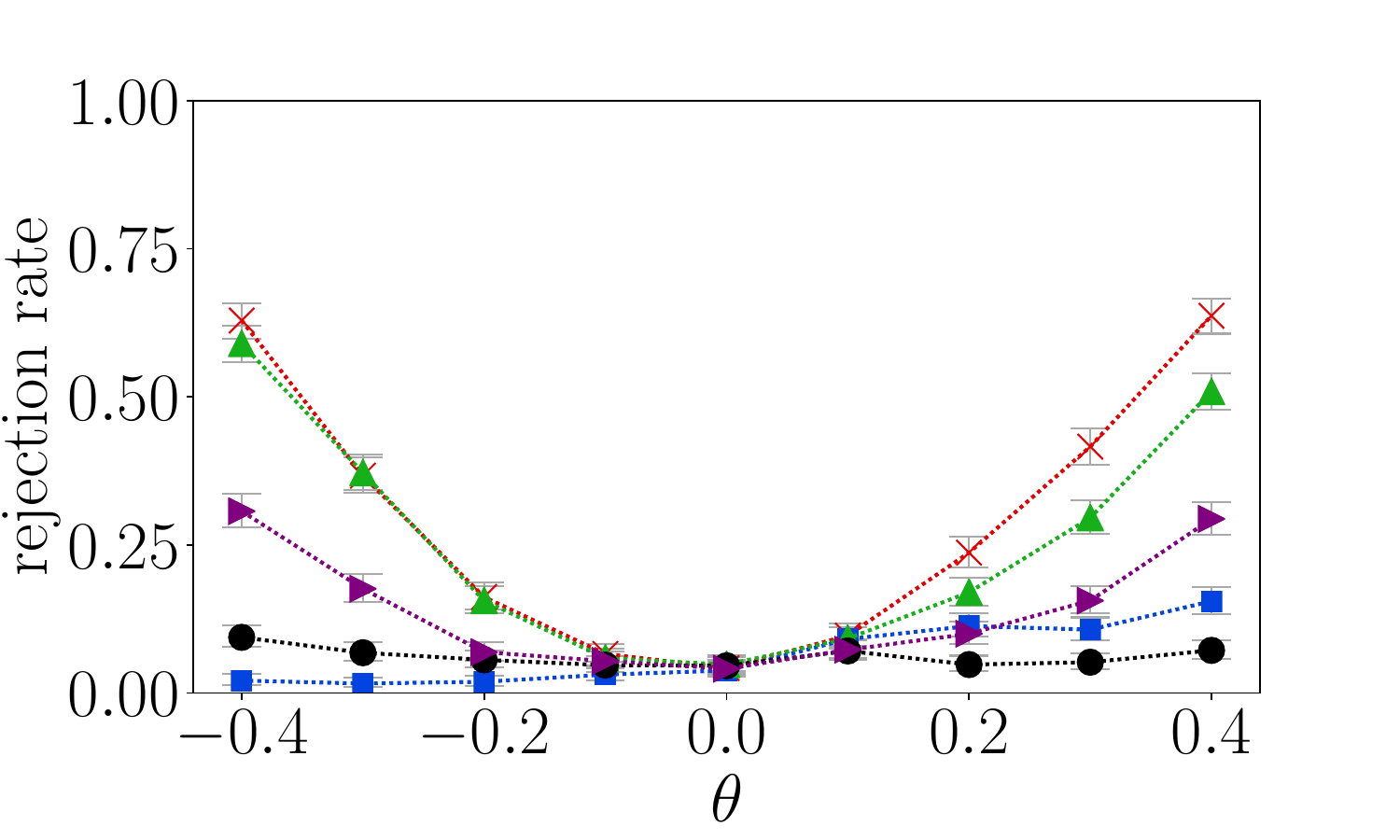}
\end{subfigure}
\begin{subfigure}{.5\textwidth}
  \centering
  \includegraphics[width=1\linewidth]{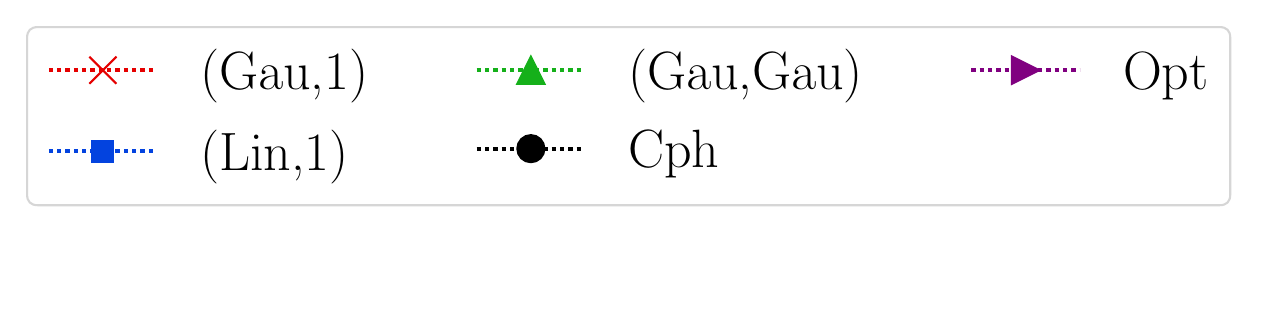}
\end{subfigure}
\caption{\label{plot:deviationsfromnull} Left: the rejection rates of the various tests against the Cph distributions. Right: the rejection rates of the various tests against the non-linear log-hazard distributions. }
\end{figure}

\begin{figure}[H]
\centering
\begin{subfigure}{.6\textwidth}
  \centering
  \includegraphics[width=1\linewidth]{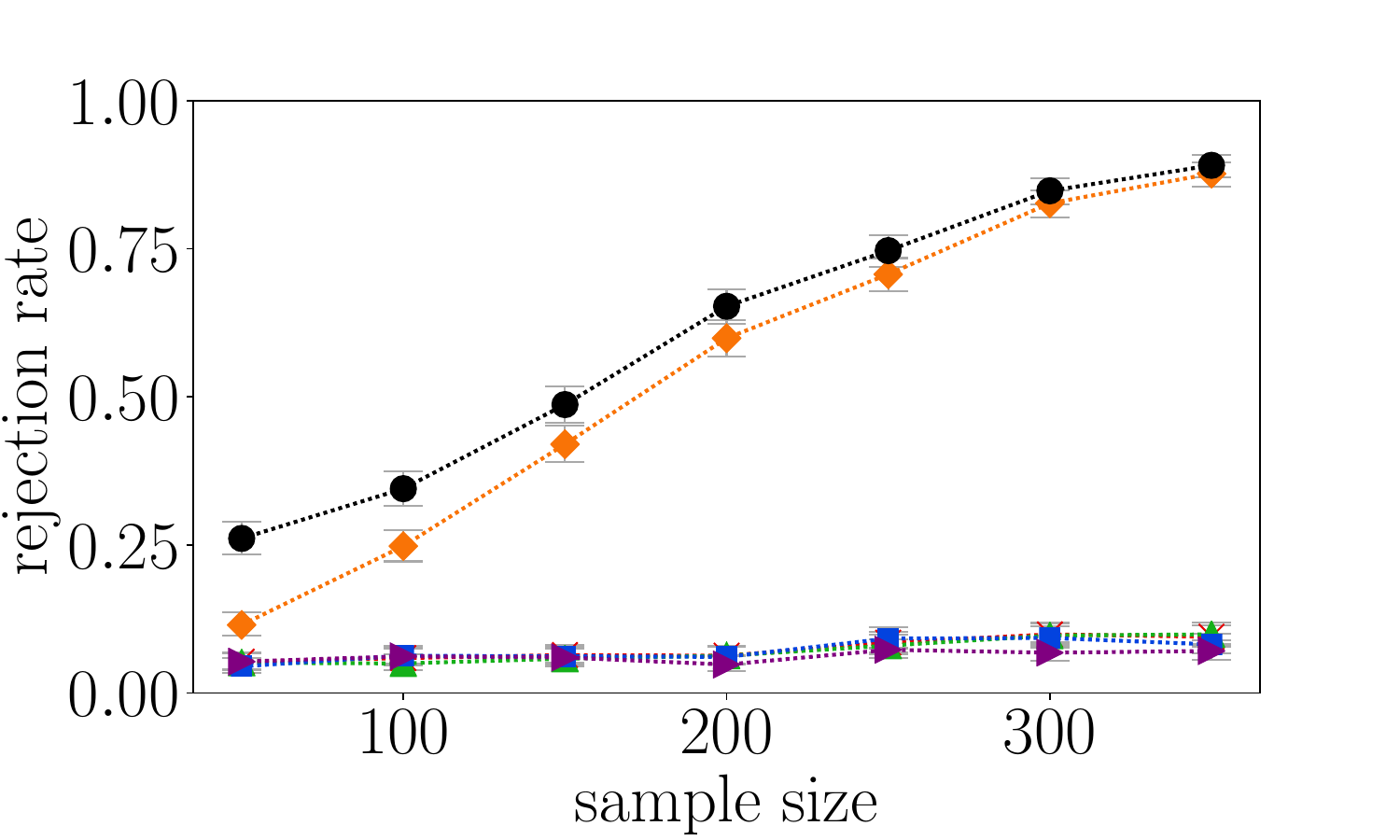}
\end{subfigure}%
\begin{subfigure}{.4\textwidth}
  \centering
  \includegraphics[width=1\linewidth]{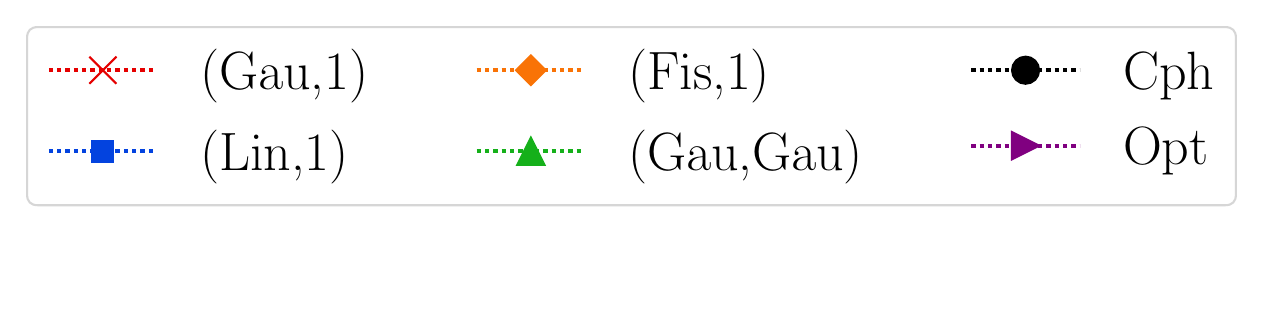}
\end{subfigure}%
\caption{\label{plot:rescalevaraibles}The rejection rate of the different methods for the distribution in which the Fisher information helps uncover the dependency.}
\end{figure}

We now provide an example of the usefulness of the Fisher kernel. In the experiments in the main text, the kernels $(\text{Lin},1)$, $(\text{Gau},1)$ and $(\text{Gau},\text{Gau})$ all performed consistently well for high dimensional covariates, and thus the usefulness of the Fisher kernel was not obvious. We now show the Fisher kernel has the ability to `standardize' the data in cases where simple centering and scaling of the individual covariate dimensions is not sufficient. To illustrate this, we consider the following example. Let $\Sigma_{11}=1/10$, $\Sigma_{ii}=1$ for $i>1$ and $\Sigma_{ij}=0$ otherwise, and let $R$ be an orthogonal rotation matrix.  Let
$X \sim \text{Normal}(0,R\Sigma R^{T})$ and let $v=(1,0,...,0)^T$.
\begin{enumerate}
    \item[] \emph{A distribution in which the Fisher information helps uncover the dependency:} Let $Z \vert X=x \sim \text{Exp}(\text{mean}=  \exp\{v^T(R^T x)\})$ and $C \vert X=x \sim \text{Exp}(\text{mean}= 1.5)$
\end{enumerate}
Figure \ref{plot:rescalevaraibles} shows that the kernel log-rank tests with kernels (Lin,1) and (Gau,1) do not detect the dependency. The Fisher kernel and CPH LR tests have power, because the inverse information matrix has the effect of standardizing the data.

\subsection{Dependent censoring and varying censoring rates \label{sec:appendix:varying_censoring}}
We now illustrate the performance of the wild bootstrap test using the kernel log-rank statistic when the censoring rate varies, and when the censoring time depends on the covariate. In particular, we parametrize the censoring distributions and vary the parameters to estimate the power of each method when 15, 30, 45, 60, 75, 90 or 100\% of the observations are uncensored. We study the type 1 error rate in Section \ref{sec:appendix:varying_censoring_type1} and the power in Section \ref{sec:appendix:varying_censoring_power}.

\subsubsection{Type 1 error under varying censoring rates \label{sec:appendix:varying_censoring_type1}}
The distributions from which we sample data in which $X\perp Z$ are given in Table \ref{table:appendix:varyingcensoring_type1_distribution}. In each case the sample size is $200$. To estimate the rejection rates we sampled 5000 times, and counted the number of times we rejected the null. Obtained rejection rates are displayed in Tables \ref{table:appendix:varyingcensoring_type1_error} and \ref{table:appendix:varyingcensoring_type1_md}. In each of the scenarios we find the type 1 error rate to be correct for each censoring percentage and for each kernel. 

\begin{table}[H]
\centering
\resizebox{\textwidth}{!}{
\begin{tabular}{  l l l l  }
\toprule
    D. &  $ Z \vert X$  & $C \vert X$  & X \\ \hline
    1   & $ \text{Exp}(\text{mean}=1) $ & $ \text{Exp}(\text{mean}=\theta) $  & \text{Unif}[-1,1] \\ 
    2   & $ \text{Exp}(\text{mean}= 2.2) $ & $ \text{Exp}(\text{mean}=\theta \exp(X)) $  & \text{Unif}[-1,1] \\ 
    3  & $ \text{Weib}(\text{shape}= 3.25) $ & $ \text{Exp}(\text{mean}=\theta X^2) $  & \text{Unif}[-1,1] \\ 
   4 & $ \mathcal{N}(\text{mean}=99,\text{var}= 5.5) $ & $ \text{Exp}(\text{mean}=82+\theta) $  & \text{Unif}[-1,1] \\ 
    5 & $ \text{Exp}(\text{mean}=1)$ & $ \text{Exp}(\text{mean}=\theta) $   & $\mathcal{N}_{10}(0,\text{cov}=\Sigma_{10})$  \\
    6  & $ \text{Exp}(\text{mean}=1) $ & $ \text{Exp}(\text{mean}=\theta \exp(1^T X/30)) $  & $\mathcal{N}_{10}(0,\text{cov}=\Sigma_{10})$   \\
    7  & $ \text{Exp}(\text{mean}=\exp(0.5) $ & $ \text{Exp}(\text{mean}=\theta\exp(X^2_2)/20) $  & $\mathcal{N}_{10}(0,\text{cov}=\Sigma_{10})$   \\ \bottomrule
\end{tabular}}\caption{\label{table:appendix:varyingcensoring_type1_distribution}The parametrized distributions to test the type 1 error rate under different censoring rates.  Here $\Sigma_{10}=MM^T$ where $M$ is a $10\times 10$ matrix of i.i.d. standard normal entries. The parameter $\theta$ is varies such that $15,30,45,60,75,90,100\%$ of the individuals are observed (i.e. $\Delta=1$). The sample size is $200$ in each case.  }    
\end{table}

\begin{table}[H]
\centering
\begin{tabular}{lllllllll}
\toprule
& & \multicolumn{7}{l}{ \% Observed} \\ 
\cmidrule(r){3-9} 

D. &   Method & 15\% & 30\% & 45\% & 60\% & 75\% & 90\% & 100\% \\ \hline
1 & Cph  & 0.053 & 0.050 & 0.047 & 0.049 & 0.052 & 0.050 & 0.046 \\
 & (Lin,1) & 0.051 & 0.046 & 0.046 & 0.046 & 0.049 & 0.048 & 0.044 \\
 & (Gau,1) & 0.046 & 0.045 & 0.047 & 0.050 & 0.053 & 0.048 & 0.044 \\
 & (Gau,Gau) & 0.045 & 0.046 & 0.044 & 0.046 & 0.056 & 0.053 & 0.045 \\
 & Opt & 0.048 & 0.047 & 0.052 & 0.045 & 0.053 & 0.046 & 0.045 \\ \midrule
2 & Cph & 0.050 & 0.050 & 0.049 & 0.055 & 0.050 & 0.057 & 0.046 \\
 & (Lin,1) & 0.049 & 0.049 & 0.047 & 0.053 & 0.050 & 0.055 & 0.046 \\
 & (Gau,1) & 0.050 & 0.050 & 0.047 & 0.045 & 0.050 & 0.053 & 0.045 \\
 & (Gau,Gau) & 0.046 & 0.048 & 0.044 & 0.044 & 0.048 & 0.052 & 0.049 \\
 & Opt & 0.053 & 0.055 & 0.050 & 0.051 & 0.050 & 0.053 & 0.049 \\ \midrule
3 & Cph & 0.054 & 0.051 & 0.050 & 0.055 & 0.053 & 0.054 & 0.046 \\
 & (Lin,1) & 0.044 & 0.048 & 0.047 & 0.052 & 0.051 & 0.054 & 0.046 \\
 & (Gau,1) & 0.043 & 0.048 & 0.044 & 0.049 & 0.054 & 0.050 & 0.046 \\
 & (Gau,Gau) & 0.042 & 0.047 & 0.040 & 0.048 & 0.053 & 0.050 & 0.047 \\
 & Opt & 0.054 & 0.052 & 0.050 & 0.053 & 0.051 & 0.052 & 0.050 \\ \midrule
4 & Cph & 0.055 & 0.051 & 0.054 & 0.050 & 0.054 & 0.052 & 0.050 \\
 & (Lin,1) & 0.053 & 0.047 & 0.049 & 0.047 & 0.051 & 0.050 & 0.049 \\
 & (Gau,1) & 0.049 & 0.045 & 0.048 & 0.047 & 0.050 & 0.051 & 0.051 \\
 & (Gau,Gau) & 0.044 & 0.047 & 0.046 & 0.044 & 0.052 & 0.050 & 0.048 \\
 & Opt & 0.050 & 0.044 & 0.049 & 0.047 & 0.055 & 0.053 & 0.049 \\
 \bottomrule
\end{tabular}
\caption{\label{table:appendix:varyingcensoring_type1_error} The type 1 error rates of the various methods under different censoring rates. D.1-4 are defined in Table \ref{table:appendix:varyingcensoring_type1_distribution}. The covariates are 1-dimensional. There are no rejection rates larger than 0.057. }
\end{table}

\begin{table}[H]
\centering
\begin{tabular}{lllllllll}
\toprule
& & \multicolumn{7}{l}{ \% Observed} \\ 
\cmidrule(r){3-9} 
 D. & Method & 15\% & 30\% & 45\% & 60\% & 75\% & 90\% & 100\% \\ \hline
5 & Cph & \textit{0.063} &  \textit{0.061} &  \textit{0.068} &  0.056 &  \textit{0.062} &  \textit{0.058} &  \textit{0.062} \\
 & (Lin,1) & 0.046 & 0.043 & 0.056 & 0.047 & 0.048 & 0.048 & 0.049 \\
 & (Fis,1) & 0.048 & 0.047 & 0.057 & 0.047 & 0.049 & 0.046 & 0.050 \\
 & (Gau,1) & 0.044 & 0.044 & 0.056 & 0.046 & 0.046 & 0.048 & 0.050 \\
 & (Gau,Gau) & 0.040 & 0.047 & 0.053 & 0.048 & 0.046 & 0.046 & 0.049 \\
 & Opt & 0.053 & 0.045 &  \textit{0.059} & 0.048 & 0.048 & 0.049 & 0.054 \\ \midrule
6 & Cph &  \textit{0.061} &  \textit{0.066} &  \textit{0.066} &  \textit{0.064} &  \textit{0.063} &  \textit{0.066} &  \textit{0.063} \\
 & (Lin,1)  & 0.051 & 0.054 & 0.056 & 0.048 & 0.050 & 0.052 & 0.050 \\
 & (Fis,1) & 0.049 & 0.050 & 0.051 & 0.049 & 0.053 & 0.052 & 0.051 \\
 & (Gau,1) & 0.048 & 0.049 & 0.055 & 0.050 & 0.052 & 0.052 & 0.052 \\
 & (Gau,Gau) & 0.050 & 0.048 & 0.057 & 0.048 & 0.050 & 0.048 & 0.053 \\
 & Opt & 0.055 & 0.051 &  \textit{0.058} & 0.049 & 0.048 & 0.052 & 0.053 \\ \midrule
7 & Cph &  \textit{0.068} &  \textit{0.068} &  \textit{0.067} &  \textit{0.064} &  \textit{0.064} &  \textit{0.064} &  \textit{0.063} \\
 & (Lin,1) & 0.048 & 0.054 & 0.055 & 0.048 & 0.049 & 0.050 & 0.050 \\
 & (Fis,1) & 0.051 & 0.055 & 0.055 & 0.046 & 0.050 & 0.053 & 0.051 \\
 & (Gau,1) & 0.047 & 0.049 & 0.053 & 0.047 & 0.050 & 0.050 & 0.052 \\
 & (Gau,Gau) & 0.047 & 0.048 &  \textit{0.059} & 0.048 & 0.051 & 0.049 & 0.053 \\
 & Opt & 0.047 & 0.049 & 0.053 & 0.054 & 0.051 & 0.051 & 0.053 \\ \bottomrule
\end{tabular}
\caption{The type 1 error rate of the various methods under different censoring rates. D.5-7 are found in Table \ref{table:appendix:varyingcensoring_type1_distribution}. The covariates are 10-dimensional. Rejection rates above 0.057 are displayed in italics.  \label{table:appendix:varyingcensoring_type1_md} }
\end{table}
\npnoround

\subsubsection{Power under varying censoring rates \label{sec:appendix:varying_censoring_power}}

We now estimate the power when the alternative hypothesis holds. To estimate power we sample 1000 times from each distribution of Table \ref{table:appendix:varyingcensoring_power_distributions} and count the number of rejections. In each case the sample size equals $200$. Obtained rejection rates are displayed in Tables \ref{table:appendix:varyingcensoring_power_values} and \ref{table:appendix:varyingcensoring_power_values_md}.

\begin{table}[H]
\centering
\resizebox{\textwidth}{!}{
\begin{tabular}{  l l l l  }
\toprule
    D. &  $ Z \vert X$  & $C \vert X$  & X \\ \hline
    8   & $ \text{Exp}(\text{mean}=\exp(X/3)) $ & $ \text{Exp}(\text{mean}=\theta) $  & \text{Unif}[-1,1] \\ 
    9   & $ \text{Exp}(\text{mean}=\exp(X^2)) $ & $ \text{Exp}(\text{mean}=\theta \exp(X)) $  & \text{Unif}[-1,1] \\ 
    10  & $ \text{Weib}(\text{shape}= 1.5X+3.25) $ & $ \text{Exp}(\text{mean}=\theta X^2) $  & \text{Unif}[-1,1] \\ 
   11 & $ \mathcal{N}(\text{mean}=100-X,\text{var}= 1.5X+5.5) $ & $ \text{Exp}(\text{mean}=82+\theta) $  & \text{Unif}[-1,1] \\ 
    12 & $ \text{Exp}(\text{mean}=\exp(1^T X/30))$ & $ \text{Exp}(\text{mean}=\theta) $   & $\mathcal{N}_{10}(0,\text{cov}=\Sigma_{10})$  \\
    13  & $ \text{Exp}(\text{mean}=\exp(X_4/7)) $ & $ \text{Exp}(\text{mean}=\theta \exp(1^T X/30)) $  & $\mathcal{N}_{10}(0,\text{cov}=\Sigma_{10})$   \\
    14  & $ \text{Exp}(\text{mean}=\exp(X_4^2/20)) $ & $ \text{Exp}(\text{mean}=\theta\exp(X^2_2)/20) $  & $\mathcal{N}_{10}(0,\text{cov}=\Sigma_{10})$   \\ \bottomrule
\end{tabular}}\caption{\label{table:appendix:varyingcensoring_power_distributions}The parametrized distributions to test the power under different censoring rates.  Here $\Sigma_{10}=MM^T$ where $M$ is a $10\times 10$ matrix of i.i.d. standard normal entries. The parameter $\theta$ is varies such that $15,30,45,60,75,90,100\%$ of the individuals are observed (i.e. $\Delta=1$). The sample size is $200$ in each case.  }    
\end{table}

\begin{table}[H]
\centering
\begin{tabular}{lllllllll}
\toprule
& & \multicolumn{7}{l}{ \% Observed} \\ 
\cmidrule(r){3-9} 

D. &   Method & 15\% & 30\% & 45\% & 60\% & 75\% & 90\% & 100\% \\ \hline
8    & Cph                                              & \textbf{0.16} & \textbf{0.30} & \textbf{0.44} & \textbf{0.56} & \textbf{0.67} & \textbf{0.72} & \textbf{0.76}  \\
     & (Lin,1)                                          & \textbf{0.16} & 0.28 & \textbf{0.44} & 0.55 & 0.66 & 0.71 & \textbf{0.76}  \\
     & (Gau,1)                                          & 0.14 & 0.24 & 0.36 & 0.48 & 0.58 & 0.64 & 0.68  \\
     & (Gau,Gau)                                        & 0.12 & 0.20 & 0.31 & 0.40 & 0.48 & 0.54 & 0.57  \\
     & Opt                                              & \textbf{0.16} & 0.29 & 0.41 & 0.52 & 0.62 & 0.68 & 0.71  \\ \hline
9    & Cph                                              & 0.06 & 0.10 & 0.10 & 0.10 & 0.08 & 0.08 & 0.06  \\
     & (Lin,1)                                          & 0.09 & 0.13 & 0.14 & 0.14 & 0.12 & 0.11 & 0.09  \\
     & (Gau,1)                                          & \textbf{0.22} & \textbf{0.41} & \textbf{0.56} & \textbf{0.68} & \textbf{0.79} & \textbf{0.85} & \textbf{0.90}  \\
     & (Gau,Gau)                                        & 0.16 & 0.31 & 0.42 & 0.54 & 0.66 & 0.77 & 0.84  \\
     & Opt                                              & 0.08 & 0.13 & 0.19 & 0.26 & 0.40 & 0.52 & 0.63  \\ \hline
10    & Cph                                              & 0.11 & 0.06 & 0.08 & 0.12 & 0.17 & 0.21 & 0.24  \\
     & (Lin,1)                                          & 0.10 & 0.05 & 0.06 & 0.09 & 0.14 & 0.19 & 0.22  \\
     & (Gau,1)                                          & 0.10 & 0.04 & 0.05 & 0.08 & 0.12 & 0.16 & 0.16  \\
     & (Gau,Gau)                                        & \textbf{0.38} & \textbf{0.54} & \textbf{0.68} & \textbf{0.76} & \textbf{0.84} & 0.88 & \textbf{0.87}  \\
     & Opt                                              & 0.22 & 0.16 & 0.18 & 0.34 & 0.63 & \textbf{0.89} & 0.86  \\ \hline
11    & Cph                                              & 0.29 & 0.12 & 0.08 & 0.05 & 0.06 & 0.12 & 0.14  \\
     & (Lin,1)                                          & 0.27 & 0.11 & 0.06 & 0.04 & 0.06 & 0.10 & 0.12  \\
     & (Gau,1)                                          & 0.22 & 0.09 & 0.07 & 0.04 & 0.05 & 0.09 & 0.11  \\
     & (Gau,Gau)                                        & 0.42 & \textbf{0.51} & \textbf{0.62} & \textbf{0.76} & \textbf{0.86} & 0.94 & 0.95  \\
         & Opt                                              & \textbf{0.54} & 0.42 & 0.39 & 0.57 & 0.85 & \textbf{0.98} & \textbf{0.99}  \\ 
 \bottomrule
\end{tabular}
\caption{The power of the various methods under different censoring rates. The Distributions 8-11 are found in Table \ref{table:appendix:varyingcensoring_power_distributions}. The covariates are 1-dimensional. The highest rejection rate for each scenario and percentage of observed events is in bold. If an elevated type 1 error rate was found for a method under the censoring distribution in question - see Table \ref{table:appendix:varyingcensoring_type1_error}, then the rejection rate is displayed in italics.  \label{table:appendix:varyingcensoring_power_values}}
\end{table}

\begin{table}[H]
\centering
\begin{tabular}{lllllllll}
\toprule
& & \multicolumn{7}{l}{ \% Observed} \\ 
\cmidrule(r){3-9} 
 D. & Method & 15\% & 30\% & 45\% & 60\% & 75\% & 90\% & 100\% \\ \hline
12    & Cph                                             & \textit{\textbf{0.22}} & \textit{0.40} & \textit{\textbf{0.58}} & \textit{0.68} & \textit{0.83} & \textit{0.88} & \textit{0.92}  \\
     & (Lin,1)                                          & 0.21 & \textbf{0.44} & \textbf{0.58} & \textbf{0.70} & \textbf{0.84} & \textbf{0.90} & \textbf{0.94}  \\
     & (Fis,1)                                          & 0.17 & 0.34 & 0.53 & 0.64 & 0.79 & 0.86 & 0.90  \\
     & (Gau,1)                                          & 0.20 & 0.40 & 0.55 & 0.67 & 0.82 & 0.88 & 0.91  \\
     & (Gau,Gau)                                        & 0.16 & 0.30 & 0.46 & 0.56 & 0.73 & 0.78 & 0.83  \\
     & Opt                                              & 0.18 & 0.33 & 0.49 & 0.59 & 0.74 & 0.82 & 0.91  \\ \hline
13   & Cph                                              & \textit{0.34} & \textit{0.60} & \textit{0.81} & \textit{0.92} & \textit{0.96} & \textit{0.98} & \textit{\textbf{1.00}}  \\
     & (Lin,1)                                          & \textbf{0.39} & \textbf{0.64} & \textbf{0.87} & \textbf{0.95} & \textbf{0.98} & \textbf{0.99} & \textbf{1.00}  \\
     & (Fis,1)                                          & 0.29 & 0.54 & 0.76 & 0.90 & 0.96 & 0.97 & 0.99  \\
     & (Gau,1)                                          & 0.33 & 0.59 & 0.81 & 0.92 & 0.97 & 0.97 & 0.99  \\
     & (Gau,Gau)                                        & 0.24 & 0.49 & 0.73 & 0.84 & 0.93 & 0.95 & 0.98  \\
     & Opt                                              & 0.30 & 0.53 & 0.74 & 0.86 & 0.95 & 0.96 & 0.99  \\ \hline
14    & Cph                                              & \textit{0.06} & \textit{0.08} & \textit{0.07} & \textit{0.08} & \textit{0.09} & \textit{0.08} & \textit{0.11}  \\
     & (Lin,1)                                          & 0.06 & 0.06 & 0.08 & 0.10 & 0.15 & 0.15 & 0.18  \\
     & (Fis,1)                                          & 0.04 & 0.07 & 0.06 & 0.07 & 0.08 & 0.08 & 0.10  \\
         & (Gau,1)                                          & \textbf{0.08} & \textbf{0.14} & \textbf{0.28} & \textbf{0.48} & \textbf{0.76} & \textbf{0.91} & \textbf{0.96}  \\
     & (Gau,Gau)                                        & 0.07 & 0.11 & 0.18 & 0.29 & 0.52 & 0.65 & 0.75  \\
     & Opt                                              & 0.05 & 0.06 & 0.08 & 0.14 & 0.25 & 0.40 & 0.81  \\ \bottomrule
\end{tabular}
\caption{The power of the various methods under different censoring rates. The Distributions 12-14 are found in Table \ref{table:appendix:varyingcensoring_power_distributions}. The covariates are 10-dimensional. The highest rejection rate for each scenario and percentage of observed events is in bold. If an elevated type 1 error rate was found for a method under the censoring distribution in question (see Table  \ref{table:appendix:varyingcensoring_type1_md}), then the rejection rate is displayed in italics. \label{table:appendix:varyingcensoring_power_values_md}}
\end{table}
\npnoround

\subsection{Varying bandwidths of the Gaussian Kernel \label{sec:appendix:bandwidths}}
We now study how the rejection rate of the kernel log-rank test with kernel $(\text{Gau},1)$ or $(\text{Gau},\text{Gau})$ is affected by the choice of the bandwidth. Note that data is scaled so that each component has mean zero and unit variance. In this section $n=200$ and $60$ or $75 \%$ of the observations are observed. In Sections \ref{sec:appendix:bandwidths_gau1_type1} and \ref{sec:appendix:bandwidths_gau1_power} we study the type 1 error rate and power of $(\text{Gau},1)$ respectively, and in Sections \ref{sec:appendix:bandwidths_gaugau_type1} and \ref{sec:appendix:bandwidths_gaugau_power} we study the type 1 error rate and power of $(\text{Gau},\text{Gau})$ respectively. For estimation of the type 1 error rate we sample 5000 times from each distribution, and for estimation of the power we sample 1000 times from each distribution.

\subsubsection{Type 1 error rate of the kernel (Gau,1) for different bandwidths \label{sec:appendix:bandwidths_gau1_type1}}

Obtained rejection rates for the distributions in Table \ref{table:appendix:varyingcensoring_type1_distribution} are displayed in Table \ref{table:appendix:bandwidths_type1_gau1}. We find that when the bandwidth is too small relative to the median heuristic, the type 1 error rate is inflated. 

\begin{table}[H]
\centering
\begin{tabular}{@{}lllllllllll@{}}
\toprule
& & \multicolumn{9}{l}{ $\sigma_X=$} \\ 
\cmidrule(r){3-11} 
 D.&  $\Delta=1$          & $\sigma_{\text{med}}$ & 0.1  & 0.2  & 0.5  & 1    & 2    & 5    & 10   & 20   \\ \midrule
1 & $60\%$   & 0.049 & 0.044 & 0.044 & 0.051 & 0.050 & 0.047 & 0.050 & 0.048 & 0.045 \\
2 & $60\%$  & 0.050 & 0.047 & 0.045 & 0.047 & 0.053 & 0.046 & 0.055 & 0.050 & 0.047 \\
3 & $75\%$  & 0.050 & 0.051 & 0.049 & 0.043 & 0.049 & 0.046 & 0.056 & 0.048 & 0.053 \\
4 & $75\%$  & 0.050 & 0.047 & 0.050 & 0.046 & 0.053 & 0.053 & 0.048 & 0.053 & 0.042 \\
5 & $75\%$  & 0.048 & \textit{0.068} & \textit{0.108} & 0.049 & 0.043 & 0.050 & 0.050 & 0.046 & 0.048 \\
6 & $60\%$  & 0.049 & \textit{0.079} & \textit{0.154} & 0.054 & 0.048 & 0.045 & 0.052 & 0.056 & 0.052 \\
7 & $75\%$  & 0.046 & \textit{0.075} & \textit{0.127} & 0.046 & 0.043 & 0.052 & 0.050 & 0.051 & 0.050 \\\bottomrule
\end{tabular}
\caption{The type 1 error rate of the kernel log-rank test with kernel $(\text{Gau},1)$ with various bandwidths. Distributions can be found in Table \ref{table:appendix:varyingcensoring_type1_distribution}. For distributions D.1-4 it holds that  $\sigma_{\text{med}}=0.72$ and for D.5-7 it holds that $\sigma_{\text{med}}=2.96$. The third column lists the rejection rate when the sigma median heuristic is used as bandwidth. The remaining columns list rejection rates when the bandwidths $0.1-20$ are used. Rejection rates above 0.57 are displayed in italics. \label{table:appendix:bandwidths_type1_gau1} }
\end{table}

\subsubsection{Power of the kernel (Gau,1) for different bandwidths \label{sec:appendix:bandwidths_gau1_power}}

\begin{table}[H]
\centering
\begin{tabular}{@{}lllllllllll@{}}
\toprule
& & \multicolumn{9}{l}{ $\sigma_X=$} \\ 
\cmidrule(r){3-11} 
 D.&  $\Delta=1$          & $\sigma_{\text{med}}$ & 0.1  & 0.2  & 0.5  & 1    & 2    & 5    & 10   & 20   \\ \midrule
8  & $60\%$   & 0.48                         & 0.21 & 0.32 & 0.44 & 0.51 & \textbf{0.54} & 0.53 & \textbf{0.54} & 0.53 \\
9  & $60\%$   & 0.68                         & 0.50 & 0.62 & \textbf{0.69} & 0.60 & 0.29 & 0.15 & 0.14 & 0.15 \\
10 & $75\%$  & 0.12 & 0.10 & 0.11 & 0.11 & 0.13 & 0.14 & 0.14 & \textbf{0.16} & 0.14 \\
11 & $75\%$ & 0.05 & \textbf{0.07} & 0.06 & 0.07 & 0.05 & 0.06 & 0.05 & 0.05 & 0.05 \\
12 & $75\%$   &0.59 & \textit{0.09} & \textit{0.17} & 0.07 & 0.27 & 0.52 & \textbf{0.66} & 0.61 & 0.63 \\
13  & $60\%$ & 0.85 & \textit{0.09} & \textit{0.17} & 0.09 & 0.42 & 0.79 & 0.86 & \textbf{0.89} & 0.87 \\
14  & $75\%$ & 0.67 & \textit{0.03} & \textit{0.07} & 0.09 & 0.59 & \textbf{0.75} & 0.40 & 0.19 & 0.13 \\\bottomrule
\end{tabular}
\caption{The power of the kernel log-rank test with kernel $(\text{Gau},1)$ with various bandwidths. Distributions are found in Table \ref{table:appendix:varyingcensoring_power_distributions}. For distributions D.8-11 it holds that  $\sigma_{\text{med}}=0.72$ and for D.12-13 it holds that $\sigma_{\text{med}}=2.96$. The third column lists the rejection rate when the median heuristic is used as bandwidth. The remaining columns list rejection rates when the bandwidths $0.1-20$ are used. The highest rejection rate is in boldface. In the event of an elevated type 1 error rate (above 0.057) in Table \ref{table:appendix:bandwidths_type1_gau1}, the rejection rate is displayed in italics.  }
\end{table}

\subsubsection{Type 1 error rate of the kernel (Gau, Gau) for different bandwidths \label{sec:appendix:bandwidths_gaugau_type1} }

\begin{table}[H]
\centering
\begin{tabular}{@{}llllllllll@{}}
\toprule
& & \multicolumn{8}{l}{ $\sigma_X=$} \\ 
\cmidrule(r){3-10} 
 &     & 0.72 & 0.10 & 0.20 & 0.50 & 1.00 & 2.00 & 5.00 & 10.00 \\ \hline
\multirow{8}*{$\sigma_Z=$} & 0.16  & 0.050 & 0.044 & 0.050 & 0.049 & 0.050 & 0.043 & 0.045 & 0.046  \\
 & 0.10  & 0.051 & 0.045 & 0.050 & 0.051 & 0.045 & 0.044 & 0.049 & 0.052  \\
 & 0.20  & 0.050 & 0.045 & 0.052 & 0.050 & 0.051 & 0.051 & 0.054 & 0.050  \\
 & 0.50  & 0.045 & 0.049 & 0.053 & 0.051 & 0.052 & 0.044 & 0.045 & 0.047  \\
 & 1.00  & 0.048 & 0.052 & 0.047 & 0.052 & 0.045 & 0.048 & 0.051 & 0.054  \\
 & 2.00  & 0.045 & 0.045 & 0.046 & 0.048 & 0.052 & 0.049 & 0.046 & 0.052  \\
 & 5.00  & 0.051 & 0.043 & 0.044 & 0.049 & 0.050 & 0.047 & 0.051 & 0.050  \\
 & 10.00 & 0.046 & 0.049 & 0.049 & 0.046 & 0.049 & 0.047 & 0.050 & 0.045 \\ \bottomrule
\end{tabular}
\caption{Type 1 error rate of the kernel log-rank test with various combinations of bandwidths for D.3 of Table \ref{table:appendix:varyingcensoring_type1_distribution} when $75\%$ of events is observed. The median heuristic yields $\sigma_X=0.72$ (first column) and $\sigma_Z=0.16$ (first row) resulting in a rejection rate of 0.050. Each combination of bandwidths has the correct type 1 error rate of around $\alpha=0.05$. \label{table:appendix:type1_bandwidths_both_kernels:d3} }
\end{table}

\begin{table}[H]
\centering
\begin{tabular}{@{}llllllllll@{}}
\toprule
& & \multicolumn{8}{l}{ $\sigma_X=$} \\ 
\cmidrule(r){3-10} 
 &     & 0.72 & 0.10 & 0.20 & 0.50 & 1.00 & 2.00 & 5.00 & 10.00 \\ \hline
\multirow{8}*{$\sigma_Z=$}  & 0.69  & 0.046 & 0.046 & 0.049 & 0.042 & 0.049 & 0.050 & 0.048 & 0.047  \\
 & 0.10  & 0.047 & 0.039 & 0.044 & 0.047 & 0.052 & 0.043 & 0.045 & 0.052  \\
 & 0.20  & 0.050 & 0.038 & 0.045 & 0.049 & 0.049 & 0.055 & 0.051 & 0.049  \\
 & 0.00  & 0.044 & 0.047 & 0.047 & 0.050 & 0.049 & 0.050 & 0.049 & 0.048  \\
 & 1.00  & 0.052 & 0.046 & 0.043 & 0.050 & 0.047 & 0.050 & 0.052 & 0.050  \\
 & 2.00  & 0.050 & 0.046 & 0.050 & 0.052 & 0.047 & 0.052 & 0.050 & 0.054  \\
 & 5.00  & 0.042 & 0.046 & 0.044 & 0.046 & 0.045 & 0.048 & 0.047 & 0.051  \\
 & 10.000 & 0.049 & 0.045 & 0.048 & 0.045 & 0.047 & 0.051 & 0.052 & 0.057  \\ \bottomrule
\end{tabular}
\caption{Type 1 error rate of the kernel log-rank test with various combinations of bandwidths for D.4 of Table \ref{table:appendix:varyingcensoring_type1_distribution} when $75\%$ of events are observed. The median heuristic yields $\sigma_X=0.72$ (first column) and $\sigma_Z=0.69$ (first row) resulting in a rejection rate of 0.050. Each combination of bandwidths has the correct type 1 error rate of around $\alpha=0.05$ - none are above 0.057. \label{table:appendix:type1_bandwidths_both_kernels:d4} }
\end{table}

\begin{table}[H]
\centering
\begin{tabular}{@{}llllllllll@{}}
\toprule
& & \multicolumn{8}{l}{ $\sigma_X=$} \\ 
\cmidrule(r){3-10} 
 &     & 2.94 & 0.10 & 0.20 & 0.50 & 1.00 & 2.00 & 5.00 & 10.00 \\ \hline
\multirow{8}*{$\sigma_Z=$}  & 0.49  & 0.047 & \textit{0.097} & \textit{0.183} & 0.047 & 0.044 & 0.046 & 0.046 & 0.047  \\
 & 0.10  & 0.048 & \textit{0.120} & \textit{0.206} & 0.018 & 0.034 & 0.054 & 0.051 & 0.047  \\
 & 0.20  & 0.048 & \textit{0.108} & \textit{0.192} & 0.028 & 0.038 & 0.045 & 0.053 & 0.046  \\
 & 0.50  & 0.047 & \textit{0.102} & \textit{0.172} & 0.046 & 0.043 & 0.046 & 0.049 & 0.052  \\
 & 1.00  & 0.045 & \textit{0.085} & \textit{0.150} & 0.049 & 0.046 & 0.051 & 0.052 & 0.049  \\
 & 2.00  & 0.049 & \textit{0.078} & \textit{0.131} & 0.050 & 0.044 & 0.049 & 0.050 & 0.052  \\
 & 5.00  & 0.052 & \textit{0.070} & \textit{0.129} & 0.053 & 0.047 & 0.055 & 0.053 & 0.053  \\
 & 10.00 & 0.048 & \textit{0.079} & \textit{0.120} & 0.047 & 0.049 & 0.046 & 0.046 & 0.053  \\ \bottomrule
\end{tabular}
\caption{Type 1 error rate of the kernel log-rank test with various combinations of bandwidths for D.7 of Table \ref{table:appendix:varyingcensoring_type1_distribution} when $75\%$ of events are observed. The median heuristic yields $\sigma_X=2.94$ (first column) and $\sigma_Z=0.49$ (first row) resulting in a rejection rate of 0.053. We note that bandwidths $\sigma_X$, which are small compared to the median heuristic of $2.94$, lead to an inflated type 1 error rate. Type 1 error rates above 0.057 are displayed in italics. \label{table:appendix:type1_bandwidths_both_kernels:d7} }
\end{table}

\subsubsection{Power of the kernel (Gau,Gau) for different bandwidths \label{sec:appendix:bandwidths_gaugau_power}}

\begin{table}[H]
\centering
\begin{tabular}{@{}llllllllll@{}}
\toprule
& & \multicolumn{8}{l}{ $\sigma_X=$} \\ 
\cmidrule(r){3-10} 
 &     & 0.72 & 0.10 & 0.20 & 0.50 & 1.00 & 2.00 & 5.00 & 10.00 \\ \hline
\multirow{8}*{$\sigma_Z=$}   & 0.69  & 0.86 & 0.44 & 0.65 & 0.80 & 0.86 & 0.90 & 0.89 & 0.89  \\
 & 0.10  & 0.64 & 0.27 & 0.39 & 0.62 & 0.68 & 0.71 & 0.70 & 0.73  \\
 & 0.20  & 0.77 & 0.37 & 0.55 & 0.74 & 0.81 & 0.84 & 0.83 & 0.82  \\
 & 0.50  & 0.86 & 0.48 & 0.67 & 0.84 & 0.88 & 0.88 & \textbf{0.91} & \textbf{0.91}  \\
 & 1.00  & 0.76 & 0.37 & 0.50 & 0.70 & 0.81 & 0.83 & 0.82 & 0.82  \\
 & 2.00  & 0.29 & 0.15 & 0.22 & 0.25 & 0.31 & 0.35 & 0.37 & 0.40  \\
 & 5.00  & 0.15 & 0.10 & 0.10 & 0.13 & 0.13 & 0.17 & 0.16 & 0.18  \\
 & 10.00 & 0.11 & 0.09 & 0.10 & 0.11 & 0.12 & 0.14 & 0.15 & 0.15   \\ \bottomrule
\end{tabular}
\caption{Power of the kernel log-rank test with various combinations of bandwidths for D.10 of Table \ref{table:appendix:varyingcensoring_power_distributions}  when $75\%$ of events are observed. The median heuristic yields $\sigma_X=0.72$ (first column) and $\sigma_Z=0.69$ (first row) resulting in a rejection rate of 0.86. The highest rejection rate 0.91 (in bold) is achieved when $(\sigma_Z,\sigma_X)=(0.5,5),(0.5,10)$. In Table \ref{table:appendix:type1_bandwidths_both_kernels:d3} we found that for this censoring distribution, each of the bandwidths led to a correct Type 1 error rate. }
\end{table}

\begin{table}[H]
\centering
\begin{tabular}{@{}llllllllll@{}}
\toprule
& & \multicolumn{8}{l}{ $\sigma_X=$} \\ 
\cmidrule(r){3-10} 
 &     & 0.72 & 0.10 & 0.20 & 0.50 & 1.00 & 2.00 & 5.00 & 10.00 \\ \hline
\multirow{8}*{$\sigma_Z=$}  & 0.67  & 0.87 & 0.43 & 0.62 & 0.82 & 0.91 & \textbf{0.95} & \textbf{0.95} & 0.92  \\
 & 0.10  & 0.68 & 0.30 & 0.41 & 0.64 & 0.74 & 0.76 & 0.79 & 0.78  \\
 & 0.20  & 0.82 & 0.43 & 0.58 & 0.77 & 0.87 & 0.88 & 0.90 & 0.88  \\
 & 0.50  & 0.90 & 0.47 & 0.69 & 0.85 & 0.92 & 0.94 & \textbf{0.95} & 0.94  \\
 & 1.00  & 0.77 & 0.31 & 0.48 & 0.70 & 0.82 & 0.89 & 0.88 & 0.88  \\
 & 2.00  & 0.29 & 0.11 & 0.14 & 0.24 & 0.34 & 0.39 & 0.36 & 0.37  \\
 & 5.00  & 0.09 & 0.07 & 0.08 & 0.07 & 0.09 & 0.07 & 0.10 & 0.09  \\
 & 10.00 & 0.06 & 0.06 & 0.07 & 0.06 & 0.06 & 0.08 & 0.07 & 0.07 \\ \bottomrule
\end{tabular}
\caption{Power of the kernel log-rank test with various combinations of bandwidths for D.11 of Table \ref{table:appendix:varyingcensoring_power_distributions} when $75\%$ of events are observed. The median heuristic yields $\sigma_X=0.72$ (first column) and $\sigma_Z=0.67$ (first row) resulting in a rejection rate of 0.87. The highest rejection rate 0.94 (in bold) is achieved when $(\sigma_Z,\sigma_X)=(0.67,2),(0.67,5),(0.5,5)$. In Table \ref{table:appendix:type1_bandwidths_both_kernels:d4} we found that for this censoring distribution, each of the bandwidths led to a correct type 1 error rate.  }
\end{table}

\begin{table}[H]
\centering
\begin{tabular}{@{}llllllllll@{}}
\toprule
& & \multicolumn{8}{l}{ $\sigma_X=$} \\ 
\cmidrule(r){3-10} 
 &    & 2.96 & 0.10 & 0.20 & 0.50 & 1.00 & 2.00 & 5.00 & 10.00 \\ \midrule
\multirow{8}*{$\sigma_Z=$}  & 0.32  & 0.48 & \textit{0.07} & \textit{0.12} & 0.08 & 0.42 & 0.57 & 0.26 & 0.13  \\
 & 0.10  & 0.34 & \textit{0.08} & \textit{0.15} & 0.03 & 0.28 & 0.41 & 0.20 & 0.10  \\
 & 0.20  & 0.43 & \textit{0.08} & \textit{0.11} & 0.06 & 0.42 & 0.49 & 0.24 & 0.13  \\
 & 0.50  & 0.55 & \textit{0.07} & \textit{0.09} & 0.06 & 0.50 & 0.64 & 0.31 & 0.14  \\
 & 1.00  & 0.62 & \textit{0.07} & \textit{0.10} & 0.10 & 0.55 & 0.70 & 0.35 & 0.17  \\
 & 2.00  & 0.63 & \textit{0.04} & \textit{0.07} & 0.08 & 0.58 & 0.73 & 0.37 & 0.19  \\
 & 5.00  & 0.67 & \textit{0.05} & \textit{0.07} & 0.08 & 0.60 & 0.75 & 0.43 & 0.17  \\
 & 10.00 & 0.67 & \textit{0.05} & \textit{0.07} & 0.09 & 0.60 & \textbf{0.76} & 0.41 & 0.19  \\ \bottomrule
\end{tabular}
\caption{Power of the kernel log-rank test with various combinations of bandwidths for D.14 \ref{table:appendix:varyingcensoring_power_distributions} when $75\%$ of events are observed. The median heuristic yields $\sigma_X=2.96$ (first column) and $\sigma_Z=0.32$ (first row) corresponding to a rejection rate of 0.48. The highest rejection rate $0.76$ is achieved when $(\sigma_Z,\sigma_X)=(10,2)$. Rejection rates for bandwidths that led to an increased type 1 error rate in Table \ref{table:appendix:type1_bandwidths_both_kernels:d7}, which featured the same censoring distribution, are displayed in italics. }
\end{table}

\section{Preliminary results}
In this section, and in order for this paper to be self-contained, we review some preliminary results that will be used in our proofs.

\subsection{Counting processes}
We state some results from the counting process literature that are frequently used in our paper.  Recall that $Y(t)=\sum_{i=1}^nY_i(t)$ denotes the pooled risk function, where $Y_i(t) = \ind_{\{T_i \geq t\}}$,  $\tau = \sup\left\{t:S_T(t)>0\right\}$, $\tau_n=\max\{T_1,\ldots,T_n\}$, and that $Z$ and $C$ are continuous random variables.

\begin{proposition}\label{lemma:supConverSH}
The following holds a.s.
\begin{enumerate}
\item $\lim_{n \to \infty}\sup_{t \leq \tau} |Y(t)/n-S_T(t)| = 0$
\item $\sup_{t\leq t^{\star}} \left |\frac{n}{Y(t)}-\frac{1}{S_T(t)}\right| \to 0.$ for any $t^{\star}\in(0,\tau)$
\end{enumerate}
\end{proposition}
The proof of Part 1 is due to the Glivenko Cantelli's Theorem, and Part 2 is a consequence of  Part 1. Also, notice that under the null hypothesis $S_T(t)=S_Z(t)\int_{\R^d}S_{C|X=x}(t)dF_X(x)$ since we assume non-informative censoring, i.e. $Z\perp C|X$.

\begin{proposition}\label{prop:ProbBounds}
Let $\beta \in (0,1)$, then
\begin{enumerate}
\item $\Prob\left(Y(t)/n \leq \beta^{-1}S_T(t),\quad\forall t\leq\tau_n\right)\geq1-\beta,$\label{prop:ProbBound2}
\item $\Prob\left(Y(t)/n \geq \beta S_T(t),\quad\forall t\leq\tau_n\right)\geq1-e(1/\beta) e^{-1/\beta}.$\label{prop:ProbBound3}
\end{enumerate}
\end{proposition}
The proof of Part 1 follows from \cite[Theorem 3.2.1.]{gill1980censoring} and Part 2 is due to \cite{gill1983large}.

\subsection{Dominated convergence in probability}
\begin{lemma}\label{lemma:integralConvergence2}
Let $(\mathcal X, \mathcal B, \mu)$ be a measurable space and $(\Omega, \mathcal F, \Prob)$ be a probability space. Consider a sequence of  non-negative stochastic processes $R_n: \Omega\times \mathcal X \to \R$. Suppose that
\begin{enumerate}
\item[i)] For each $\alpha\in(0,1)$, there exists an event $A_{\alpha}$ with $\Prob(A_{\alpha})\geq 1-\alpha$, such that $R_n(\omega,x)\to0$ for all $(\omega,x)\in A_{\alpha}\times\mathcal{X}$.
\item[ii)] For each $\beta \in (0,1)$, there exists a non-negative function $R_{\beta} \in L_1(\mathcal X, \mathcal B, \mu)$ and $N_0$ large enough such that for each $n \geq N_0$ there exists an event $B_{n,\beta}$ with $\Prob(B_{n,\beta}) \geq 1-\beta$ and 
$$R_n(\omega,x) \leq R_\beta(x)$$
for $(\omega,x) \in B_{n,\beta} \times \mathcal X$.
\end{enumerate}
Then $I_n(\omega) = \int_{\mathcal X}R_n(\omega,x)\mu(dx) \to 0$ in $\Prob$-probability.
\end{lemma}
The proof follows by noticing that in a set of probability tending to one, we can apply dominated convergence. 
	
\subsection{Lenglart-Rebolledo Inequality}\label{sec:LenReb}
Let $X(t)$ and $X'(t)$ be right-continuous adapted processes. We say that $X$ is majorised by $X'$ if for all bounded stopping times $T$, it holds that $\E(|X(T)|)\leq \E(X'(T))$.
\begin{lemma}[Theorem 3.4.1. of \cite{Flemming91}]\label{lemma:LRInequality}
Let $X$ be a right-continuous adapted
process, and $X'$ a non-decreasing predictable process with $X'(0) = 0$ such that $X$ is majorised by $X'$. For any stopping time $T$ (even unbounded), and any $\varepsilon, \delta > 0$, 

$$\Prob\left(\sup_{t \leq T} |X(t)|>\varepsilon\right) \leq \frac{\delta}{\varepsilon}+\Prob(X'(T) \geq \delta).$$
\end{lemma}
In our setting $X(t)$ is a sub-martingale $W_n(t)$ (based on $n$ data points) and $X'(t)$ is the corresponding compensator $A_n(t)$. We consider the stopping time $\tau_n = \max\{T_1,\ldots,T_n\}$, and apply the previous Lemma to prove
\begin{align}
&A_n(\tau_n) = o_p(1) \Rightarrow \sup_{t\leq \tau_n}  W_n(t) = o_p(1).\label{eqn:LenglartApp2}
\end{align}

\subsection{Double martingale integrals}\label{sec:specialMartingales}
Next we state results regarding double integrals with respect to a $\mathcal{F}_t$- counting process martingale $M(t)$, introduced in \cite{fernandez2018kaplan}. Consider $W(t) = \int_{C_t} h(x,y)dM(x)dM(y)$, where $C_t = \{(x,y): 0< x <y\leq t\}$. The following results state conditions under which $W(t)$ defines a proper martingale with respect the filtration $(\mathcal F_t)_{t \geq 0}$.

\begin{definition}
\label{def:predicDsigmaAlgebra}
We define the predictable $\sigma$- algebra $\mathcal{P}$ as the $\sigma$-algebra generated by the sets of the form
\begin{align*}
\sigma\{\{(a_1,b_1]\times(a_2,b_2]\times X:0\leq  a_1\leq b_1 < a_2 \leq b_2,X\in \mathcal{F}_{a_2}\}\cup \{(0,0)\times X: X\in\mathcal{F}_0\}\}.
\end{align*}

\end{definition}

Let $C = \{(x,y): 0< x <y<\infty\}$. A function $h:C\times\Omega \to\R$ is called \emph{``elementary predictable"} if it can be written as a finite sum of indicator functions of sets belonging to the predictable $\sigma$-algebra $\mathcal{P}$. On the other hand, if a function $h$ is $\mathcal{P}$-measurable, then it is the limit of elementary predictable functions. 

\begin{proposition}\label{coll:h predictable}
Any deterministic measurable function is $\mathcal{P}$-measurable. Additionally the functions $C\times \Omega \to \R$: $Y(x)Y(y)$ and $\ind_{\{x<y\leq \tau_n\}}$ are $\mathcal{P}$-measurable.
\end{proposition}

\begin{remark}
Note that any deterministic function of the covariates $X_1,\ldots,X_n$, individual risk functions $Y_i(t)$, and indicator function $\ind_{\{t\leq \tau_n\}}$ are $\mathcal{P}$-measurable as in Definition \ref{def:predicDsigmaAlgebra}. 
\end{remark}

\begin{theorem}\label{thm:doubleMartingale}
Let $h(x,y)$ be a $\mathcal{P}$-measurable function (see Definition \ref{def:predicDsigmaAlgebra}) and define $C_t=\{(x,y):0<x<y\leq t\}$ with $t\geq0$. Let $M(t)$ and $M'(t)$ be  $\mathcal{F}_t$-martingales, and suppose that for all $t\geq 0$, it holds
\begin{align}\label{eqn:DoubleMcond2}
\E\left(\int_{C_t} |h||dM(x)dM'(y)|\right)&<\infty.
\end{align} 
Then, the process 
\begin{align*}
W(t) &= \int_{C_t} h(x,y)dM(x)dM'(y)
\end{align*}
is a martingale on $t\geq 0$ with respect the filtration $\mathcal F_t$.
\end{theorem}

\subsection{Forward and Backward operators}\label{appe:BackFoward}

Consider the forward and backward operators introduced in \cite{Efron1990}, $A:\mathcal{L}_2(Z\times C\times X)\to \mathcal{L}_2(Z\times C\times X)$, and $B:\mathcal{L}_2(Z\times C\times X)\to \mathcal{L}_2(Z\times C\times X)$, respectively,  defined as 
\begin{align*}
(Af)(z,c,x)=f(z,c,x)-\frac{1}{S_Z(z)}\int_z^\infty f(s,c,x)dF_Z(s)
\end{align*}
and
\begin{align*}
(Bf)(z,c,x)=f(z,c,x)-\int_0^z f(s,c,x)d\Lambda_Z(s).
\end{align*}
It was proved in \cite{Efron1990} that the previous operators satisfy the following properties. For any $f,g\in\mathcal{L}_2(Z\times C\times X)$:
\begin{enumerate}
\item $\E((Af)g|C_1,X_1)=\E(f(Bg)|C_1,X_1)$
\item $ABf=f$
\item $BAf=f-\E(f|C_1,X_1)$
\item $\E(Bf|C_1,X_1)=0$
\item For any $f(z,c,x)=\ind_{\{z\leq c\}}g(z,c,x)$, $$(Bf)(z,c,x)=\int_{\R_+}g(s,c,x)dm_{z,c}(s),$$ where $dm_{z,c}(s)=\ind_{z\leq c}\delta_z(s)-\ind_{\{\min\{z,c\}\geq s\}}d\Lambda_Z(s)$.
\end{enumerate}

\subsection{Kernel results}
We state the definition of a characteristic kernel, which is defined in Section 2 in  \cite{Univer}.
\begin{definition}[characteristic kernel]
Given the set of all Borel probability measures defined on the topological space $\mathcal{X}$, a measurable and bounded kernel $K$ is said to be characteristic if
$$\mathbb{P}\to\int_{\mathcal{X}}K(\cdot,x)d\mathbb{P}(x)$$
is injective, that is, the probability measure $\mathbb{P}$ is embedded to a unique element $\int_{\mathcal{X}}K(\cdot,x)d\mathbb{P}(x)\in\mathcal{F}$, where $\mathcal{F}$ is the RKHS induced by the kernel $K$.
\end{definition}
An example of a characteristic kernel is the exponentiated quadratic kernel, which is defined as $K(x,y)=\exp\{-(x-y)^\intercal(x-y)/\sigma^2\}$ for any $x,y\in\R^d$ and $\sigma>0$. 

We state the definition of $c_0$-universality, which is introduced in Proposition 2 in \cite{Univer}.

\begin{proposition}\label{prop:uni}($c_0$-universality and RKHS embedding of measures) Suppose that $\mathcal X$ is a locally compact Hausdorff space and the kernel $K$ is a $c_0$-kernel. Then $K$ is $c_0$- universal if and only if the embedding 
\begin{align*}
\mu\to\int_\mathcal{X}K(\cdot,x)d\mu(x),\quad\mu\in M_b(\mathcal{X})
\end{align*}
is injective, where $M_b(\mathcal{X})$ denotes the space of all finite signed Radon measures on $\mathcal{X}$.
\end{proposition}
\begin{remark}We say that $K$ is a $c_0$-kernel if $K$ is such that $K(\cdot,x)\in\mathcal{C}_0$ for any $x\in\mathcal{X}$, where $C_0(\mathcal{X})$ denotes the class of continuous functions on $\mathcal{X}$ that vanish at infinity. 
\end{remark}

Finally, we state the result which relates translation invariant kernels with the notion of $c_0$-universality. This result is presented in Section  3.2 in \cite{Univer}

\begin{proposition}\label{prop:trans} Bounded, continuous, translation invariant, $c_0$-kernels are $c_0$-universal if and only if they are characteristic.
\end{proposition}

Note that by combining the two previous results, under the same assumptions,  we obtain that bounded, continuous, characteristic, translation invariant, $c_0$-kernels provide injective embeddings of finite signed Radon measures.

\subsection{Wild-Bootstrap for degenerate $V$-statistics}

The following result appears in Theorem 3.1 in \cite{dehling1994random}.

\begin{theorem}\label{Thm:ConsistencyWild}Let $X_1,\ldots,X_n$ be a collection of i.i.d. random variables, and let $W_1,\ldots,W_n$ be a collection of i.i.d. Rademacher random variables which are independent of $X_1,\ldots,X_n$. Suppose $h$ is a symmetric degenerate kernel satisfying $\E(|h(X_1,X_1)|)<\infty$ and $\E(h^2(X_1,X_2))<\infty$. Then there exists a probability distribution $\mathcal{L}$ such that 
\begin{align*}
\frac{1}{n}\sum_{i=1}^n\sum_{j=1}^n h(X_i,X_j)\overset{\mathcal{D}}{\to}\mathcal{L},
\end{align*}
and for almost every sequence $(x_i)_{i\geq 0}$ sampled from $(X_i)_{i\geq 0}$
\begin{align*}
\frac{1}{n}\sum_{i=1}^n\sum_{j=1}^n h(x_i,x_j)W_iW_j \overset{\mathcal{D}}{\to}\mathcal{L},
\end{align*}
as $n$ grows to infinity.

\end{theorem}

\newpage
\section{Auxiliary results}
In this section we introduce some auxiliary results that will be used in the proofs of our main results. The proofs of these results are given in Section \ref{sec:proofsaux}.

\begin{itemize}
\item Proposition \ref{prop:ess sup} is a technical result from which we deduce $\tau_x\leq\tau$ for almost all $x\in\R^d$.
\item Proposition \ref{prop:nu_0approx}  is used in the proof of Theorem \ref{thm:Logrankconistency}.
\item Theorem \ref{Thm:Eigen} is used in the proof of Theorem \ref{Thm:SimplerKernel}
\item Lemma \ref{Lemma:ReplacementYP} is used in the proof of Lemma \ref{lemma:consistency}.
\end{itemize}

\begin{proposition}\label{prop:ess sup}
Let $\tau'$ be the essential supremum defined as $\tau'=\text{ess }\sup \tau_x=\inf\{t:F_X(\{x:\tau_x> t\})=0\}$. Then $\tau=\tau'$.
\end{proposition}

\begin{proposition}\label{prop:nu_0approx}Let $\omega:\R_+\times\R^d\to\R$ be a bounded measurable function. Then
\begin{align*}
\int_{\R^d}\int_{0}^\tau \omega(s,x) d\nu_0^n(s,x)
&=\frac{1}{n^2}\sum_{j=1}^n\sum_{i=1}^n\Delta_j\omega(T_j,X_i)\frac{Y_i(T_j)}{S_T(T_j)}+o_p(1).
\end{align*}
\end{proposition}

\begin{theorem}\label{Thm:Eigen}
Let $Q:(\R_+\times\R_+\times \R^d)^2\to\R$, $T^Q:\mathcal{L}_2(Z\times C\times X)\to \mathcal{L}_2(Z\times C\times X)$ and $T^J:\mathcal{L}_2(T\times\Delta\times X)\to \mathcal{L}_2(T\times\Delta\times X)$ be given by
\begin{align*}
Q((z_1,c_1,x_1),(z_2,c_2,x_2))&=\ind_{\{z_1\leq c_1\}}\ind_{\{z_2\leq c_2\}}\bar{\mathfrak{K}}((z_1,x_1),(z_2,x_2)),
\end{align*}
\begin{align*}
(T^Q f)(Z_1,C_1,X_1)=\E_2\left(Q((Z_1,C_1,X_1),(Z_2,C_2,X_2))f(Z_2,C_2,X_2)\right),
\end{align*}
and
\begin{align}
(T^J f)(T_1,\Delta_1,X_1)=\E_2\left(J((T_1,\Delta_1,X_1),(T_2,\Delta_2,X_2))f(T_2,\Delta_2,X_2)\right),\label{eqn:IOTJ}
\end{align}
respectively, where $J$ is the function defined in Equation \eqref{eqn:Jkernel} in the main document, $\E_2(\cdot)=\E(\cdot|(T_1,\Delta_1,X_1))$, and $(Z_1,C_1,X_1)$ and $(Z_2,C_2,X_2)$ are independent copies of our random variables in the augmented space. 

Then, $T^J=BT^QA$, and $T^J$ and $T^Q$ have the same set of non-zero eigenvalues, including multiplicities.

\end{theorem}

\begin{lemma}\label{Lemma:ReplacementYP}
\begin{align}
\frac{1}{n}\int_0^\tau\left|1-\frac{Y(t)/n}{S_T(t)}\right|^kdN(t)&=o_p(1),\label{eqn:intdN}
\end{align}
for $k=1$ and $k=2$.
\end{lemma}

\newpage
\section{Main proofs}
In this section we prove the main results of our paper.

\subsection{Proofs of Section \ref{sec:testConstruction}}

\subsubsection{Proof of Theorem 3.1}
The result follows from proving
\begin{align*}
\int_{\R^d}\int_0^\tau\omega(s,x)d\nu_k^n(s,x)\overset{\Prob}{\to}\int_{\R^d}\int_0^\tau\omega(s,x)d\nu_k(t,x)
\end{align*}
for $k\in\{0,1\}$.

Note that the result for $k=1$ follows from a direct application of the law of large numbers. We continue proving the result for $k=0$.  By Proposition \ref{prop:nu_0approx}, it holds
\begin{align*}
\int_{\R^d}\int_0^\tau\omega(s,x)d\nu_0^n(s,x)
&=\frac{1}{n^2}\sum_{j=1}^n\sum_{i=1}^n\Delta_j\omega(T_j,X_i)\frac{Y_i(T_j)}{S_T(T_j)}+o_p(1).
\end{align*}
The sum in the left-hand-side of the previous equation can be decomposed into two sums, one over indices such that $\{i\neq j\}$ and the other over indices such that $\{i= j\}$. Notice that the sum over $\{i\neq j\}$ satisfies 
\begin{align*}
\frac{1}{n^2}\sum_{i\neq j}^n\Delta_j\omega(T_j,X_i)\frac{Y_i(T_j)}{S_T(T_j)}\overset{\Prob}{\to}\E\left(\Delta_2\omega(T_2,X_1)\frac{Y_1(T_2)}{S_T(T_2)}\right)=\int_{\R^d}\int_0^\tau\omega(s,x)d\nu_0(s,x)
\end{align*}
by the law of large numbers for  U-statistics. Thus, to conclude the proof, we just need to prove that the sum over indices $\{i=j\}$ converges to zero in probability. To see this, notice that $U_i=S_T(T_i)$ are i.i.d. uniform random variables, and that 
\begin{align*}
\frac{1}{n^2}\sum_{i=1}^n\Delta_i\omega(T_i,X_i)\frac{1}{S_T(T_i)}&\leq \frac{C}{n^2}\sum_{i=1}^n\frac{1}{S_T(T_i)}\leq \frac{C}{n^2}\sum_{i=1}^n\frac{1}{U_i}=o_p(1),
\end{align*}
where the first inequality follows from the assumption that $\omega$ is bounded by some constant $C$, and the last equality follows from $n^{-2}\sum_{i=1}^nU_i^{-1}=o_p(1)$, which  we now proceed to prove.

Define the event $\mathcal{B}_{n}=\cup_{i=1}^n\left\{1/U_i\geq n^{3/2}\right\}$, and note that $\Prob(\mathcal{B}_n)\leq  n\Prob(U_i\leq 1/n^{3/2})=n^{-1/2}$. Then, for any $\epsilon>0$, it holds that
\begin{align*}
\Prob\left(\frac{C}{n^2}\sum_{i=1}^n\frac{1}{U_i}>\epsilon\right)&=\Prob\left(\left\{\frac{C}{n^2}\sum_{i=1}^n\frac{1}{U_i}>\epsilon\right\}\cap \mathcal{B}_n\right)+\Prob\left(\left\{\frac{C}{n^2}\sum_{i=1}^n\frac{1}{U_i}>\epsilon\right\}\cap \mathcal{B}_n^c\right)\\
&\leq \Prob\left(\mathcal{B}_n\right)+\Prob\left(\frac{1}{n}\sum_{i=1}^n\frac{\ind_{\{1/U_i<n^{3/2}\}}}{U_i}>\frac{n\epsilon}{C}\right)\\
&\leq \frac{1}{n^{1/2}}+\frac{C}{n\epsilon }\E\left(\frac{\ind_{\{1/U_i<n^{3/2}\}}}{U_i}\right)\\
&\leq \frac{1}{n^{1/2}}+\frac{C}{n\epsilon}\int_{n^{-3/2}}^1\frac{1}{u}du\\
&\leq \frac{1}{n^{1/2}}+\frac{3C}{2\epsilon}\frac{\log(n)}{n}\to0,
\end{align*}
for any $\epsilon>0$ as n grows to infinity.

\subsubsection{Proof of Proposition \ref{Prop: nu iff indep}}

$(\Longleftarrow)$ Assume $Z\perp X$. Note that $\tau=\tau_{x}$ for almost all $x\in\R^d$, since if this did not hold, then $Z\not\perp X$. Consider arbitrary $t\leq\tau$ and measurable $A\subseteq \R^d$, then it holds that
\begin{align*}
\nu_1((0,t]\times A)&=\int_0^t\int_{x\in A}S_{C|X=x}(s)dF_{Z}(s)dF_{X}(x),
\end{align*}
and
\begin{align*}
\nu_0((0,t]\times A)&=\int_{0}^t\int_{x\in\R^d}S_{C|X=x}(s)\frac{\int_{x'\in A}S_{Z}(s)S_{C|X=x'}(s)dF_{X}(x')}{\int_{x''\in\R^d}S_{Z}(s)S_{C|X=x''}(s)dF_{X}(x'')}dF_{Z}(s)dF_X(x)\\
(Z\perp X)&=\int_0^t\int_{x\in A}S_{C|X=x}(s)dF_{Z}(s)dF_{X}(x).
\end{align*}
Because $\nu_1$ and $\nu_0$ agree on generating sets, they agree on any measurable set in $\R_+\times\R^d$. 

$(\Longrightarrow)$ 
We first prove that $\nu_1=\nu_0$ implies that $\tau=\tau_x$ for almost all $x\in\R^d$. For this result, we observe that Assumption \ref{assumption:no_complete_censoring} implies $\tau_x=\sup\{t:S_{T|X=x}(t)>0\}=\sup\{t:S_{Z|X=x}(t)>0\}$, and  by Proposition \ref{prop:ess sup}, $\tau_x\leq \tau$ for almost all $x\in\R^d$.  We now proceed by contradiction. Assume there exists $\tau'<\tau$ such that the set $E=\{x\in\R^d:\tau_x\leq\tau'\}$  satisfies $\Prob(X\in E)>0$, otherwise $\tau_x=\tau$ for almost all $x\in\R^d$. Let $\omega(s,x)=\ind_{\{s<\tau_x\}}S_{T|X=x}(s)^{-1}$, and note that $\omega(s,x)<\infty$ for all $(s,x)\in(0,\tau)\times\R^d$. By the hypothesis $\nu_0=\nu_1$,
\begin{align}
\int_{A}\int_0^t\omega(s,x)S_{C|X=x}(s)dF_{ZX}(s,x)&=\int_{A}\int_0^t\omega(s,x)S_{T|X=x}(s)d\alpha(s) dF_{X}(x)\label{eqn:equalityprop2.4}
\end{align}
for any $t\leq\tau$, measurable set $A\subseteq\R^d$, and measurable function $\omega(s,x)$.  This leads to a contradiction, however. Indeed, on one hand,
the left-hand-side of equation \eqref{eqn:equalityprop2.4}, when evaluated at $\omega(s,x)=\ind_{\{s<\tau_x\}}S_{T|X=x}(s)^{-1}$, $A=E$ and $t=\tau'$ satisfies
\begin{align*}
\int_{E}\int_0^{\tau'}\omega(s,x)S_{C|X=x}(s)dF_{ZX}(s,x)&=\int_{E}\int_0^{\tau'} \ind_{\{s<\tau_x\}}d\Lambda_{Z|X=x}(s)dF_{X}(x)\\
(\tau_x<\tau'\text{ in }E)&=\int_{E}\Lambda_{Z|X=x}(\tau_x)dF_{X}(x)\\
&=\infty,
\end{align*}
since $\Prob(X\in E)>0$ and the cumulative hazard function $\Lambda_{Z|X=x}(t)$ diverges as $t$ tends to $\tau_x$. 

On the other hand, the right-hand-side of equation \eqref{eqn:equalityprop2.4}, when evaluated at $\omega(s,x)=\ind_{\{s<\tau_x\}}S_{T|X=x}(s)^{-1}$, $A=E$ and $t=\tau'$,
satisfies 
\begin{align*}
\int_{E}\int_0^{\tau'}\omega(s,x)S_{T|X=x}(s)d\alpha(s) dF_{X}(x)&=\int_{E}\int_0^{\tau'}\ind_{\{s\leq \tau_x\}}d\alpha(s) dF_{X}(x)\\
&\leq \Prob(X\in E)\alpha((0,\tau'))<\infty,
\end{align*}
since 
\begin{align*}
\alpha((0,\tau'))&=\int_0^{\tau'}\frac{\int_{x\in\R^d}S_{C|X=x}(s)dF_{ZX}(s,x)}{S_T(s)}\leq \frac{1}{S_T(\tau')}<\infty,
\end{align*}
due to the fact that $S_T(t)>0$ for all $t<\tau$ and $\tau'<\tau$. This finally leads to a contradiction, meaning that such a $\tau'$ cannot exist, and thus $\tau_x=\tau$ for almost all $x\in\R^d$.

We  proceed to prove that $\nu_0=\nu_1$ implies $Z\perp X$. Let $\omega(s,x)=S_{T|X=x}(s)^{-1}$, let $A\subseteq\R^d$ be a measurable set, and let $t<\tau$. Note that the function $\omega$ is well-defined $F_X$-almost surely for $t<\tau$ as $\Prob(X\in\{x:\R^d:\tau_x\neq \tau\})=0$.  By substituting $\omega$ in equation \eqref{eqn:equalityprop2.4}, we obtain
\begin{align}
\int_{0}^t\int_{A} \frac{S_{C|X=x}(s)}{S_{T|X=x}(s)}dF_{ZX}(s,x)&=\int_{0}^t\int_A d\alpha(s)dF_X(x),\label{eqn:prop2.4=>}
\end{align}
and by further choosing $A=\R^d$, we obtain
\begin{align*}
\int_{\R^d} \int_{0}^t\frac{dF_{Z|X=x}(s)}{S_{Z|X=x}(s)}dF_{X}(x)&=\int_{\R^d} \Lambda_{Z|X=x}(t) dF_{X}(x)=\Lambda_{Z}(t)=\int_{0}^t d\alpha(t),
\end{align*}
which implies $d\alpha(s)=d\Lambda_{Z}(s)$ for almost all $s\in\R_+$. Substituting the previous equality in equation \eqref{eqn:prop2.4=>}, we obtain for arbitrary measurable $A$,
\begin{align*}
\int_{0}^t\int_{A} \frac{S_{C|X=x}(s)}{S_{T|X=x}(s)}dF_{ZX}(s,x)&=\int_{A} \Lambda_{Z|X=x}(t)dF_{X}(x)=\int_A \Lambda_Z(t)dF_X(x),
\end{align*}
from which we conclude that $d\Lambda_{Z|X}(s)=d\Lambda_{Z}(s)$ $F_X$-almost surely for all $s<\tau$. Note that Assumption \ref{assumption:no_complete_censoring} implies that $\tau=\sup\{t:S_{Z}(t)>0\}$ from which we deduce $Z\perp X$.

\subsubsection{Proof of Theorem \ref{Thm:closed form}}
Let $\omega^\star=\|\phi_0^n-\phi_1^n\|_{\mathcal{H}}^{-1}(\phi_0^n-\phi_1^n)$, clearly $\|\omega^\star\|^2_{\mathcal{H}}=1$ and $$\langle\omega^\star,\phi_0^n-\phi_1^n\rangle_{\mathcal{H}}=\|\phi_0^n-\phi_1^n\|_{\mathcal{H}}\leq \Psi_n.$$
On the other hand
\begin{align*}
\Psi_n&=\sup_{\omega\in\mathcal{H}:\|\omega\|_{\mathcal{H}}^2\leq1}\langle\omega,\phi_0^n-\phi_1^n\rangle_{\mathcal{H}}\leq \sup_{\omega\in\mathcal{H}:\|\omega\|_{\mathcal{H}}^2\leq1}\|\omega\|_{\mathcal{H}}\|\phi_0^n-\phi_1^n\|_{\mathcal{H}}=\|\phi_0^n-\phi_1^n\|_{\mathcal{H}},
\end{align*}
from which we deduce the desired result $\Psi_n=\|\phi_0^n-\phi_1^n\|_{\mathcal{H}}$. 

Observe 
\begin{align*}
\|\phi_0^n-\phi_1^n\|_{\mathcal{H}}^2&=\left\|\int_{\R^d}\int_{\R_+}L(t,\cdot)K(x,\cdot)(d\nu_1^n(t,x)-d\nu_0^n(t,x))\right\|_{\mathcal{H}}^2\\
&=\left\|\frac{1}{n}\sum_{i=1}^n\int_{\R_+}L(t,\cdot)\left(K(X_i,\cdot)-\sum_{j=1}^nK(X_j,\cdot)\frac{Y_j(t)}{Y(t)}\right)dN_i(t)\right\|_{\mathcal{H}}^2\\
&=\left\|\frac{1}{n}\sum_{i=1}^n\Delta_i L(T_i,\cdot)\left(K(X_i,\cdot)-\sum_{j=1}^nK(X_j,\cdot)\frac{Y_j(T_i)}{Y(T_i)}\right)\right\|_{\mathcal{H}}^2.
\end{align*}

Define $\boldsymbol{\Phi^X}=(K(X_1,\cdot),\ldots, K(X_n,\cdot))^{\intercal}$ and $\boldsymbol{\Phi^T}=(\Delta_1L(T_1,\cdot),\ldots,\Delta_nL(T_n,\cdot))^{\intercal}$. Then 
\begin{align*}
\|\phi_0^n-\phi_1^n\|_{\mathcal{H}}^2&=\left\|\frac{1}{n}\sum_{i=1}^n\Phi^T_i((\boldsymbol{I}-\boldsymbol{A})\boldsymbol{\Phi^X})_i\right\|_{\mathcal{H}}^2\\
&=\frac{1}{n^2}\sum_{i=1}^n\sum_{i'=1}^n\left\langle\Phi^T_i((\boldsymbol{I}-\boldsymbol{A})\boldsymbol{\Phi^X})_i,\Phi^T_{i'}((\boldsymbol{I}-\boldsymbol{A})\boldsymbol{\Phi^X})_{i'}\right\rangle_{\mathcal{H}}\\
&=\frac{1}{n^2}\sum_{i=1}^n\sum_{i'=1}^n(\boldsymbol{L^\Delta})_{i,i'}((\boldsymbol{I}-\boldsymbol{A})\boldsymbol{K}(\boldsymbol{I}-\boldsymbol{A})^{\intercal})_{i,i'}\\
&=\frac{1}{n^2}\tr(\boldsymbol{L^\Delta}(\boldsymbol{I}-\boldsymbol{A})\boldsymbol{K}(\boldsymbol{I}-\boldsymbol{A})^{\intercal}).
\end{align*}

\subsection{Proofs of Section \ref{sec:asymptotics}}
\subsubsection{Proof of  Proposition \ref{Prop:martingalerepre}}
Observe that
\begin{align*}
(\phi^n_0-\phi^n_1)(\cdot)&=\int_{\R_+}\int_{\R^d}\mathfrak{K}((t,x),\cdot)(d\nu_1^n(t,x)-d\nu_0^n(t,x))\\
&=\sum_{i=1}^n\int_{\R_+}\left(\mathfrak{K}((t,X_i),\cdot)-\sum_{j=1}^n\mathfrak{K}((t,X_j),\cdot)\frac{Y_j(t)}{Y(t)}\right)dN_i(t)
\end{align*}
follows from the definition of the embeddings, $\phi_0^n$ and $\phi_1^n$, and of the empirical measures, $\nu_1^n$ and $\nu_0^n$.

Assume that the process $N_i$ can be replaced by the martingale $M_i$ in the previous equation, that is,
\begin{align}
(\phi^n_0-\phi^n_1)(\cdot)&=\sum_{i=1}^n\int_{\R_+}\left(\mathfrak{K}((t,X_i),\cdot)-\sum_{j=1}^n\mathfrak{K}((t,X_j),\cdot)\frac{Y_j(t)}{Y(t)}\right)dM_i(t).\label{eqn:prove4.1}
\end{align} 
Then, using Theorem \ref{Thm:closed form} and Equation \eqref{eqn:prove4.1}, it holds that
\begin{align*}
\Psi^2_n&=\|\phi^n_0-\phi^n_1\|_{\mathcal{H}}^2=\left\|\sum_{i=1}^n\int_{\R_+}\left(\mathfrak{K}((t,X_i),\cdot)-\sum_{j=1}^n\mathfrak{K}((t,X_j),\cdot)\frac{Y_j(t)}{Y(t)}\right)dM_i(t)\right\|_{\mathcal{H}}^2,
\end{align*}
where the main result is deduced by computing the previous term using the reproducing property.

To end the proof, we just need to prove Equation \eqref{eqn:prove4.1}. Since $dM_i(t)=dN_i(t)-Y_i(t)d\Lambda_Z(t)$, the result follows from proving
\begin{align*}
\sum_{i=1}^n\int_{\R_+}\left(\mathfrak{K}((t,X_i),\cdot)-\sum_{j=1}^n\mathfrak{K}((t,X_j),\cdot)\frac{Y_j(t)}{Y(t)}\right)Y_i(t)d\Lambda_Z(t)=0.
\end{align*}
Observe
\begin{align*}
&\sum_{i=1}^n\int_{\R_+}\left(\mathfrak{K}((t,X_i),\cdot)-\sum_{j=1}^n\mathfrak{K}((t,X_j),\cdot)\frac{Y_j(t)}{Y(t)}\right)Y_i(t)d\Lambda_Z(t)\\
&=\sum_{i=1}^n\int_{\R_+}\mathfrak{K}((t,X_i),\cdot)Y_i(t)d\Lambda_Z(t)-\sum_{i=1}^n\int_{\R_+}\sum_{j=1}^n\mathfrak{K}((t,X_j),\cdot)\frac{Y_j(t)}{Y(t)}Y_i(t)d\Lambda_Z(t)\\
&=\sum_{i=1}^n\int_{\R_+}\mathfrak{K}((t,X_i),\cdot)Y_i(t)d\Lambda_Z(t)-\sum_{j=1}^n\int_{\R_+}\mathfrak{K}((t,X_j),\cdot)Y_j(t)d\Lambda_Z(t)\\
&=0,
\end{align*}
which completes the result.

\subsubsection{Proof of Lemma \ref{lemma:rkhs approx}}

Recall the definition of the log-rank test statistic $\text{LR}_n(\omega)$ given in Equation \eqref{eqn:logrank} in the main document,
\begin{align*}
\text{LR}_n(\omega)=\frac{1}{n}\sum_{i=1}^n\int_{\R_+}(\omega(t,X_i)-\bar{\omega}_n(t))dN_i(t),
\end{align*}
where $\bar{\omega}_n(t)=\sum_{j=1}^n\omega(t,X_j)\frac{Y_j(t)}{Y(t)}$. Additionally, note that
\begin{align*}
\frac{1}{n}\sum_{i=1}^n\int_{\R_+}(\omega(t,X_i)-\bar{\omega}_n(t))Y_i(t)d\Lambda_Z(t)=0.
\end{align*}
Then, by substracting the previous term to $\text{LR}_n(\omega)$, we obtain
\begin{align*}
\text{LR}_n(\omega)=\frac{1}{n}\sum_{i=1}^n\int_{\R_+}(\omega(t,X_i)-\bar{\omega}_n(t))dM_i(t),
\end{align*}
and thus, by the definition of the log-rank test-statistic, it follows that
\begin{align}
\Psi_n
&=\sup_{\omega\in\mathcal{H}:\|\omega\|^2_{\mathcal{H}}\leq 1}\frac{1}{n}\sum_{i=1}^n\int_{\R_+}\left(\omega(t,X_i)-\bar\omega_n(t)\right)dM_i(t).\label{eqn:RRq1}
\end{align}

Let $\{(T_i',\Delta_i', X_i')\}_{i=1}^n$ be an independent copy of the data $\mathcal{D}=\{(T_i,\Delta_i,X_i)\}_{i=1}^n$, let $Y_i'(t)=\ind_{\{T_i'\geq t\}}$ and let $\E'(\cdot)=\E(\cdot|\mathcal{D})$. The desired result is then obtained by showing that we can replace $\bar{\omega}_n(t)$ in Equation \eqref{eqn:RRq1} by its population limit (under the null hypothesis), given by 
\begin{align*}
\frac{\E'(\omega(t,X_1')Y_1'(t))}{S_T(t)}&=\frac{\int_{\R^d}\omega(t,x)S_{C|X=x}(t)dF_X(x)}{S_C(t)},
\end{align*}
up to an error of order $o_p(n^{-1/2})$. This result is not obvious because of the supremum.

Let's us make this replacement, and define
\begin{align*}
\bar\Psi_n
&=\sup_{\omega\in\mathcal{H}:\|\omega\|^2_{\mathcal{H}}\leq 1}\frac{1}{n}\sum_{i=1}^n\int_{\R_+}\left(\omega(t,X_i)-\frac{\E'(\omega(t,X_1')Y_1'(t))}{S_T(t)}\right)dM_i(t)\\
&=\sup_{\omega\in\mathcal{H}:\|\omega\|^2_{\mathcal{H}}\leq 1}\left\langle\omega,\frac{1}{n}\sum_{i=1}^n\int_{\R_+}\left(\mathfrak{K}((t,X_i),\cdot)-\frac{\E'(\mathfrak{K}((t,X_1'),\cdot)Y_1'(t))}{S_T(t)}\right)dM_i(t)\right\rangle.
\end{align*}
Note that since the supremum is taken over the unit ball of a reproducing kernel Hilbert space, it is straightforward that
\begin{align*}
\bar{\Psi}_n^2&=\left\|\frac{1}{n}\sum_{i=1}^n\int_{\R_+}\left(\mathfrak{K}((t,X_i),\cdot)-\frac{\E'(\mathfrak{K}((t,X_1'),\cdot)Y_1'(t))}{S_T(t)}\right)dM_i(t)\right\|^2_{\mathcal{H}}\\
&=\frac{1}{n^2}\sum_{i=1}^n\sum_{l=1}^n\int_{\R_+}\int_{\R_+}\bar{\mathfrak{K}}((t,X_i),(t',X_l))dM_i(t)dM_l(t'),
\end{align*}
where $\bar{\mathfrak K}$ is the population limit of $\bar{\mathfrak K}_n$ (under the null) given in Equation \eqref{eqn:popkernel} in the main document. Thus, the desired result follows from proving $\Psi_n^2=\bar{\Psi}_n^2+o_p(n^{-1})$.

Define the error term
\begin{align*}
\Upsilon_n&= \sup_{\omega\in\mathcal{H}:\|\omega\|^2_{\mathcal{H}}\leq 1}\frac{1}{n}\sum_{i=1}^n\int_{\R_+}\left(\frac{\E'(\omega(t,X_1')Y_1'(t))}{nS_T(t)}-\bar{\omega}_n(t)\right)dM_i(t),
\end{align*}
and note that
\begin{align*}
\Upsilon_n&= \sup_{\omega\in\mathcal{H}:\|\omega\|^2_{\mathcal{H}}\leq 1}\frac{1}{n}\sum_{i=1}^n\int_{\R_+}\left(\frac{\E'(\omega(t,X_1')Y_1'(t))}{nS_T(t)}-\bar{\omega}_n(t)\right)dM_i(t)\\
&=\sup_{\omega\in\mathcal{H}:\|\omega\|^2_{\mathcal{H}}\leq 1}\frac{1}{n}\int_{\R_+}\left(\frac{\E'(\omega(t,X_1')Y_1'(t))}{nS_T(t)}-\sum_{j=1}^n\omega(t,X_j)\frac{Y_j(t)}{Y(t)}\right)\sum_{i=1}^ndM_i(t)\\
&= \sup_{\omega\in\mathcal{H}:\|\omega\|^2_{\mathcal{H}}\leq 1}\frac{1}{n}\int_{\R_+}\sum_{j=1}^n\left(\frac{\E'(\omega(t,X_j')Y_j'(t))}{nS_T(t)}-\omega(t,X_j)\frac{Y_j(t)}{Y(t)}\right)dM(t).
\end{align*}
Also, note that by the triangular inequality, 
\begin{align*}
\Psi_n\leq \bar{\Psi}_n+\Upsilon_n\quad\text{and}\quad\Psi_n\geq \bar{\Psi}_n-\Upsilon_n.
\end{align*}
Then, if we assume that $\Upsilon_n=o_p(n^{-1/2})$, the result follows by taking square in the previous inequalities,   
\begin{align*}
n\Psi_n^2=(n^{1/2}\bar{\Psi}_n+o_p(1))^2=n\bar{\Psi}^2_n+o_p(1),
\end{align*}
where we note that $n^{1/2}\bar{\Psi}_no_p(1)=o_p(1)$ since $n^{1/2}\bar \Psi_n$ converges in distribution to some random variable (which is proved in Theorem \ref{thm:rkhsLimit}). 
  
The rest of this proof is concerned with showing that $\Upsilon_n=o_p(n^{-1/2})$ holds under Assumption  \ref{Assu:BoundedKernelSep}.  Again, observe that since the supremum in $\Upsilon_n$ is taken over the unit ball of a reproducing kernel Hilbert space, $\Upsilon_n^2$ satisfies
\begin{align}
\Upsilon^2_n&=\left\|\frac{1}{n}\int_{\R_+}\sum_{j=1}^n\left(\frac{\E'(\mathfrak{K}((t,X_j'),\cdot)Y_j'(t))}{nS_T(t)}-\mathfrak{K}((t,X_j),\cdot)\frac{Y_j(t)}{Y(t)}\right)dM(t)\right\|^2_{\mathcal{H}}.
\end{align}
Under Assumption \ref{Assu:BoundedKernelSep}, $\mathfrak{K}((t,x),(t',x'))=L(t,t')K(x,x')$, and by using the reproducing property, we get
\begin{align*}
\Upsilon^2_n
&=\frac{1}{n^2}\int_{\R_+}\int_{\R_+}L(t,s)\sum_{j=1}^n\sum_{i=1}^n\kappa_{i,j}(t,s)dM(t)dM(s),
\end{align*}
where
\begin{align*}
\kappa_{i,j}(t,s)
&=\frac{\E'(K(X_i',X_j')Y_i'(t)Y_j'(s))}{n^2S_T(t)S_T(s)}-\frac{\E'(K(X_i',X_j)Y_i'(t))Y_j(s)}{nS_T(t)Y(s)}\\
&\quad -\frac{\E'(K(X_i,X_j')Y_j'(s))Y_i(t)}{nS_T(s)Y(t)}+\frac{K(X_i,X_j)Y_i(t)Y_j(s)}{Y(t)Y(s)}.
\end{align*}

Define the process $(R(z))_{z\geq 0}$ by
\begin{align*}
R(z)=\frac{1}{n}\int_{0}^z\int_{0}^zL(t,s)\sum_{j=1}^n\sum_{i=1}^n\kappa_{i,j}(t,s)dM(t)dM(s),
\end{align*}
and note that 
\begin{align*}
n\Upsilon_n^2=R(\tau_n).
\end{align*}

We will prove that there exists an increasing predictable process $(A(z))_{z\geq 0}$ with $A(0)=0$ such that $\E(R(T))\leq \E(A(T))$ for any finite stopping time, i.e., the process $A$ majorises $R$. Then, we will apply the Lenglart-Rebolledo inequality, given in Lemma \ref{lemma:LRInequality}, to show that $A(\tau_n)=o_p(1)$ implies $\sup_{z\leq \tau_n}R(z)=o_p(1)$, and thus we will deduce the desired result from $n\Upsilon^2_n=R(\tau_n)$.

To find the process $A$, note that
\begin{align*}
R(z)=Z_D(z)+Z_O(z),
\end{align*}
where $(Z_D(z))_{z\geq 0}$ and $(Z_O(z))_{z\geq 0}$ are the processes defined by 
\begin{align*}
Z_D(z)&=\frac{1}{n}\int_{0}^{z}L(t,t)\sum_{i=1}^n\sum_{j=1}^n\kappa_{i,j}(t,t) dM(t)dM(t)
\end{align*}
and
\begin{align*}
Z_O(z)&=\frac{2}{n}\int_{0}^{z}\int_{(0,t)}L(t,s)\sum_{i=1}^n\sum_{j=1}^n\kappa_{i,j}(t,s)dM(s)dM(t).
\end{align*}
By Theorem \ref{thm:doubleMartingale}, since the functions $L(t,s)$ and $\kappa_{i,j}(t,s)$ (for any $i,j\in\{1,\ldots,n\}$) are both predictable functions in the sense of definition \ref{def:predicDsigmaAlgebra}, the process $Z_O(z)$ is a zero-mean square-integrable $(\mathcal{F}_z)$-martingale. By the Optional Stopping Theorem, $\E(Z_O(T))=0$ for any bounded stopping time $T$, and thus
\begin{align}
\E(R(T))=\E(Z_D(T)),\label{eqn:Majorize}
\end{align}
for any bounded stopping time $T$. 

Note that
\begin{align*}
Z_D(z)&=\frac{1}{n}\int_{0}^{z}L(t,t)\sum_{i=1}^n\sum_{j=1}^n\kappa_{i,j}(t,t) dM(t)dM(t)\\
&=\frac{1}{n}\int_{0}^{z}L(t,t)\sum_{i=1}^n\sum_{j=1}^n\kappa_{i,j}(t,t) dN(t),
\end{align*}
where the second equality holds since $dM(t)^2=dN(t)$, which follows from the fact the the $T_i'$s are continuous random variables. Additionally, note that $Z_D(0)=0$, and that $Z_D(z)\geq 0$, since
\begin{align*}
L(t,t)\sum_{i=1}^n\sum_{j=1}^n\kappa_{i,j}(t,t)&=\left\|\sum_{j=1}^n\left(\frac{\E(\mathfrak{K}((t,X_j'),\cdot)Y_j'(t))}{nS_T(t)}-\mathfrak{K}((t,X_j),\cdot)\frac{Y_j(t)}{Y(t)}\right)\right\|^2_{\mathcal{H}}\geq 0.
\end{align*}
The process $Z_D(z)$ is adapted and increasing, and it can be compensated by 
\begin{align*}
\langle Z_D(z)\rangle&=\frac{1}{n}\int_{0}^{z}L(t,t)\sum_{i=1}^n\sum_{j=1}^n\kappa_{i,j}(t,t) Y(t)d\Lambda_Z(t).
\end{align*}
Since  $Z_D-\langle Z_D\rangle$ is a martingale, then $\E(Z_D(T)-\langle Z_D(T)\rangle)=0$ for any finite stopping time $T$. Using Equation \eqref{eqn:Majorize}, we conclude that 
\begin{align*}
E(R(T))=\E(\langle Z_D(T)\rangle),
\end{align*} 
for any finite stopping time. That is, the process $\langle Z_D(z)\rangle$ is a predictable increasing process with $\langle Z_D(0)\rangle=0$, and $\langle Z_D(z)\rangle$ majorizes $R(z)$.

The final step is to prove that $\langle Z_D(\tau_n)\rangle=o_p(1)$, and to use the Lenglart-Rebolledo inequality to conclude the main result. Note that
\begin{align*}
\langle Z_D(\tau_n)\rangle&=O_p(1)\int_{0}^{z}L(t,t)\sum_{i=1}^n\sum_{j=1}^n\kappa_{i,j}(t,t) S_C(t)dF_Z(t),
\end{align*}
since by by Proposition \ref{prop:ProbBounds}.\ref{prop:ProbBound2}, $Y(t)/n=O_p(1)S_T(t)$ uniformly for all $t\leq\tau_n$. Recall that 
\begin{align*}
L(t,t)\sum_{i=1}^n\sum_{j=1}^n\kappa_{i,j}(t,t)&=\left\|\sum_{j=1}^n\left(\frac{\E(\mathfrak{K}((t,X_j'),\cdot)Y_j'(t))}{nS_T(t)}-\mathfrak{K}((t,X_j),\cdot)\frac{Y_j(t)}{Y(t)}\right)\right\|^2_{\mathcal{H}}.\\
\end{align*}
and note that
\begin{align*}
&L(t,t)\sum_{i=1}^n\sum_{j=1}^n\kappa_{i,j}(t,t)\\
&\leq \sum_{j=1}^n\left\|\frac{\E(\mathfrak{K}((t,X_j'),\cdot)Y_j'(t))}{nS_T(t)}-\mathfrak{K}((t,X_j),\cdot)\frac{Y_j(t)}{Y(t)}\right\|^2_{\mathcal{H}}\\
&=L(t,t)\sum_{j=1}^n\left(\frac{\E(K(X_j',X_j')Y_j'(t))}{n^2S_T(t)^2}-\frac{2\E'(K(X_j',X_j)Y_j'(t))Y_j(t)}{nS_T(t)Y(t)}+\frac{K(X_j,X_j)Y_j(t)}{Y(t)^2}\right)\\
&\leq C\sum_{j=1}^n\left(\frac{1}{n^2S_T(t)}+\frac{2Y_j(t)}{nY(t)}+\frac{Y_j(t)}{Y(t)^2}\right)\\
&\leq C\left(\frac{1}{nS_T(t)}+\frac{2}{n}+\frac{1}{Y(t)}\right),
\end{align*}
where the first equality is due to the reproducing property and Assumption \ref{Assu:BoundedKernelSep}, and the second inequality follows also from Assumption \ref{Assu:BoundedKernelSep}, in which we assume the kernel is bounded by some constant $C>0$. 

Using the previous upper bound, we obtain
\begin{align*}
\langle Z_D(\tau_n)\rangle&=O_p(1)\int_{0}^{\tau_n}\left(\frac{1}{nS_T(t)}+\frac{2}{n}+\frac{1}{Y(t)}\right) S_C(t)dF_Z(t)\\
&=O_p(1)\int_{0}^{\tau_n}\frac{1}{nS_T(t)}S_C(t)dF_Z(t)\\
&=O_p(1)\frac{1}{n}\Lambda_Z(\tau_n).
\end{align*}

Note that $\tau_n=\max\{T_1,\ldots,T_n\}\leq\max\{Z_1,\ldots,Z_n\}$, and thus
\begin{align*}
\frac{1}{n}\Lambda_Z(\tau_n)\leq \frac{1}{n}\max_{i\in\{1,\ldots,n\}}\Lambda_Z(Z_i).
\end{align*}
Note also that the random variables $\Lambda_Z(Z_i)$ are i.i.d. exponential random variables with mean 1, as 
\begin{align*}
\Prob(\Lambda_Z(Z_1)>x)=\Prob(Z_1>\Lambda_Z^{-1}(x))=S_Z(\Lambda_Z^{-1}(x))=\exp\{-\Lambda_Z(\Lambda_Z^{-1}(x))\}=\exp\{-x\}.
\end{align*}
We finally deduce the result in consequence of
\begin{align*}
P\left(\max_{i\in\{1,\ldots,n\}}\Lambda_Z(Z_i)\geq 2\log(n)\right)&\leq nP\left(\Lambda_Z(Z_1)\geq 2\log(n)\right)=n\exp\{\log (n^{-2})\}=\frac{1}{n}.
\end{align*}
Thus in a set of probability tending to 1, $\max_{i\in\{1,\ldots,n\}}\Lambda_Z(Z_i)\leq 2\log(n)$, and thus
\begin{align*}
\langle Z_D(\tau_n)\rangle&=O_p(1)\frac{1}{n}\Lambda_Z(\tau_n)=o_p(1).
\end{align*}

\subsubsection{Proof of Theorem \ref{thm:rkhsLimit}}
\begin{proof}
By Lemma \ref{lemma:rkhs approx}, 
\begin{align*}
\Psi_n^2&=\frac{1}{n^2}\sum_{i=1}^n\sum_{l=1}^nJ((T_i,\Delta_i,X_i),(T_l,\Delta_l,X_l))+o_p(n^{-1})
\end{align*}
under the null hypothesis. The non-negligible part in the right-hand side of the previous equation is a degenerate $V$-statistic of order 2. 

The degeneracy property can be easily verified, since 
\begin{align*}
J((T_i,\Delta_i,X_i),(t,r,x))&=\int_{\R_+}\left(\int_{\R_+}\bar{\mathfrak{K}}((s,
X_i),(s',x))dm_{t,r}(s')\right)dM_i(s)\\
&=\int_{\R_+}f_i(s,t,r,x)dM_i(s),
\end{align*}
where $f_i(s,t,r,x)=\int_{\R_+}\bar{\mathfrak{K}}((s,X_i),(s',x))dm_{t,r}(s')$. Then, note that for any fixed triple $(t,r,x)\in\R_+\times\{0,1\}\times\R^d$, the process $(f_i(s,t,r,x))_{s\geq 0}$ is $(\mathcal{F}_t)$-predictable, and thus $\E(J(T_i,\Delta_i,X_i),(t,r,x))=\E\left(\int_{\R_+}f_i(s,t,r,x)dM_i(s)\right)=0$ for any $(t,r,x)\in\R_+\times\{0,1\}\times\R^d$, since $\int_{0}^zf_i(s,t,r,x)dM_i(s)$, $z\geq 0$, is a $\mathcal{F}_z$-martingale. 

Using the previous observation, and from the classical theory of $V$-statistics \cite{serfling2009approximation}, we obtain
\begin{align}
n\Psi^2&\overset{\mathcal{D}}{\to}E(J((T_i,\Delta_i,X_i),(T_i,\Delta_i,X_i)))+\mathcal{Y},
\end{align} 
where $\mathcal{Y}=\sum_{i=1}^n\lambda_i(\xi_i^2-1)$, $\xi_1,\xi_2,\ldots$ are independent standard normal random variables, and $\lambda_1,\lambda_2,\ldots$ are the eigenvalues associated to the integral operator of $J$, $T^J$, defined in Equation \eqref{eqn:IOTJ}. 

Finally, note that
\begin{align*}
&\E(J((T_i,\Delta_i,X_i),(T_i,\Delta_i,X_i)))\\
&=\E\left(\int_{0}^{T_i}\int_{0}^{T_i}\ind_{\{t\neq t'\}}\bar{\mathfrak{K}}((t,X_i),(t',X_i))dM_{i}(t)dM_{i}(t')\right)+\E\left(\bar{\mathfrak{K}}((T_i,X_i),(T_i,X_i))\Delta_i\right)\\
&=\int_{x\in\R^d}\int_{0}^{\tau}\bar{\mathfrak{K}}((t,x),(t,x))S_{C|X=x}(t)dF_{Z}(t)dF_X(x),
\end{align*}
where the expectation of the first term is equal to zero by the double martingale Theorem \ref{thm:doubleMartingale}.
\end{proof}

\subsubsection{Proof of Theorem \ref{Thm:SimplerKernel}}
\begin{proof}
Let $Q:(\R_+\times\R_+\times\R^d)^2\to\R$ be given by
\begin{align*}
Q((z_1,c_1,x_1),(z_2,c_2,x_2))&=\ind_{\{z_1\leq c_1\}}\ind_{\{z_2\leq c_2\}}\bar{\mathfrak{K}}((z_1,x_1),(z_2,x_2)).
\end{align*}
Note that $\Delta_i\Delta_j \bar{\mathfrak{K}}((T_i,X_i),(T_j,X_j))=Q((Z_i,C_i,X_i),(Z_j,C_j,X_j))$,
and thus the asymptotic distribution of $\frac{1}{n}\sum_{i=1}^n\sum_{j=1}^n\Delta_i\Delta_j \bar{\mathfrak{K}}((T_i,X_i),(T_j,X_j))$ can be found by studying the asymptotic distribution of the degenerate $V$-statistic,
\begin{align*}
\frac{1}{n}\sum_{i=1}^n\sum_{j=1}^n Q((Z_i,C_i,X_i),(Z_j,C_j,X_j)).
\end{align*}

Let $Q_{ij}=Q((Z_i,C_i,X_i),(Z_j,C_j,X_j))$. Then $\E(|Q_{ii}|)<\infty$ and $\E(Q_{ij}^2)<\infty$ for $i\neq j$, since, by Assumption \ref{Assu:BoundedKernelSep}, the kernel $\mathfrak{K}$ is bounded. By the classical theory of $U/V$-statistics \cite{serfling2009approximation},
\begin{align*}
\frac{1}{n}\sum_{i=1}^n\sum_{j=1}^nQ_{ij}\overset{\mathcal{D}}{\to}\int_{\R^d}\int_0^\tau \bar{\mathfrak{K}}((t,x),(t,x))S_{C|X=x}(t)dF_Z(t)dF_X(t)+\mathcal{Y}',
\end{align*}
where $\mathcal{Y}'=\sum_{i=1}^\infty \lambda_i'(\xi^2_i-1)$, $\xi_1,\xi_2,\ldots$ are independent standard normal random variables and $\lambda_1,',\lambda_2',\ldots$ are the eigenvalues associated to the integral operator $T^Q$. By Theorem \ref{Thm:Eigen}, $T^J$ (the integral operator of $J$) and $T^Q$ share the same set of non-zero eigenvalues (including multiplicities), and thus the result holds.
\end{proof}

\subsubsection{Proof of lemma \ref{lemma:consistency}}
Observe that
\begin{align*}
\|\phi_0^n-\phi_1^n\|_{\mathcal{H}}^2&=V_{0,0}-2V_{0,1}+V_{1,1},
\end{align*}
where 
\begin{align}
V_{0,0}&=\frac{1}{n^2}\sum_{i=1}^n\sum_{j=1}^n\sum_{l=1}^n\sum_{k=1}^n\mathfrak{K}((T_i,X_j),(T_l,X_k))\frac{Y_j(T_i)}{Y(T_i)}\frac{Y_k(T_l)}{Y(T_l)}\Delta_i\Delta_l,\label{eqnV00}\\
V_{0,1}&=\frac{1}{n^2}\sum_{i=1}^n\sum_{j=1}^n\sum_{l=1}^n\mathfrak{K}((T_i,X_j),(T_l,X_l))\Delta_i\Delta_l\frac{Y_j(T_i)}{Y(T_i)},\label{eqnV01}\\
V_{1,1}&=\frac{1}{n^2}\sum_{i=1}^n\sum_{l=1}^n\mathfrak{K}((T_i,X_i),(T_l,X_l))\Delta_i\Delta_l\label{eqnV11},
\end{align}
hold by the reproducing property. The desired result 
\begin{align*}
\|\phi_0^n-\phi_1^n\|_{\mathcal{H}}^2\overset{\Prob}{\to}\|\phi_0-\phi_1\|_{\mathcal{H}}^2,
\end{align*}
follows by proving 
\begin{itemize}
\item[i)]$V_{0,0}\overset{\Prob}{\to}\|\phi_0\|_{\mathcal{H}}^2$,
\item[ii)]$V_{0,1}\overset{\Prob}{\to}\langle\phi_0,\phi_1\rangle_{\mathcal{H}}$,
\item[iii)]$V_{1,1}\overset{\Prob}{\to}\|\phi_1\|_{\mathcal{H}}^2$.
\end{itemize}

We start with iii). Note that $V_{1,1}$ is a V-statistic of order 2, and that, by Assumption \ref{Assu:BoundedKernelSep}, $\E(|\mathfrak{K}((T_i,X_i),(T_i,X_i))\Delta_i|)<\infty$ and $\E(|\mathfrak{K}((T_i,X_i),(T_j,X_j))\Delta_i\Delta_j|)<\infty$ for $i\neq j$. Then, by the  classical theory of V-statistics, it holds
\begin{align*}
V_{1,1}&\overset{\Prob}{\to}\E(\mathfrak{K}((T_1,X_1),(T_2,X_2))\Delta_1\Delta_2)\\
&=\int_0^\tau\int_{x\in\R^d}\int_0^\tau\int_{x'\in\R^d}\mathfrak{K}((t,x),(t',x'))S_{C|X=x}(t)S_{C|X=x'}(t')dF_{ZX}(t',x')dF_{ZX}(t,x)\\
&=\|\phi_1\|^2_{\mathcal{H}}.
\end{align*}

We next prove i) and ii). Since the proofs of $i)$ and $ii)$ use the same arguments, it sufficesto prove the result for $i)$. Rewrite $V_{0,0}$ as
\begin{align*}
V_{0,0}=\frac{1}{n^2}\sum_{i=1}^n\sum_{j=1}^n\sum_{l=1}^n\sum_{k=1}^nf(i,j,l,k),
\end{align*}
where
\begin{align*}
f(i,j,l,k)&=\mathfrak{K}((T_i,X_j),(T_l,X_k))\frac{Y_j(T_i)}{Y(T_i)}\frac{Y_k(T_l)}{Y(T_l)}\Delta_i\Delta_l.
\end{align*}
Note that 
\begin{align*}
f(i,j,l,k)\leq C\frac{Y_j(T_i)}{Y(T_i)}\frac{Y_k(T_l)}{Y(T_l)},
\end{align*}
for some constant $C\geq 0$ due to Asssumption \ref{Assu:BoundedKernelSep}. 

We begin by proving that the sum over elements in $V_{0,0}$ that contain the repetition of one index $(i,j,l,k)$ converge to zero. Define the set $\mathcal{S}=\{(i,j,l,k)\in[n]^4:|\{i,j,l,k\}|\leq 3\}$, e.g. $(1,1,2,3)\in\mathcal{S}$, and let $\mathcal{S}_{i=l}$ be the set containing all the indices $(i,j,l,k)$ such that $i=l$. It is not hard to see that $\mathcal{S}\subseteq \mathcal{S}_{i=l}\cup\mathcal{S}_{i=j}\cup\mathcal{S}_{i=k}\cup\mathcal{S}_{l=k}\cup\mathcal{S}_{l=j}\cup\mathcal{S}_{j=k}$ (note that the intersection of these sets may not be empty). 

Consider the set $\mathcal{S}_{i=l}$, and note that the sum over the elements of $\mathcal{S}_{i=l}$ satisfies
\begin{align*}
\frac{1}{n^2}\sum_{(i,j,k,l)\in\mathcal{S}_{i=l}} f(i,j,l,k)\leq\frac{1}{n^2}\sum_{i=1}^n\sum_{j=1}^n\sum_{k=1}^n C\frac{Y_j(T_i)}{Y(T_i)}\frac{Y_k(T_i)}{Y(T_i)}\leq\frac{1}{n^2}\sum_{i=1}^nC\to0.
\end{align*}
as $n$ tends to infinity. 

The sum over the elements in $\mathcal{S}_{i=j}$ (or, by symmetry, over the elements $\mathcal{S}_{l=k}$) satisfies 
\begin{align*}
\frac{1}{n^2}\sum_{(i,j,k,l)\in\mathcal{S}_{i=j}} f(i,j,l,k)&\leq\frac{1}{n^2}\sum_{i=1}^n\sum_{l=1}^n\sum_{k=1}^n C\frac{Y_i(T_i)}{Y(T_i)}\frac{Y_k(T_l)}{Y(T_l)}\leq\frac{C}{n}\sum_{i=1}^n\frac{1}{Y(T_i)}\\
&\leq \frac{C}{n}\sum_{i=1}^n\frac{1}{n-i+1}\leq \frac{C}{n}\sum_{k=1}^n\frac{1}{k}\leq C\frac{\log(n)+1}{n}\to 0,
\end{align*}
as n grows to infinity.

The sum over the elements of $\mathcal{S}_{i=k}$ (or, by symmetry, over $\mathcal{S}_{l=j}$) satisfies
\begin{align*}
\frac{1}{n^2}\sum_{(i,j,k,l)\in\mathcal{S}_{i=k}}f(i,j,l,k)&\leq\frac{1}{n^2}\sum_{i=1}^n\sum_{j=1}^n\sum_{l=1}^nC\frac{Y_j(T_i)}{Y(T_i)}\frac{Y_i(T_l)}{Y(T_l)}\\
&\leq \frac{1}{n^2}\sum_{i=1}^n\sum_{l=1}^nC\frac{Y(T_i)}{Y(T_i)}\frac{Y_i(T_l)}{Y(T_l)}\leq \frac{1}{n^2}\sum_{l=1}^nC\to0.
\end{align*}
Finally, the sum over the elemnts of  $\mathcal{S}_{j=k}$, satisfies
\begin{align*}
\frac{1}{n^2}\sum_{(i,j,k,l)\in\mathcal{S}_{j=k}} f(i,j,l,k)&\leq\frac{1}{n^2}\sum_{i=1}^n\sum_{l=1}^n\sum_{k=1}^nC\frac{Y_k(T_i)}{Y(T_i)}\frac{Y_k(T_l)}{Y(T_l)}\leq\frac{C}{n}\sum_{l=1}^n\frac{1}{Y(T_l)}\to0,\\
\end{align*}
as $n$ grows to infinity.

The previous results imply
\begin{align*}
V_{0,0}=\frac{1}{n^2}\sum_{(i,j,k,l)\in\mathcal{S}^c}\mathfrak{K}((T_i,X_j),(T_l,X_k))\frac{Y_j(T_i)}{Y(T_i)}\frac{Y_k(T_l)}{Y(T_l)}\Delta_i\Delta_l+o(1),
\end{align*} 
where $\mathcal{S}^c=\{(i,j,l,k)\in[n]^4:|\{i,j,l,k\}|=4\}$.

We continue by assuming that $Y(t)/n$ can be replaced by its limit, $S_T(t)$, in the previous equation. By doing this, we obtain
\begin{align}
V_{0,0}=\frac{1}{n^4}\sum_{\mathcal{S}^c}\mathfrak{K}((T_i,X_j),(T_l,X_k))\frac{Y_j(T_i)}{S_T(T_i)}\frac{Y_k(T_l)}{S_T(T_l)}\Delta_i\Delta_l+o_p(1),\label{eqn:assV00}
\end{align} 
which is a U-statistic of order 4. Then, by using the law of large numbers for U-statistics \cite{serfling2009approximation}, it follows that
\begin{align*}
V_{0,0}&=\E\left(\mathfrak{K}((T_1,X_3),(T_2,X_4))\frac{Y_3(T_1)}{S_T(T_1)}\frac{Y_4(T_2)}{S_T(T_2)}\Delta_1\Delta_2\right)+o_p(1)\\
&=\int_0^\tau\int_{x\in\R^d}\int_0^\tau\int_{x'\in\R^d}K((t,x),(t',x'))d\nu_0(t',x')d\nu_0(t,x)+o_p(1), \\
&=\|\phi_0\|^2_{\mathcal{H}}+o_p(1)
\end{align*} 
from which we conclude the desired result. 

We finalise our proof by justifying the replacement of $Y(t)$ by $nS_T(t)$ made in equation  \eqref{eqn:assV00}. By the triangular inequality, it holds
\begin{align}
&\left|\frac{1}{n^2}\sum_{\mathcal{S}^c}\mathfrak{K}((T_i,X_j),(T_l,X_k))\left(\frac{Y_j(T_i)}{Y(T_i)}\frac{Y_k(T_l)}{Y(T_l)}-\frac{Y_j(T_i)}{nS_T(T_i)}\frac{Y_k(T_l)}{nS_T(T_l)}\right)\Delta_i\Delta_l\right|\nonumber\\
&\quad\leq \left|\frac{1}{n^2}\sum_{\mathcal{S}^c}\mathfrak{K}((T_i,X_j),(T_l,X_k))\frac{Y_j(T_i)}{Y(T_i)}\left(\frac{Y_k(T_l)}{Y(T_l)}-\frac{Y_k(T_l)}{nS_T(T_l)}\right)\Delta_i \Delta_l\right|\label{eqn381}\\
&\quad\quad+\left|\frac{1}{n^2}\sum_{\mathcal{S}^c}\mathfrak{K}((T_i,X_j),(T_l,X_k))\frac{Y_k(T_l)}{S_T(T_l)}\left(\frac{Y_j(T_i)}{Y(T_i)}-\frac{Y_j(T_i)}{nS_T(T_i)}\right)\Delta_i \Delta_l\right|\label{eqn382}.
\end{align}

We prove that equations \eqref{eqn381} and \eqref{eqn382} are both $o_p(1)$ under  Assumption \ref{Assu:BoundedKernelSep}. Since the proofs use the same arguments, we only show the result for equation \eqref{eqn381}. 

Note that under Assumption \ref{Assu:BoundedKernelSep}, it holds
\begin{align*}
&\left|\frac{1}{n^2}\sum_{(i,j,k,l)\in\mathcal{S}^c}\mathfrak{K}((T_i,X_j),(T_l,X_k))\frac{Y_j(T_i)}{Y(T_i)}\left(\frac{Y_k(T_l)}{Y(T_l)}-\frac{Y_k(T_l)}{nS_T(T_l)}\right)\Delta_i \Delta_l\right|\\
&\quad\leq \frac{C}{n^2}\sum_{i=1}^n\sum_{j=1}^n\sum_{l=1}^n\sum_{k=1}^n\frac{Y_j(T_i)}{Y(T_i)}\left|\frac{Y_k(T_l)}{Y(T_l)}-\frac{Y_k(T_l)}{nS_T(T_l)}\right|\Delta_l\\
&\quad\leq \frac{C}{n}\sum_{l=1}^n\sum_{k=1}^n\left|\frac{Y_k(T_l)}{Y(T_l)}-\frac{Y_k(T_l)}{nS_T(T_l)}\right|\Delta_l\\
&\quad\leq \frac{C}{n}\int_0^\tau\sum_{k=1}^n\frac{Y_k(t)}{Y(t)}\left|1-\frac{Y(t)/n}{S_T(t)}\right|dN(t)\\
&\quad\leq \frac{C}{n}\int_0^\tau\left|1-\frac{Y(t)/n}{S_T(t)}\right|dN(t)=o_p(1),
\end{align*}
where the last equality follows from Lemma \ref{Lemma:ReplacementYP}.

\subsubsection{Proof of Theorem \ref{Thm:Power}}

By Lemma \ref{lemma:consistency} under Assumption \ref{Assu:BoundedKernelSep}, we have $$\Psi^2_n\overset{\Prob}{\to}\|\phi_1-\phi_0\|^2_{\mathcal{H}},$$ where
\begin{align*}
\phi_i(s,y)=\int_{0}^\tau\int_{x\in\R^d}L(s,t)K(y,x)d\nu_i(t,x),\quad i\in\{0,1\}.
\end{align*}
Then, the result follows from proving $\|\phi_1-\phi_0\|^2_{\mathcal{H}}=0$ if and only if the null hypothesis holds, since if  $\|\phi_1-\phi_0\|^2_{\mathcal{H}}>0$, then $n\Psi_n^2=n\|\phi_1-\phi_0\|^2_{\mathcal{H}}\to\infty$ as $n\to\infty$.

Suppose that the null hypothesis holds. Then, under Assumption \ref{assumption:no_complete_censoring}, Proposition \ref{Prop: nu iff indep} implies that $\nu_1=\nu_0$, and thus $\phi_0=\phi_1$.

We proceed to show that $\phi_1-\phi_0=0$ implies $\nu_1=\nu_0$. Note that by Proposition \ref{Prop: nu iff indep}, the latter implies that the null hypothesis holds. 

To show this result, we repeat the proof in \cite{gretton2015simpler}. Recall that
\begin{align}
(\phi_1-\phi_0)(s,y)=\int_0^\tau\int_{x\in\R^d}K(y,x)L(s,t)(d\nu_1(t,x)-d\nu_0(t,x))\label{eqn:funDiff}.
\end{align}  
 
Note that by our assumptions both kernels $K$ and $L$ are characteristic, translation invariant, $c_0$-kernels. Thus, an application of proposition $\ref{prop:trans}$ allows us to deduce that both $K$ and $L$ are $c_0$-universal kernels.

Let $\mathcal{H}_K$ and $\mathcal{H}_L$ be the RKHS induced by the kernels $K$ and $L$, respectively. For every $\kappa \in\mathcal{H}_K$ define the signed measure $\gamma_\kappa$ given by
\begin{align*}
\gamma_\kappa(B)&=\int_{B}\int_{\R^d}\langle K(\cdot,x),\kappa\rangle_{\mathcal{H}_K}(d\nu_1(t,x)-d\nu_0(t,x)),
\end{align*}
for any measurable $B\subseteq (0,\tau)$. Consider the embedding of the above measure onto $\mathcal{H}_L$, $\mu_{\gamma_\kappa}$ given by
\begin{align*}
\mu_{\gamma_\kappa}(s)&=\int_0^\tau L(s,t)d\gamma_\kappa(t).
\end{align*}
Following some computations, we obtain
\begin{align*}
\mu_{\gamma_\kappa}(s)&=\int_0^\tau L(s,t)\int_{\R^d}\langle K(\cdot,x),\kappa\rangle_{\mathcal{H}_K}(d\nu_1(t,x)-d\nu_0(t,x))\\
&=\left\langle\int_0^\tau\int_{\R^d} L(s,t)K(\cdot,x)(d\nu_1(t,x)-d\nu_0(t,x)),\kappa\right\rangle_{\mathcal{H}_K}\\
&=\left\langle(\phi_1-\phi_0)(s,\cdot),\kappa\right\rangle_{\mathcal{H}_K}\\
&=0,
\end{align*}
since $(\phi_1-\phi_0)(s,y)=0$ for all $(s,y)\in \R_+\times\R_d$ by hypothesis.

By Proposition \ref{prop:uni}, since $L$ is $c_0$-universal, the embedding $\mu_{\gamma_\kappa}$ is injective, and thus $\gamma_\kappa(B)=0$ for all $B$ and $\kappa\in \mathcal{H}_K$. Since this is true for all $\kappa\in\mathcal{H}_K$, it is true in particular for $K(\cdot,y)\in\mathcal{H}_K$, and thus for $\kappa(\cdot)=K(\cdot,y)$, we have
\begin{align*}
\gamma_{\kappa}(B)&=\int_{B}\int_{\R^d}\langle K(\cdot,x),K(\cdot,y)\rangle_{\mathcal{H}_K}(d\nu_1(t,x)-d\nu_0(t,x))\\
&=
\int_0^\tau\int_{\R^d}K(y,x)\ind_{B}(t)(d\nu_1(t,x)-d\nu_0(t,x))=0.
\end{align*}
for all measurable $B\subseteq \R$. For any fixed measurable $B\subseteq (0,\tau)$, define the signed measure $\tilde\gamma_B(A)=\int_0^\tau\int_A \ind_{B}(t)(d\nu_1(t,x)-d\nu_0(t,x))$ for any measurable $A\subseteq \R^d$. The above equation can be interpreted as the embeding of $\tilde\gamma_B$ to $\mathcal{H}_K$, that is,
\begin{align*}
\mu_{\tilde\gamma_B}(y)=\int_0^\tau\int_{\R^d}K(y,x)\ind_{B}(t)(d\nu_1(t,x)-d\nu_0(t,x))=0.
\end{align*}
which is also injective (by Proposition \ref{prop:uni}, since $K$ is $c_0$-universal).  We conclude that
$$\tilde\gamma_B(A)=\int_0^\tau\int_A\ind_{B}(t)(d\nu_1(t,x)-d\nu_0(t,x))=0$$
for all Borel sets $A,B$, and hence $\nu_1-\nu_0=0$.

\subsection{Proofs of Section \ref{sec:wildbootstrap}}
\subsubsection{Proof of Lemma \ref{lemma:Wild1} }
The desired result is obtained by showing that 
\begin{align*}
\sup_{\omega\in\mathcal{H}:\|\omega\|_{\mathcal{H}}\leq 1}\frac{1}{n}\sum_{i=1}^n\int_0^\tau W_i\left(\bar\omega_n(t)-\bar{\omega}(t)\right)dN_i(t)&=o_p(n^{-1/2}),
\end{align*}
where $\bar\omega(t)$, given by $\bar\omega(t)=\int_{\R^d} \omega(t,x)\frac{S_{T|X=x}(t)}{S_T(t)}dF_X(x),$ is the population version of $\bar\omega_n(t)=\sum_{j=1}^n\omega(t,X_j)\frac{Y_j(t)}{Y(t)}$.

Let $\omega^\star(t)=\frac{1}{nS_T(t)}\sum_{j=1}^n\omega(t,X_j)Y_j(t)$. By the triangular inequality, 
\begin{align*}
\sup_{\omega\in\mathcal{H}:\|\omega\|_{\mathcal{H}}\leq 1}\frac{1}{n}\sum_{i=1}^n\int_0^\tau W_i\left(\bar\omega_n(t)-\bar{\omega}(t)\right)dN_i(t)&\leq E_1+E_2,
\end{align*}
where 
\begin{align}
E_1&=\sup_{\omega\in\mathcal{H}:\|\omega\|_{\mathcal{H}}\leq 1}\frac{1}{n}\sum_{i=1}^n\int_0^\tau W_i\left(\bar\omega_n(t)-\omega^\star(t)\right)dN_i(t)\nonumber\\
&=\sup_{\omega\in\mathcal{H}:\|\omega\|^2_{\mathcal{H}}\leq 1}\frac{1}{n}\sum_{i=1}^n\int_0^\tau W_i\sum_{j=1}^n\omega(t,X_j)Y_j(t)R_n(t)dN_i(t),\label{eqn:diffLRW}
\end{align}
where $R_n(t)=\frac{1}{Y(t)}-\frac{1}{nS_T(t)}$, and
\begin{align}
E_2&=\sup_{\omega\in\mathcal{H}:\|\omega\|_{\mathcal{H}}\leq 1}\frac{1}{n}\sum_{i=1}^n\int_0^\tau W_i\left(\bar\omega_n(t)-\omega^\star(t)\right)dN_i(t)\nonumber\\
&=\sup_{\omega\in\mathcal{H},\|\omega\|^2\leq 1}\frac{1}{n}\sum_{i=1}^n\int_0^\tau \frac{W_i}{
nS_T(t)}\left(\sum_{j=1}^n\omega(t,X_j)Y_j(t)-g(t)\right)dN_i(t),\label{eqn:secondlimit}
\end{align}
where $g(t)=\int_{\R^d}\omega(t,x)S_{T|X=x}(t)dF_X(x)$. The result follows by proving that both $E_1$ and $E_2$ are $o_p(n^{-1/2})$. 

We proceed to prove $E_1=o_p(n^{-1/2})$. Let $\beta\in(0,1)$, and define the set $B_{n,\beta}$ by 
\begin{align}
B_{n,\beta}=\left\{\frac{Y(t)/n}{S_T(t)}\leq\beta^{-1},\quad \forall t\leq\tau_n\right\}.\label{eqn:betaset}
\end{align}
By Proposition \ref{prop:ProbBounds}.\ref{prop:ProbBound2}, the set $B_{n,\beta}$ satisfies $\Prob(B_{n,\beta})\geq 1-\beta$. Note that we only need to prove that $n^{1/2}E_1$ converges to zero in probability in the set $B_{n,\beta}$, as $\beta$ can be chosen arbitrarily small. Observe 
\begin{align*}
nE_1^2&=\left(\sup_{\omega\in\mathcal{H}:\|\omega\|^2_{\mathcal{H}}\leq 1}\frac{1}{n^{1/2}}\sum_{i=1}^n\int_0^\tau W_i\sum_{j=1}^n\omega(t,X_j)Y_j(t)R_n(t)dN_i(t)\right)^2\\
&=\frac{1}{n}\sum_{i,l=1}^nW_iW_l\int_0^\tau\int_0^\tau \sum_{j,k=1}^n\mathfrak{K}((t,X_j),(s,X_k))Y_j(t)Y_k(s)R_n(t)R_n(s)dN_i(t)dN_l(s)\\
&\leq \frac{C}{n}\sum_{i,l=1}^nW_iW_l\int_0^\tau\int_0^\tau Y(t)Y(s)|R_n(t)||R_n(s)|dN_i(t)dN_l(s),
\end{align*}
where the first equality is due to the fact that we are taking supremum in the unit ball of a RKHS, and the inequality in the second line follows from Assumption \ref{Assu:BoundedKernelSep}; that is, the kernel $\mathfrak{K}$ is bounded by some constant $C>0$.

Note that
\begin{align}
\E(nE_1^2\ind_{\{B_{n,\beta}\}})
&\leq\E\left(\frac{C}{n}\sum_{i=1}^n\left(\int_0^\tau Y(t)|R_n(t)|dN_i(t)\right)^2\ind_{\{B_{n,\beta}\}}\right)\label{eqn:WB1},
\end{align}
since $W_1,\ldots,W_n$ are i.i.d. Rademacher random variables which are independent of our data $((T_i,\Delta_i,X_i))_{i=1}^n$. We will prove that the expectation in the right-hand side of Equation \eqref{eqn:WB1} converges to zero. To do this, we use dominated convergence in sets of arbitrarily large probability. First, note that $$Y(t)|R_n(t)|=\left|1-\frac{Y(t)/n}{S_T(t)}\right|\leq 1+\beta^{-1}$$ uniformly for all $t\leq\tau_n$ in the set $B_{n,\beta}$, and thus
\begin{align}
\frac{C}{n}\sum_{i=1}^n\left(\int_0^\tau Y(t)|R_n(t)|dN_i(t)\right)^2\ind_{\{B_{n,\beta}\}}\leq C(1+\beta^{-1})^2.
\end{align}
Second, note that
\begin{align*}
\frac{C}{n}\sum_{i=1}^n\left(\int_0^\tau Y(t)|R_n(t)|dN_i(t)\right)^2
&=\frac{C}{n}\int_0^\tau\left|1-\frac{Y(t)/n}{S_T(t)}\right|^2dN(t)=o_p(1),
\end{align*}
by Lemma \ref{Lemma:ReplacementYP}. By the dominated convergence theorem we deduce  $\E(nE_1^2\ind_{\{B_{n,\beta}\}})\to0$ as $n$ grows to infinity,  for which follows the desired result $E_1=o_p(n^{-1/2})$.

We continue by proving that $E_2=o_p(n^{-1/2})$. Note that it is enough to prove the result in the set $B_{n,\beta}$ defined in Equation \eqref{eqn:betaset}, as $\beta$ can be chosen arbitrarily small. Observe that
\begin{align*}
nE_2^2&=\left(\sup_{\omega\in\mathcal{H},\|\omega\|^2\leq 1}\frac{1}{n^{1/2}}\sum_{i=1}^n\int_0^\tau \frac{W_i}{
nS_T(t)}\left(\sum_{j=1}^n\omega(t,X_j)Y_j(t)-g(t)\right)dN_i(t)\right)^2\\
&=\frac{1}{n}\sum_{i=1}^n\sum_{l=1}^n\int_0^\tau\int_0^\tau \frac{W_iW_l}{n^2S_T(t)S_T(s)}\mathfrak K^\star(t,s)dN_i(t)dN_l(s),
\end{align*}
where 
\begin{align}
\mathfrak{K}^\star(t,s)&=\sum_{j=1}^n\sum_{k=1}^n\left(\mathfrak{K}((t,X_j),(s,X_k))Y_j(t)Y_k(s)\right.\nonumber\\
&\quad\left.-\int_{x'\in\R^d}\mathfrak{K}((t,X_j),(s,x'))S_{T|X=x'}(s)dF_X(s)Y_j(t)\right.\nonumber\\
&\quad\left.-\int_{x\in\R^d}\mathfrak{K}((t,x),(s,X_k))S_{T|X=x}(t)dF_X(x)Y_k(s)\right.\nonumber\\
&\quad\left.+\int_{x'\in\R^d}\int_{x\in\R^d}\mathfrak{K}((t,x),(s,x'))S_{T|X=x}(t)S_{T|X=x'}(s)dF_X(x)dF_X(x')\right)\label{eqn:kstar},
\end{align}
follows from the fact we are taking supremum in the unit ball of an RKHS, and from the definition of the function $g(t)$. Note that
\begin{align}
\E(nE^2_2\ind{\{B_{n,\beta}\}})&=\E\left(\frac{1}{n^3}\sum_{i=1}^n\int_0^\tau\int_0^\tau \frac{1}{S_T(t)S_T(s)}\mathfrak{K}^\star(t,s)dN_i(t)dN_i(s)\ind_{\{B_{n,\beta}\}}\right)\nonumber\\
&=\E\left(\frac{1}{n^3}\sum_{i=1}^n\int_0^\tau \frac{1}{S_T(t)^2}\mathfrak{K}^\star(t,t)dN_i(t)\ind_{\{B_{n,\beta}\}}\right),\label{Eqn:W22}
\end{align}
where the first line follows since $W_1,\ldots,W_n$ are i.i.d. Rademacher random variables independent from the data, and the second line follows from the fact the process $N_i(t)$ only jumps once.

We use dominated convergence in sets of large probability to show that the expectation in Equation \eqref{Eqn:W22} converges to 0. To verify the conditions of the dominated convergence theorem, we verify that the term
\begin{align}
\Gamma_n=\frac{1}{n^3}\sum_{i=1}^n\int_0^\tau \frac{1}{S_T(t)^2}\mathfrak{K}^\star(t,t)dN_i(t)\ind_{\{B_{n,\beta}\}}\label{eqn:boundedk}
\end{align}

is bounded, and that it converges to zero in probability.

We first check that $\Gamma_n$ is bounded.  Note that by Assumption \ref{Assu:BoundedKernelSep},
\begin{align}
|\mathfrak{K}^\star(t,t)|&\leq C(Y(t)+nS_T(t))^2\label{eqn:Kstarbound},
\end{align}
where $C>0$ is some constant. Thus,
\begin{align*}
\Gamma_n
&\leq\frac{C}{n}\sum_{i=1}^n\int_0^\tau\left(\frac{Y(t)/n}{S_T(t)}+1\right)^2dN_i(t)\ind_{\{B_{n,\beta}\}}\leq C(\beta^{-1}+1)^2.
\end{align*}

Finally, we need to check that $\Gamma_n=o_p(1)$. Let $\epsilon>0$, and let $t_{\epsilon}>0$ be such that $S_T(t_{\epsilon})=\epsilon$. Observe that
\begin{align*}
\Gamma_n
&=\frac{1}{n^3}\sum_{i=1}^n\int_0^{t_{\epsilon}} \frac{1}{S_T(t)^2}\mathfrak{K}^\star(t,t)dN_i(t)\ind_{\{B_{n,\beta}\}}+\frac{1}{n^3}\sum_{i=1}^n\int_{t_{\epsilon}}^\tau \frac{1}{S_T(t)^2}\mathfrak{K}^\star(t,t)dN_i(t)\ind_{\{B_{n,\beta}\}}.
\end{align*}

We prove that each of the sums in the right-hand side of the previous equation converges to zero in probability. For the first sum, observe that
\begin{align*}
\frac{1}{n^3}\sum_{i=1}^n\int_0^{t_{\epsilon}}\frac{1}{S_T(t)^2}\mathfrak K^\star(t,t)dN_i(t)\ind_{\{B_n\}}&\leq \frac{1}{\epsilon^2}\frac{1}{n^3}\sum_{i=1}^n\int_0^{t_{\epsilon}}\mathfrak K^\star(t,t)dN_i(t)\ind_{\{B_n\}}\\
&\leq\frac{1}{\epsilon^2}\frac{1}{n^3}\sum_{i=1}^n\mathfrak K^\star(T_i,T_i)\\
&=o_p(1),
\end{align*}
where the first inequality follows from the definition of $t_{\epsilon}$, and  by
\begin{align*}
\mathfrak{K}^\star(t,t)&=\left(\sup_{\omega\in\mathcal{H}:\|\omega\|^2_{\mathcal{H}}\leq 1}\sum_{j=1}^n\left(\omega(t,X_j)-\int_{\R^d}\omega(t,x)S_{T|X=x}(t)dF_X(x)\right)\right)^2\geq 0,
\end{align*}
and the first equality follows from $\frac{1}{n^3}\sum_{i=1}^n\mathfrak{K}^\star(T_i,T_i)$ being a bounded $V$-statistic of order 3 (see Equation \eqref{eqn:kstar}) whose diagonal part converges to zero, and with zero mean on the off-diagonal part.

For the second integral, by Equation \eqref{eqn:Kstarbound} and the definition of the set $B_{n,\beta}$, observe
\begin{align*}
\frac{1}{n^3}\sum_{i=1}^n\int_{t_{\epsilon}}^\tau \frac{1}{S_T(t)^2}\mathfrak{K}^\star(t,t)dN_i(t)\ind_{\{B_{n,\beta}\}}
&\leq \frac{C}{n}\sum_{i=1}^n\int_{t_{\epsilon}}^\tau \left(\frac{Y(t)/n}{S_T(t)}+1\right)^2 dN_i(t)\ind_{\{B_{n,\beta}\}}\\
&\leq C(\beta^{-1}+1)^2\frac{1}{n}\sum_{i=1}^n\ind_{\{T_i\geq t_{\epsilon}\}}\\
&=O_p(1)C(\beta^{-1}+1)^2\epsilon,
\end{align*}
where the last equality follows from the definition of $t_{\epsilon}$. Since $\epsilon>0$ can be chosen arbitrarily small, we conclude that the second integral converges to zero in probability. Thus, by the dominated convergence theorem, we conclude the desired result $E_2=o_p(n^{-1/2})$.

\subsubsection{Proof of Theorem \ref{thm:Wild1}}
By Lemma \ref{lemma:Wild1}, under Assumption \ref{Assu:BoundedKernelSep},
\begin{align*}
n(\Psi_n^W)^2
&=\frac{1}{n}\sum_{i=1}^n\sum_{j=1}^n W_iW_jh((T_i,\Delta_i,X_i),(T_j,\Delta_j,X_j))+o_p(1),
\end{align*}
where 
\begin{align*}
h((T_i,\Delta_i,X_i),(T_j,\Delta_j,X_j))&=\Delta_i\Delta_j\bar{\mathfrak{K}}((T_i,X_i),(T_j,X_j)), 
\end{align*}
and $\bar{\mathfrak{K}}$ is defined in Equation \eqref{eqn:popkernel} in the main document. 

A simple computation shows that the kernel $h$ is degenerate under the null hypothesis: that is,
\begin{align*}
\E(h((T_1,\Delta_1,X_1),(t,\delta,x)))&=\delta\E(\Delta_1\bar{\mathfrak{K}}((T_1,X_1),(t,x)))=0
\end{align*}
holds for all $(t,\delta,x)\in(0,\tau)\times\{0,1\}\times\R^d$  under the null. Also, note that under Assumption \ref{Assu:BoundedKernelSep}, $\bar{\mathfrak{K}}$ is bounded, and thus $\E(|h((T_1,\Delta_1,X_1),(T_1,\Delta_1,X_1))|)<\infty$ and $\E(h((T_1,\Delta_1,X_1),(T_2,\Delta_2,X_2))^2)<\infty$.  Then, by Theorem \ref{Thm:ConsistencyWild}, there exists a distribution $\mathcal{L}$ such that
\begin{align*}
\frac{1}{n}\sum_{i=1}^n\sum_{j=1}^n\Delta_i\Delta_j\bar{\mathfrak{K}}((T_i,X_i),(T_j,X_j))\overset{\mathcal{D}}{\to}\mathcal{L}
\end{align*}
under the null hypothesis, and for almost every sequence $(t_i,\delta_i,x_i)_{i\geq 1}$ sampled from $(T_i,\Delta_i,X_i)_{i\geq 1}$,
\begin{align*}
\frac{1}{n}\sum_{i=1}^n\sum_{j=1}^nW_iW_j\delta_i\delta_j\bar{\mathfrak{K}}((t_i,x_i),(t_j,x_j))\overset{\mathcal{D}}{\to}\mathcal{L}.
\end{align*}

Finally, note that by Theorem \ref{Thm:SimplerKernel}, $n\Psi^2_n$ and $$\frac{1}{n}\sum_{i=1}^n\sum_{j=1}^n\Delta_i\Delta_j\bar{\mathfrak{K}}((T_i,X_i),(T_j,X_j))$$ converge to the same asymptotic distribution under the null hypothesis, which implies the desired result.

\subsubsection{Proof of Theorem \ref{Thm:PowerW}}
This proof is similar to the proof of consistency of HSIC in \cite{rindtconsistency}.
 Given any $\gamma>0$
\begin{align*}
\Prob(n\Psi_n^2< Q^W_{n,M})&=\Prob(n\Psi_n^2< Q^W_{n,M},n\Psi_n^2\leq \gamma)+\Prob(n\Psi_n^2< Q^W_{n,M},n\Psi_n^2> \gamma)\\
&\leq\Prob(n\Psi_n^2\leq \gamma)+\Prob(\gamma< Q^W_{n,M}).
\end{align*}
The first term is dealt with by Theorem \ref{Thm:Power}, from which we deduce $\limsup_{n\to\infty}\Prob(n\Psi_n^2\leq \gamma)\to0$ (since $n\Psi_n^2\to\infty$). 

For the second term, observe 
\begin{align*}
\Prob(\gamma< Q^W_{n,M})&\leq \Prob\left(\bigcup_{m\in\{1,\ldots,M\}}\left\{\gamma<(n(\Psi_n^W)^{2})_m\right\}\right)\leq  M \Prob\left(\gamma<n(\Psi_n^W)^{2}\right).
\end{align*}
By Markov's inequality 
\begin{align*}
\Prob\left(\gamma<n(\Psi_n^W)^{2}\right)\leq \frac{\E(n(\Psi_n^W)^{2})}{\gamma}.
\end{align*}
Recall that 
\begin{align*}
n(\Psi_n^W)^2
&=\frac{1}{n}\sum_{i=1}^n\sum_{j=1}^n W_iW_j\Delta_i\Delta_j\bar{\mathfrak{K}_n}((T_i,X_i),(T_j,X_j)),
\end{align*}
where $\bar{\mathfrak{K}}_n$ is the kernel defined in Theorem \ref{Thm:closed form} in the main document. Then,
\begin{align*}
&\Prob\left(\gamma<n(\Psi_n^W)^{2}\right)\\
&\leq \frac{1}{\gamma}\E(n(\Psi_n^W)^{2})\\
&=\frac{1}{\gamma}\E\left(\left.\E\left(\frac{1}{n}\sum_{i=1}^n\sum_{j=1}^n W_iW_j\Delta_i\Delta_j\bar{\mathfrak{K}_n}((T_i,X_i),(T_j,X_j))\right|(T_i,\Delta_i,X_i)_{i=1}^n\right)\right)\\
&=\frac{1}{\gamma}\E\left(\frac{1}{n}\sum_{i=1}^n \Delta_i\bar{\mathfrak{K}_n}((T_i,X_i),(T_i,X_i))\right),
\end{align*}
since $W_1,\ldots,W_n$ are Rademacher random variables. By Assumption \ref{Assu:BoundedKernelSep}, we have that  $\bar{\mathfrak{K}}_n((t,x),(t',x'))\leq C$ for all $(t,x),(t',x')\in \R\times\R^d$ for some constant $C>0$, and thus 
\begin{align*}
\Prob\left(\gamma<n(\Psi_n^W)^{2}\right)&\leq \frac{C}{\gamma}.
\end{align*}
Let $\delta>0$. By choosing $\gamma=MC/\delta$, it holds 
\begin{align*}
\Prob(\gamma<Q_{n,M}^W)\leq M\Prob\left(\gamma<n(\Psi_n^W)^{2}\right)\leq \frac{C\delta}{MC}=\delta.
\end{align*}
Since $\delta$ can be chosen arbitrarily small, and since $\limsup_{n\to\infty}\Prob(n\Psi_n^2\leq\gamma)=0$ for any $\gamma>0$, the result holds.

\subsubsection{Proof of Proposition \ref{proposition:computational_time}}
As to the computational cost of $n\Psi_n^2,$ note first that \begin{align*}
    \tr(\boldsymbol L^\Delta( \boldsymbol I-\boldsymbol A)\boldsymbol K(\boldsymbol I-\boldsymbol A)^{\intercal}) = \sum_{i,j=1}^n [\{ \boldsymbol L^\Delta(\boldsymbol I-\boldsymbol A) \} \circ \{ (\boldsymbol I-\boldsymbol A)\boldsymbol K]_{ij} ,
\end{align*}
where $\circ$ denotes the elementswise product. The matrix $\boldsymbol{AK}$ can be computed in $\mathcal{O}(n^2)$ because, after sorting the data, $\boldsymbol A$ is upper triangular, with the nonzero entries equal along each row. The same observation allows for fast computation of the matrix products in $n\Psi_n^2$ and $(n\Psi_n^2)^{W}$.

\newpage
\section{Proofs auxiliary results}\label{sec:proofsaux}
\subsection{Proof of Proposition \ref{prop:ess sup}}
Recall $\tau_x=\sup\{t:S_{T|X=x}(t)>0\}$ and $\tau=\sup\left\{t:S_T(t)>0\right\}$.  We start by proving $\tau'\leq\tau$. Assume otherwise, $\tau<\tau'$, then there exists $\epsilon>0$ such that $B=\{x:\tau_x>\tau+\epsilon\}$ and $F_X(B)>0$. Let 
\begin{align*}
B_{1/n}&=\left\{x:S_{T|X=x}(\tau+\epsilon)>\frac{1}{n}\right\},
\end{align*}
and observe $B=\bigcup_{n\geq 1}B_{1/n}$. Since $F_{X}(B)>0$, by the union bound, we deduce that there exists $n\geq 1$ such that $F_X(B_{1/n})>0$. We then arrive at the contradiction
\begin{align*}
0=\int_{\R^d}S_{T|X=x}(\tau+\epsilon)dF_{X}(x)
&\geq \int_{B_{1/n}}\frac{1}{n} dF_{X}(x)\geq \frac{1}{n}F_{X}(B_{1/n})>0,
\end{align*}
from which we deduce that $\tau'\leq \tau$.

We continue by proving $\tau'\geq \tau$. Assume that $\tau'<\tau$, then
\begin{align*}
0<S_T(\tau')=\int_{\R^d}S_{T|X=x}(\tau')dF_{X}(x)=0,
\end{align*}
from which we deduce $\tau\leq\tau'$, and thus $\tau'=\tau$.

\subsection{ Proof of Proposition \ref{prop:nu_0approx}}
\begin{proof}
Recall $d\nu_0^n(s,x)=\frac{1}{n}\sum_{j=1}^n\sum_{i=1}^n\Delta_j\delta_{T_j}(s)\frac{Y_i(s)}{Y(s)}\delta_{X_i}(x)$. Then, the result follows from proving

\begin{align*}
D_n=\frac{1}{n^2}\sum_{j=1}^n\sum_{i=1}^n\Delta_j\omega(T_j,X_i)Y_i(T_j)\left(\frac{n}{Y(T_j)}-\frac{1}{S_T(T_j)}\right)=o_p(1).
\end{align*}

Notice that $D_n$ can be written as
\begin{align*}
D_n=\frac{1}{n^2}\int_{0}^{\tau_n}\sum_{i=1}^n\omega(s,X_i)Y_i(s)\left(\frac{n}{Y(s)}-\frac{1}{S_T(s)}\right)dN(s).
\end{align*}
Let $\epsilon>0$, and define  $t_{\epsilon}\geq 0$ such that $S_{T}(t_{\epsilon})=\epsilon$ (notice that $t_{\epsilon}\geq 0$ exists since the distribution over the times is continuous). We decompose $D_n$ into two integrals, one considering integration over $\{s\leq t_{\epsilon}\}$ and the other one over $\{s>t_{\epsilon}\}$, and we prove that both integrals converge to zero in probability.

For the first integral, observe that
\begin{align*}
&\frac{1}{n^2}\int_{0}^{t_\epsilon}\sum_{i=1}^n\omega(s,X_i)Y_i(s)\left(\frac{n}{Y(s)}-\frac{1}{S_T(s)}\right)dN(s)\\
&\quad\leq \sup_{s\leq t_{\epsilon}}\left|\frac{n}{Y(s)}-\frac{1}{S_T(s)}\right|\frac{1}{n^2}\sum_{j=1}^n\sum_{i=1}^n|\omega(T_j,X_i)|=o_p(1),
\end{align*}
where the last equality follows by noticing the supremum converges to zero by Proposition \ref{lemma:supConverSH}.2., and that  $Y_i(s)\leq 1$ and $\omega$ are bounded.

For the second integral, it holds
\begin{align*}
&\frac{1}{n^2}\int_{t_{\epsilon}}^{\tau_n}\sum_{i=1}^n\omega(s,X_i)Y_i(s)\left(\frac{n}{Y(s)}-\frac{1}{S_T(s)}\right)dN(s)\\
&\leq \frac{C}{n}\int_{t_{\epsilon}}^{\tau_n}\sum_{i=1}^n\frac{Y_i(s)}{Y(s)}\left|1-\frac{Y(s)/n}{S_T(s)}\right|dN(s)\\
&\quad=O_p(1)\frac{1}{n}\int_{t_{\epsilon}}^{\tau_n}dN(s)=O_p(1)\frac{1}{n}\sum_{j=1}^n\ind_{\{t_{\epsilon}\geq T_j\}}=O_p(1)\epsilon,
\end{align*}
where the second line follows from the assumption $|\omega|\leq C$ for some constant $C>0$, the first equality in the third line is due to Proposition \ref{prop:ProbBounds}.\ref{prop:ProbBound2}, and the last equality is due to the definition of $t_{\epsilon}$. Since $\epsilon>0$ can be chosen arbitrarily small, the result holds.
\end{proof}

\subsection{Proof of Theorem \ref{Thm:Eigen}}

\begin{proof}
Consider a re-parametrisation of the kernel $J$ (given in Equation \eqref{eqn:Jkernel} in the main document) in terms of the augmented space $(Z,C,X)$, as $J:(\R_+\times\R_+\times\R^d)^2\to\R$ given by
\begin{align*}
J((Z_1,C_1,X_1),(Z_2,C_2,X_2))&=\int_{\R_+}\int_{\R_+}\bar{\mathfrak{K}}((s_1,X_1),(s_2,X_2))dM_1(s_1)dM_2(s_2),
\end{align*} 
where $dM_1(s_1)=\ind_{\{Z_1\leq C_1\}}\delta_{Z_1}(s_1)-\ind_{\{\min\{Z_1,C_1\}\}\geq s_1}d\Lambda_Z(s_1)$.  Notice that to evaluate $J$ in the previous definition we do not need extra information, it only needs $(T_1,\Delta_1,X_1)$ and $(T_2,\Delta_2,X_2)$ to be evaluated, but it is more convenient theoretically.

From the definition of the backward operator $B$, and Property \ref{appe:BackFoward}.5  (see Section \ref{appe:BackFoward}), we deduce  
\begin{align}
J((Z_1,C_1,X_1),(Z_2,C_2,X_2))=(B_1B_2 Q)((Z_1,C_1,X_1),(Z_2,C_2,X_2)),
\end{align}
where  $B_1$ and $B_2$ denote the operator $B$ applied to $(Z_1,C_1,X_1)$ and $(Z_2,C_2,X_2)$, respectively; the same definition holds for $A_1$ and $A_2$. Thus, 
\begin{align*}
(T^Jf)(Z_1,C_1,X_1)&=\E_2\left((B_1B_2Q)((Z_1,C_1,X_1),(Z_2,C_2,X_2))f(Z_2,C_2,X_2)\right).
\end{align*}

Using property \ref{appe:BackFoward}.1, and the linearity of $B_1$, we find that
\begin{align*}
(T^Jf)(Z_1,C_1,X_1)&=\E_2\left((B_1Q)((Z_1,C_1,X_1),(Z_2,C_2,X_2))(Af)(Z_2,C_2,X_2)\right)\\
&=B\E_2\left(Q((Z_1,C_1,X_1),(Z_2,C_2,X_2))(Af)(Z_2,C_2,X_2)\right)\\
&=B(T^Q(Af)),
\end{align*}
where $T^Q$ is the integral operator associated to the kernel $Q$.

Let $(\lambda_i,f_i)$, with $\lambda_i\neq 0$, be an eigenpair of $T^J$, then we claim that $(\lambda_i, Af_i)$ is a eigenpair of $T^Q$. Indeed, $$T^QAf_i=ABT^QAf_i=AT^Jf_i=\lambda_iAf_i,$$
where in the first equality we use property \ref{appe:BackFoward}.2. Also, if $(\lambda_j,f_j)$ is another eigenpair such that $\langle f_i,f_j\rangle_{\mathcal{L}_2}=0$, then $$\langle Af_i,Af_j\rangle_{\mathcal{L}_2}=\langle f_i,BAf_j\rangle_{\mathcal{L}_2}=\langle f_i,f_j-\E(f_j|C_1,X_1)\rangle_{\mathcal{L}_2}=0,$$
since $\E(f_j|C_1,X_1)=\frac{1}{\lambda_j}\E(T^Jf_j|C_1,X_1)=\frac{1}{\lambda_j}\E(B(T^QAf_j)|C_1,X_1)=0$ by property \ref{appe:BackFoward}.4. Hence, we conclude that for every different eigenpair $(\lambda_i,f_i)$ belonging to $T^J$, there exists a corresponding pair $(\lambda_i,Af_i)$ associated to $T^Q$. 

Conversely for any eigenpair $(\lambda_i', g_i)$ associated to $T^Q$, we claim that $(\lambda_i',Bg_i)$ is a eigenpair for $T^J$. To see this, observe $$T^JBg_i=BT^QABg_i=BT^Qg_i=\lambda_i'Bg_i.$$
Similarly, if $(\lambda_j',g_j)$ is another eigenpair with $\lambda_j\neq 0$, and such that $\langle g_i,g_j\rangle_{\mathcal{L}_2}=0$, then
$$\langle Bg_i,Bg_j\rangle_{\mathcal{L}_2}=\langle g_i,ABg_j\rangle_{\mathcal{L}_2}=\langle g_i,g_j\rangle_{\mathcal{L}_2}=0.$$
Hence, we conclude that for every different eigenpair $(\lambda_i',g_i)$ belonging to $T^Q$, there exists a corresponding pair $(\lambda_i',Bg_i)$ associated to $T^Q$. We conclude that $T^Q$ and $T^J$ have the same set of non-zero eigenvalues including multiplicities.
\end{proof}

\subsection{Proof of Lemma \ref{Lemma:ReplacementYP}}

\begin{proof}
We only prove the result for $k=1$ as the other result follows by using the same arguments. 

Let $\epsilon>0$ and choose $t_{\epsilon}\geq 0$ such that $S_T(t_{\epsilon})=\epsilon$. We split the integral into two integrals, one over $\{t\leq t_{\epsilon}\}$,  and other $\{t> t_{\epsilon}\}$.  For the integral, observe that
\begin{align*}
\frac{1}{n}\int_{0}^{t_\epsilon}\left|1-\frac{Y(t)/n}{S_T(t)}\right|dN(t)&\leq\sup_{t\leq t_{\epsilon}}\left|\frac{n}{Y(t)}-\frac{1}{S_T(t)}\right|\frac{1}{n}\int_{0}^{t_{\epsilon}}dN(t)=o_p(1),
\end{align*}
since $\frac{1}{n}\int_0^{t_{\epsilon}}dN(t)\leq 1$ for all $n\geq 1$, and $\sup_{t\leq t_{\epsilon}}\left|\frac{n}{Y(t)}-\frac{1}{S_T(t)}\right|\to 0$ by Proposition \ref{lemma:supConverSH}.2. For the second integral, observe that
\begin{align*}
\frac{1}{n}\int_{t_{\epsilon}}^{\tau}\left|1-\frac{Y(t)/n}{S_T(t)}\right|dN(t)&=O_p(1)\frac{1}{n}\int_{0}^{\tau_n}\ind_{\{t> t_{\epsilon}\}}dN(t)\\
&=O_p(1)\frac{1}{n}\sum_{i=1}^n\Delta_i\ind_{\{T_i> t_{\epsilon}\}}\\
&=O_p(1)\epsilon,
\end{align*}
where the first equality is due to $Y(t)/n=O_p(1)S_T(t)$ uniformly for all $t\leq \tau_n$ by Proposition \ref{prop:ProbBounds}.\ref{prop:ProbBound2}, and the third equality follows from Markov's inequality and the definition of $t_{\epsilon}$. Since $\epsilon>0$ is arbitrary, we deduce equation \eqref{eqn:intdN}.
\end{proof}


\end{document}

%% file: defsProb.tex
\theoremstyle{plain}
\newtheorem{theorem}{Theorem}[section]
\newtheorem{lemma}[theorem]{Lemma}

\newtheorem{proposition}[theorem]{Proposition}

\theoremstyle{definition}
\newtheorem{definition}[theorem]{Definition}
\newtheorem{example}[theorem]{Example}%
\theoremstyle{remark}
\newtheorem*{remark}{Remark}

\newcommand{\proofstring}{Proof}
\renewenvironment{proof}[1][]{\noindent\textbf{\proofstring\ifthenelse{\equal{#1}{}}{:}{~(#1) :}}\xspace}{\hfill$\blacksquare$\medskip\par}

\newcommand{\R}{\mathds{R}}

\newcommand{\E}{\mathds{E}}

\newcommand{\Prob}{\mathds{P}}
\newcommand{\Ind}{\mathds{1}}
\newcommand{\ind}{\Ind}


%% file: Independence-RKHS-Arxiv.bbl
\begin{thebibliography}{10}

\bibitem{aalen1978nonparametric}
Odd Aalen.
\newblock Nonparametric estimation of partial transition probabilities in
  multiple decrement models.
\newblock {\em Ann. Statist.}, 6(3):534--545, 1978.

\bibitem{meynaoui19Adaptive}
Mélisande Albert, Béatrice Laurent, Amandine Marrel, and Anouar Meynaoui.
\newblock Adaptive test of independence based on {HSIC} measures.
\newblock {\em arXiv preprint arXiv:1902.06441}, 2021.

\bibitem{Bagdonavivius2010}
Vilijandas Bagdonavi\v{c}ius, Julius Kruopis, and Mikhail~S. Nikulin.
\newblock {\em Non-parametric tests for censored data}.
\newblock ISTE, London; John Wiley \& Sons, Inc., Hoboken, NJ, 2011.

\bibitem{berlinet2011reproducing}
Alain Berlinet and Christine Thomas-Agnan.
\newblock {\em Reproducing kernel {H}ilbert spaces in probability and
  statistics}.
\newblock Kluwer Academic Publishers, Boston, MA, 2004.
\newblock With a preface by Persi Diaconis.

\bibitem{ChwGre14}
Kacper Chwialkowski and Arthur Gretton.
\newblock A kernel independence test for random processes.
\newblock In {\em ICML'14: Proceedings of the 31st International Conference on
  International Conference on Machine Learning}, page II–1422–II–1430.
  Proceedings of Machine Learning Research, 2014.

\bibitem{cox1972regression}
David~R. Cox.
\newblock Regression models and life-tables.
\newblock {\em J. Roy. Statist. Soc. Ser. B}, 34:187--220, 1972.

\bibitem{dehling1994random}
Herold Dehling and Thomas Mikosch.
\newblock Random quadratic forms and the bootstrap for {$U$}-statistics.
\newblock {\em J. Multivariate Anal.}, 51(2):392--413, 1994.

\bibitem{denk1997videoendoscopic}
Doris-Maria Denk and Alexandra Kaider.
\newblock Videoendoscopic biofeedback: a simple method to improve the efficacy
  of swallowing rehabilitation of patients after head and neck surgery.
\newblock {\em ORL}, 59(2):100--105, 1997.

\bibitem{dunkler2018weighted}
Daniela Dunkler, Meinhard Ploner, Michael Schemper, and Georg Heinze.
\newblock Weighted cox regression using the {R} package {coxphw}.
\newblock {\em Journal of Statistical Software, Articles}, 84(2):1--26, 2018.

\bibitem{Efron1990}
Bradley Efron and Iain~M. Johnstone.
\newblock Fisher's information in terms of the hazard rate.
\newblock {\em Ann. Statist.}, 18(1):38--62, 1990.

\bibitem{fernandez2018kaplan}
Tamara Fern\'{a}ndez and Nicol\'{a}s Rivera.
\newblock Kaplan-{M}eier {V}- and {U}-statistics.
\newblock {\em Electron. J. Stat.}, 14(1):1872--1916, 2020.

\bibitem{fernandez2019reproducing}
Tamara Fernández and Nicolás Rivera.
\newblock A reproducing kernel hilbert space log-rank test for the two-sample
  problem.
\newblock {\em Scandinavian Journal of Statistics}, 2021.

\bibitem{Flemming91}
Thomas~R. Fleming and David~P. Harrington.
\newblock {\em Counting processes and survival analysis}.
\newblock Wiley Series in Probability and Mathematical Statistics: Applied
  Probability and Statistics. John Wiley \& Sons, Inc., New York, 1991.

\bibitem{gares2015omnibus}
Val\'{e}rie Gar\`es, Sandrine Andrieu, Jean-Fran\c{c}ois Dupuy, and Nicolas
  Savy.
\newblock An omnibus test for several hazard alternatives in prevention
  randomized controlled clinical trials.
\newblock {\em Stat. Med.}, 34(4):541--557, 2015.

\bibitem{gill1980censoring}
R.~D. Gill.
\newblock {\em Censoring and stochastic integrals}, volume 124 of {\em
  Mathematical Centre Tracts}.
\newblock Mathematisch Centrum, Amsterdam, 1980.

\bibitem{gill1983large}
Richard Gill.
\newblock Large sample behaviour of the product-limit estimator on the whole
  line.
\newblock {\em Ann. Statist.}, 11(1):49--58, 1983.

\bibitem{gray1992}
Robert~J. Gray.
\newblock Flexible methods for analyzing survival data using splines, with
  applications to breast cancer prognosis.
\newblock {\em Journal of the American Statistical Association},
  87(420):942--951, 1992.

\bibitem{gretton2015simpler}
Arthur Gretton.
\newblock A simpler condition for consistency of a kernel independence test.
\newblock {\em arXiv preprint arXiv:1501.06103}, 2015.

\bibitem{gretton2012kernel}
Arthur Gretton, Karsten~M. Borgwardt, Malte~J. Rasch, Bernhard Sch\"{o}lkopf,
  and Alexander Smola.
\newblock A kernel two-sample test.
\newblock {\em J. Mach. Learn. Res.}, 13:723--773, 2012.

\bibitem{GreFukTeoSonetal08}
Arthur Gretton, Kenji Fukumizu, Choon~Hui Teo, Le~Song, Bernhard Sch\"{o}lkopf,
  and Alexander~J. Smola.
\newblock A kernel statistical test of independence.
\newblock In {\em Proceedings of the 20th International Conference on Neural
  Information Processing Systems}, NIPS'07, page 585–592, Red Hook, NY, USA,
  2007. Curran Associates Inc.

\bibitem{gretton2012optimal}
Arthur Gretton, Dino Sejdinovic, Heiko Strathmann, Sivaraman Balakrishnan,
  Massimiliano Pontil, Kenji Fukumizu, and Bharath~K. Sriperumbudur.
\newblock Optimal kernel choice for large-scale two-sample tests.
\newblock In F.~Pereira, C.~J.~C. Burges, L.~Bottou, and K.~Q. Weinberger,
  editors, {\em Advances in Neural Information Processing Systems}, volume~25,
  pages 1205--1213. Curran Associates, Inc., 2012.

\bibitem{harrington1982class}
David~P. Harrington and Thomas~R. Fleming.
\newblock A class of rank test procedures for censored survival data.
\newblock {\em Biometrika}, 69(3):553--566, 1982.

\bibitem{kossler2010max}
Wolfgang K\"{o}ssler.
\newblock Max-type rank tests, {$U$}-tests, and adaptive tests for the
  two-sample location problem---an asymptotic power study.
\newblock {\em Comput. Statist. Data Anal.}, 54(9):2053--2065, 2010.

\bibitem{laurie1989surgical}
John~A Laurie, Charles~G Moertel, Thomas~R Fleming, Harry~S Wieand, John~E
  Leigh, Jebal Rubin, Greg~W McCormack, James~B Gerstner, James~E Krook, and
  James Malliard.
\newblock Surgical adjuvant therapy of large-bowel carcinoma: an evaluation of
  levamisole and the combination of levamisole and fluorouracil. the north
  central cancer treatment group and the mayo clinic.
\newblock {\em Journal of Clinical Oncology}, 7(10):1447--1456, 1989.

\bibitem{le1994association}
Chap~T. Le, Patricia~M. Grambsch, and Thomas~A. Louis.
\newblock Association between survival time and ordinal covariates.
\newblock {\em Biometrics}, 50(1):213--219, 1994.

\bibitem{LiuXuLuZhaGreSut20}
Feng Liu, Wenkai Xu, Jie Lu, Guangquan Zhang, Arthur Gretton, and Danica~J.
  Sutherland.
\newblock Learning deep kernels for non-parametric two-sample tests.
\newblock In {\em ICML '20: Proceedings of the 37th International Conference on
  Machine Learning}, pages 6316--6326, 2020.

\bibitem{mantel1966evaluation}
Nathan Mantel.
\newblock Evaluation of survival data and two new rank order statistics arising
  in its consideration.
\newblock {\em Cancer chemotherapy reports}, 50(3):163, 1966.

\bibitem{mckeague1995omnibus}
Ian~W. McKeague, A.~M. Nikabadze, and Yan~Qing Sun.
\newblock An omnibus test for independence of a survival time from a covariate.
\newblock {\em Ann. Statist.}, 23(2):450--475, 1995.

\bibitem{moertel1995fluorouracil}
Charles~G Moertel, Thomas~R Fleming, John~S Macdonald, Daniel~G Haller, John~A
  Laurie, Catherine~M Tangen, James~S Ungerleider, William~A Emerson,
  Douglass~C Tormey, John~H Glick, et~al.
\newblock Fluorouracil plus levamisole as effective adjuvant therapy after
  resection of stage iii colon carcinoma: a final report.
\newblock {\em Annals of internal medicine}, 122(5):321--326, 1995.

\bibitem{neyman1933ix}
Jerzy Neyman and Egon~S. Pearson.
\newblock On the problem of the most efficient tests of statistical hypotheses.
\newblock {\em Philosophical Transactions of the Royal Society of London.
  Series A, Containing Papers of a Mathematical or Physical Character},
  231:289--337, 1933.

\bibitem{peto1972asymptotically}
Richard Peto and Julian Peto.
\newblock Asymptotically efficient rank invariant test procedures.
\newblock {\em Journal of the Royal Statistical Society. Series A (General)},
  135(2):185--207, 1972.

\bibitem{rindtconsistency}
David Rindt, Dino Sejdinovic, and David Steinsaltz.
\newblock Consistency of permutation tests of independence using distance
  covariance, hsic and dhsic.
\newblock {\em Stat}, page e364.

\bibitem{rindt2019nonparametric}
David Rindt, Dino Sejdinovic, and David Steinsaltz.
\newblock A kernel-and optimal transport-based test of independence between
  covariates and right-censored lifetimes.
\newblock {\em The International Journal of Biostatistics}, 1(ahead-of-print),
  2020.

\bibitem{serfling2009approximation}
Robert~J. Serfling.
\newblock {\em Approximation theorems of mathematical statistics}.
\newblock John Wiley \& Sons, Inc., New York, 1980.
\newblock Wiley Series in Probability and Mathematical Statistics.

\bibitem{sriperumbudur2011universality}
Bharath~K. Sriperumbudur, Kenji Fukumizu, and Gert R.~G. Lanckriet.
\newblock Universality, characteristic kernels and {RKHS} embedding of
  measures.
\newblock {\em J. Mach. Learn. Res.}, 12:2389--2410, 2011.

\bibitem{Univer}
Bharath~K. Sriperumbudur, Kenji Fukumizu, and Gert R.~G. Lanckriet.
\newblock Universality, characteristic kernels and {RKHS} embedding of
  measures.
\newblock {\em J. Mach. Learn. Res.}, 12:2389--2410, 2011.

\bibitem{SutTunStretal17}
Danica~J. Sutherland, Hsiao-Yu Tung, Heiko Strathmann, Soumyajit De, Aaditya
  Ramdas, Alex Smola, and Arthur Gretton.
\newblock Generative models and model criticism via optimized maximum mean
  discrepancy.
\newblock In {\em 5th International Conference on Learning Representations,
  {ICLR} 2017}, 2017.

\bibitem{tarone1981distribution}
Robert~E. Tarone.
\newblock On the distribution of the maximum of the logrank statistic and the
  modified wilcoxon statistic.
\newblock {\em Biometrics}, 37(1):79--85, 1981.

\bibitem{therneau2015package}
Terry~M Therneau and Thomas Lumley.
\newblock Package ‘survival’.
\newblock {\em R Top Doc}, 128(10):28--33, 2015.

\bibitem{largescaleindependence}
Qinyi Zhang, Sarah Filippi, Arthur Gretton, and Dino Sejdinovic.
\newblock Large-scale kernel methods for independence testing.
\newblock {\em Stat. Comput.}, 28(1):113--130, 2018.

\bibitem{zucker1990}
David~M. Zucker and Alan~F. Karr.
\newblock Nonparametric survival analysis with time-dependent covariate
  effects: a penalized partial likelihood approach.
\newblock {\em Ann. Statist.}, 18(1):329--353, 1990.

\end{thebibliography}
